\documentclass[12pt,preprint]{aastex}

\begin{document}
\title{Radiation Mechanisms and Physical Properties of GeV-TeV BL Lac Objects}
\author{Jin Zhang\altaffilmark{1,2}, En-Wei Liang\altaffilmark{3}, Shuang-Nan Zhang\altaffilmark{4,1}, J. M. Bai\altaffilmark{5,6}}
\altaffiltext{1}{National Astronomical Observatories, Chinese Academy of Sciences, Beijing, 100012, China;
zhang.jin@hotmail.com}\altaffiltext{2}{College of Physics and Electronic Engineering, Guangxi Teachers Education
University, Nanning, 530001, China} \altaffiltext{3}{Department of Physics and GXU-NAOC Center for Astrophysics
and Space Sciences, Guangxi University, Nanning, 530004, China} \altaffiltext{4}{Key Laboratory of Particle
Astrophysics, Institute of High Energy Physics, Chinese Academy of Sciences, Beijing 100049,
China}\altaffiltext{5}{National Astronomical Observatories/Yunnan Observatory, Chinese Academy of Sciences,
Kunming, 650011, China}\altaffiltext{6}{Key Laboratory for the Structure and Evolution of Celestial Objects, Chinese Academy of Sciences, Kunming 650011, China}

\begin{abstract}
BL Lac objects are the best candidates to study the jet properties since their spectral energy distributions (SEDs) are less contaminated by the emission from the accretion disk and external Compton processes. We compile the broadband SEDs observed with {\em Fermi}/LAT and other instruments from literature for 24 TeV BL Lac objects. Two SEDs, which are identified as a low or high state according to its flux density at $1\ {\rm TeV}$, are available for each of ten objects. These SEDs can be explained well with the synchrotron+synchrotron-self-Compton model. We constrain the magnetic filed strength ($B$) and the Doppler factor ($\delta$) of the radiation region by incorporating the $\chi^{2}$-minimization technique and the $\gamma$-ray transparency condition. Twenty-four out of the 34 SEDs can constrain $B$ and $\delta$ in 1 $\sigma$ significance level and others only present the limits for both $B$ and $\delta$. The typical values of $B$ and $\delta$ are $0.1\sim 0.6$ G and $10\sim 35$, respectively. The derived values of $\gamma_{\rm b}$ are significantly different among sources and even among the low and high states of a given source. Prominent flux variations with a clear spectral shift
are observed and a tentative correlation between the ratio of the flux density at 1 TeV and the ratio of $\gamma_{\rm b}$ in the low and
high states is presented, likely indicating that the relativistic shocks for the electron acceleration may be responsible for the flux variations and the spectral shift. A weak anti-correlation between the jet power and the mass of the central black hole is observed, i.e., $P_{\rm jet}\propto M^{-1}_{\rm BH}$, which disfavors the scenario of a pure accretion-driven jet. Implications for the blazar sequence and the intergalactic magnetic field from our results are also briefly discussed.
\end{abstract}


\section{Introduction}           
\label{sect:intro} The broadband spectral energy distributions (SEDs) of blazars are double-peaked. Generally, the bump at the
IR-optical-UV band is explained with the synchrotron process of relativistic electrons accelerated in the jets,
and the bump at the GeV-TeV gamma-ray band is due to the inverse Compton (IC) scattering of the same electron
population (e.g., Ghisellini et al. 1996; Urry 1999). The seed photon field may be from the synchrotron
radiation themselves (the so-called SSC model; Maraschi et al. 1992, Ghisellini et al. 1996) or from external
radiation fields (EC), such as the broad line region (BLR; Sikora et al. 1994), accretion disk (Dermer et al.
1992), torus (BLazejowski et al. 2000), and cosmic microwave background (CMB; Tavecchio et al. 2000). The low
energy bump is well sampled with the data from multi-wavelength campaigns at the radio, optical, and X-ray
bands. The observations with CGRO/EGRET and ground-based TeV gamma-ray telescopes sketch the feature of the high
energy bump for some sources, but they are still poorly constrained. The observations with {\em Fermi}/LAT, which covers an energy band from 20 MeV to $\sim 300$ GeV, together
with the ground-based observations at the TeV gamma-ray band, now pin down both the luminosity and
the frequency of high energy peak well in the observed SEDs for some blazars. Broadband
SEDs obtained simultaneously or quasi-simultaneously are critical to reveal the radiation mechanisms.

About forty active galactic nuclei (AGNs) have been detected in the TeV gamma-ray band since the first detection of
TeV gamma-rays from Mkn 421 with the Whipple imaging atmospheric-Cherenkov telescopes in 1992 (Punch et al.
1992) and most of them are blazars. Observations show that these sources have violent variability in multi-frequency,
especially at the X-ray and the TeV bands (Buckley et al. 1996; Takahashi et al. 2000; Sambruna et al. 2000).
The flux variations at the X-ray and TeV bands are correlated, such as that observed in Mkn 421 (Blazejowski
et al. 2005). Some sources detected by EGRET are not confirmed by {\em Fermi}/LAT, and some in the 1FGL (1st LAT
Catalog) with fluxes much over the EGRET threshold were not detected by EGRET (Abdo et al. 2010a).
Three TeV AGNs still have not be detected by {\em Fermi}/LAT in the 1FGL and in the 2FGL (Ackermann et al. 2011).
These facts indicate that AGNs should have
significant variability in the GeV-TeV band. In the high state of TeV emission, the peak frequency ($\nu_{\rm
s}$) of the synchrotron radiation in the SEDs moves to higher energies for some sources, showing a tendency
that a higher TeV flux corresponds to a harder spectrum for the emission at the X-ray and gamma-ray bands,
as observed in Mkn 501 (Anderhub et al. 2009a). It is
unclear what may be responsible for the flux variation and the corresponding spectral shift. The abundant data observed with {\em Fermi}/LAT in the MeV-GeV band, together with the TeV gamma-ray data,
now provide an excellent opportunity to reveal the physical origin of the temporal and spectral variations of
these TeV sources.

It is generally believed that the jets of AGNs are powered by the central massive black hole (BH), whereas the mechanism of the jet
production is still a mystery. The proposed models can be simplified as two types, i.e., accretion and rotation
of the BH. The connection between jet and accretion disk has been widely investigated (Ulvestad \& Ho 2001;
Marscher et al. 2002; Merloni et al. 2003; Falcke et al. 2004; King et al. 2011). With a sample of 23 blazars in
the three months of Fermi satellite survey, Ghisellini et al. (2009) found that the jet power is slightly larger
than the disk luminosity and proportional to the mass accretion rate. However, the accretion process cannot
explain the radio loudness of AGN, and thus the rotation of the BH is suggested to be the possible ingredient (Sikora et al. 2007;
Lagos et al. 2009). In fact, there is some evidence for rapidly spinning BH (Cao \& Li 2008; Wu et al. 2011).
The jets of AGNs may be powered by both the accretion process and the spin of the central BH (Fanidakis et al. 2011).
The SEDs of BL Lac objects dominated by jet emission suffer less contamination of the emission from accretion
disk and EC process. Therefore, they are the best candidates for investigating the jet properties and the
relation to the central BHs.

In this paper, we compile a sample of SEDs for TeV BL Lac objects to study their physical properties of the jets and explore the physical
reasons that result in the flux variation and the spectral shift in these sources as well as the relations of jet
production to the accretion and rotation of the central BHs. The sample selection and the observed SEDs are presented in \S 2.
 The model, the strategy, and the results of SED fitting are described in \S 3. The physical properties derived from our SED fitting are presented in \S 4. A summary for our results is in $\S$5. A simple discussion on implication of the blazar sequence and on the intergalactic magnetic field is given in \S 6.

\section{Sample Selection}
\label{sect:data}
All TeV BL Lac objects that have {\em Fermi}/LAT detections are included in our sample, except for six sources that have only upper limits of {\em Fermi}/LAT detections. Twenty-four BL Lac objects are included in our sample as listed in Table 1. We compile their broadband SEDs that were simultaneously or quasi-simultaneously observed
with {\em Fermi}/LAT and other instruments from literature. Only one broadband SED is constructed for 14 objects and two broadband SEDs are available for the other ten objects. The two well-sampled SEDs
for the ten sources are defined as a high or low state with the observed or
extrapolated flux density at 1 TeV. The observed broadband SEDs are shown in Fig. \ref{Fig:1}. The description
of each source is summarized in Appendix.

\section{Modeling the SEDs}
\subsection{Model}
 Although the contributions of EC component may be involved for some objects (e.g., 3C 66A, Abdo et al. 2011), the external photon fields are usually much weaker than the synchrotron radiation photon field for the BL Lac objects\footnote{The fitting results with the SSC+EC model for some sources reported in the literatures are summarized in Appendix.}. Multi-zone syn+SSC models are also used to explain the SEDs of some sources, such as Mkn 421 (Blazejowski et al. 2005). For seeking the consistency among objects in our statistical analysis, we only consider the simple single-zone syn+SSC model to fit the observed broadband SEDs in this paper.

In the process of the SED modeling, the radiation region is taken as a homogeneous sphere with radius $R$. The electron distribution is assumed as a broken power law with indices $p_1$ and $p_2$ below and above the break energy $\gamma_{\rm b} m_{\rm e}c^2$,
\begin{equation}
N(\gamma )= N_{0}\left\{ \begin{array}{ll}
\gamma ^{-p_1}  &  \mbox{ $\gamma_{\rm min}\leq\gamma \leq \gamma _{\rm b}$}, \\
\gamma _{\rm b}^{p_2-p_1} \gamma ^{-p_2}  &  \mbox{ $\gamma _{\rm b} <\gamma <\gamma _{\rm max} $,}
\end{array}
\right.
\end{equation}
where $p_{1,2}=2\alpha_{1,2}+1$, and $\alpha_{1,2}$ are the observed spectral indices below and above the synchrotron radiation peak.
The synchrotron radiation peak frequency of a single electron with Lorentz factor $\gamma$ is
\begin{equation}
\nu_{\rm syn}=3.7\times10^{6}B\gamma^{2}\frac{\delta}{1+z}.
\end{equation}
The synchrotron emissivity $\epsilon_{\rm s}(\nu)$
is calculated with
\begin{equation}
\epsilon_{\rm s}(\nu)=\frac{1}{4\pi}\int_{\gamma_{\rm min}}^{\gamma_{\rm max}}d
\gamma N(\gamma)P_{\rm s}(\nu, \gamma),
\end{equation}
where $P_{\rm s}(\nu, \gamma)$ is the single electron synchrotron
emissivity averaged over an isotropic distribution of pitch
angles (see e.g., Crusius \& Schlickeiser 1986;
Ghisellini et al. 1988). The synchrotron radiation field $I_{\rm s}(\nu)$ is
calculated by the transfer equation,
\begin{equation}
I_{\rm s}(\nu)=\frac{\epsilon_{\rm s}(\nu)}{k(\nu)}[1-e^{-k(\nu)R}],
\end{equation}
where $k(\nu)$ is the absorption coefficient (e.g., Ghisellini \& Svensson 1991).

In the SSC scenario, the relativistic electrons interact with
synchrotron radiation photons through the IC scattering. The IC
emissivity is calculated by
\begin{equation}\label{ec}
\epsilon_{\rm c}(\nu_{\rm c})=\frac{\sigma_{\rm T}}{4}\int_{\nu_{\rm i}^{\rm min}}^{\nu_{\rm i}^{\rm max}}
\frac{d\nu_{\rm i}}{\nu_{\rm i}}\int_{\gamma_1}^{\gamma_2}\frac{d\gamma}{\gamma^2\beta^2}N(\gamma)f(\nu_{\rm i},\nu_{\rm c})\frac{\nu_{\rm c}}{\nu_{\rm i}}I_{\rm s}(\nu_{\rm i}),
\end{equation}
where $\nu_{\rm i}$ is the frequency of the incident photons emitted by
the synchrotron radiation between $\nu_{\rm i}^{\rm min}$ and $\nu_{\rm i}^{\rm max}$,
$\beta=v/c$, $\gamma_1$ and $\gamma_2$ are the lower and upper
limits of the scattering electrons, and $f(\nu_{\rm i},\nu_{\rm c})$ is the
spectrum produced by scattering monochromatic photons of frequency
$\nu_{\rm i}$ with a single electron.
The medium is transparent for the IC radiation field, hence $I_{\rm c}(\nu_{\rm c})=\epsilon_{\rm c}(\nu_{\rm c})R$.

Assuming that $I_{\rm s,c}$ is an isotropic radiation field, the
monochromatic luminosity around the source is obtained by
\begin{equation}
L(\delta\nu)=4\pi^2 R^2 I_{\rm s,c}(\nu)\delta^3.
\end{equation}
Then the observed flux density is given by
\begin{equation}
F(\nu_{\rm obs})=\frac{4\pi^2 R^2 I_{\rm s,c}(\nu)\delta^3(1+z)}{4\pi
d_{\rm L}^{2}},
\end{equation}
where $d_{\rm L}$ is the luminosity distance of the source and
$\nu_{\rm obs}=\nu\delta/(1+z)$.

In the GeV-TeV regime, the KN effect could be significant and make the IC spectrum have a high energy
cut off. We take this effect into account in our model calculations. The high energy gamma-ray photons can
also be absorbed by EBL, yielding electron-positron pairs, and the observed spectrum in the VHE band must be steeper
than the intrinsic one. The absorption in the GeV-TeV band is considered with the EBL model as proposed by
Franceschini et al. (2008).

\subsection{Our Strategy and Results}
As shown above, the spectrum derived by the syn+SSC model is described with the following parameters, i.e., the strength of magnetic field $B$, the size of the radiating region $R$, the Doppler boosting factor $\delta$, and the electron energy spectrum ($p_1$, $p_2$, $\gamma_{\rm min}$, $\gamma_{\rm b}$, $\gamma_{\rm max}$, and the electron density
parameter $N_0$) (Tavecchio et al. 1998). Since $\delta=[\Gamma(1-\beta \cos\theta)]^{-1}$, where $\Gamma=1/\sqrt{1-\beta^2}$ and $\beta=v/c$, we take $\delta\sim \Gamma$ in this paper because the relativistic
jets of blazars are close to the line of sight ($\cos \theta\sim 0$). Besides $p_1$ and $p_2$, these parameters are not directly observable. They are coupled with each other in the framework of the syn+SSC model. Our strategy and procedure to fit the SEDs are as follows:
\begin{itemize}
\item Derive $p_1$ and $p_2$ values from the spectral indices of the observed SEDs, which
are obtained by fitting the observed data below and above the first bump with a power-law function. According to the syn+SSC model, the spectral indices for the two bumps are consistent. Therefore, if no data are available for the first bump, the observed data in the second bump is used. For six SEDs, the observed data are not enough to constrain $p_1$, we search for historical $p_1$ from the published paper as listed in Table 1.

\item Set the values of $\gamma_{\rm min}$ and $\gamma_{\rm max}$. $\gamma_{\rm min}$ is quite uncertain and is constrained with the observed SEDs via a method reported by Tavecchio et al. (2000). If no observational data are available to confine $\gamma_{\rm min}$, we take $\gamma_{\rm min}=2$ in our calculation. The $\gamma_{\rm max}$ can be constrained by the last observation point in the first bump of the SEDs, or be taken as $\gamma_{\rm max}=100\gamma_{\rm b}$ in this paper.

\item Set the radius ($R$) in the co-moving frame with the observed minimum variability timescale ($\Delta t$), i.e., $R=c\delta\Delta t/(1+z)$. The values of $\Delta t$ are listed in table 1 and the references are presented in Appendix. Rapid variabilities at timescales from  hours to intra-day are observed in BL Lacs objects (Xie et al. 2001; Fossati et al. 2008), therefore, for some sources $\Delta t$ is taken as one day when no variability timescales are available (see details in Appendix).

\item Set the lower limit of $\delta$ with the $\gamma$-ray transparency condition for pair-production absorption (Dondi \& Ghisellini 1995),
\begin{equation}
\delta>\delta_{\tau}=[\frac{\sigma_{\rm T}}{5hc^2}d_L^2(1+z)^{2\beta}\frac{F(\nu_0)}{t_{\rm var}}]^{1/(4+2\beta)},
\end{equation}
where $F(\nu_0)$ is the flux density at the target photon frequency $\nu_0=1.6\times10^{40}/\nu_{\gamma}$ Hz,
$\beta$ is the spectral index of the target photons ($\beta=\alpha_1$ for $\nu_0<\nu_{\rm s}$ and
$\beta=\alpha_2$ for $\nu_0>\nu_{\rm s}$), and $t_{\rm var}$ is the minimum variation timescale that is taken as $t_{\rm var}=\Delta t$.

\item Replace the values of $\gamma_{\rm b}$ and $N_0$ with the functions of $B$, $\delta$, the
peak frequency $\nu_{\rm s}$ and the corresponding synchrotron radiation luminosity
$L_{\rm syn}$ (e.g., see equations (2), (5) in Zhang et al. 2009) i.e., a set of $B$ and $\delta$ should reproduce the synchrotron radiation bump in a given parameters set of $p_1$, $p_2$, and $R$.

\item Constrain $B$ and $\delta$ with the $\chi^{2}$-minimization technique. Note that no error is available for some data points. We estimate the errors of these data points with the average relative errors of the data points whose errors are available. Since the relative errors of the data for the synchrotron bump are usually much smaller than that for the SSC bump, we derive the average relative errors of the data in the two bumps separately. We fit the distributions of the observed relative errors with a Gaussian function and take the central values of the Gaussian fits as the average relative errors. In case of a bow-tie in the SEDs, we use the average flux and its error at the low and high frequency ends as well as at the middle frequency of the bow-tie. We vary the values of $\delta$ and $B$ in wide ranges and calculate the corresponding $\chi^{2}$ of each parameter set of $B$ and $\delta$. Then, we derive the probability of the fit by $p\propto e^{-\chi^{2}/2}$. We make the contours of $p$ in the $B-\delta$ plane and find the best fitting parameters.
\end{itemize}
Following the above procedure, we finally get the contours of $p$ for each SED in our sample. We find that the contours are usually thin and slanted ellipses, even degenerate into an unclosed ellipse for some SEDs that their SSC peaks are poorly observed. The contours of the SEDs in our sample are classified into three groups as following:

\begin{itemize}
\item Closed ellipses in 1 $\sigma$ significance level. In this case, both $B$ and $\delta$ are constrained since the synchrotron radiation and SSC process peaks of the SED are well confined by the observation data. We get 16 such cases in our SED sample. As an example, the contours of $p$ for the SED of 3C 66A are shown in Fig. \ref{Fig:2}(a). Both $B$ and $\delta$ can be constrained in 1 $\sigma$ significance level. We fit the the distributions of $p$ for $B$ and $\delta$ with a Gaussian function and derive the value of $p$ that corresponds to 1 $\sigma$ significance level, as shown in Fig. \ref{Fig:2}(b), (c). We take the central values of the distributions of $p$ for $B$ and $\delta$ as the centers of the contours and report our results in the form as $\delta^{\delta_{\rm up}}_{\delta_{\rm low}}$ and $B^{B_{\rm up}}_{B_{\rm low}}$, where ``up" and ``low" mark the values at high and low ends of the contours in $1 \sigma $ significance level. The lower limit of $\delta$ constrained with the $\gamma$-ray transparency condition for pair-production absorption is also displayed. Therefore, the lower limit of $\delta$ should be taken as $\delta_{\rm low}=\rm max[\delta_{\tau},\delta_{\rm low}]$ with the corresponding ${B_{\rm up}}$.

\item Upward-opened ellipses in 1 $\sigma$ significance level. In this case, the distributions of $p$ for $B$ and $\delta$ still show as a hump that can be roughly fitted with a Gaussian function. The high end of $\delta$ and the corresponding low end of $B$ ($\delta_{\rm up}$ and $B_{\rm low}$) can be constrained with the $1\sigma$ contour. The low end of $\delta$ and the corresponding high end of $B$ cannot be constrained with the contour, but their limits ($\delta_{\tau}$ and $B_{\tau}$) can be obtained with the $\gamma$-ray transparency condition. We get eight such cases in our SED sample. As an example, we show the contours of Mkn 180 in Fig. \ref{Fig:2}(d). We take the parameters of $\delta$ and $B$ that present the minimum $\chi^2$ as the central values and report our results as $\delta_{>\delta_{\tau}}^{<\delta_{\rm up}}$ and $B^{<B_{\tau}}_{>B_{\rm low}}$.

\item Downward-opened or unclosed ellipses. In this case, $B$ and $\delta$ cannot be confined by the contours. There are ten such SEDs in this group. As an example, we show the contours of Mkn 421 in the high state in Fig. \ref{Fig:2}(g). Inspecting the distribution of $p$ for $\delta$, one can find that $p$ increases and comes to almost a constant at a given $\delta_{\rm c}$. The constant value $p$ roughly corresponds to the best fit to the data with the minimum $\delta$. Therefore, we fit the distribution of $p$ with a single-sided Gaussian function to obtain $\delta_{\rm c}$. Similarly, we get $B_{\rm c}$ from the distribution of $p$ for $B$. We also derive the lower limit of $\delta$ and the corresponding upper limit of $B$ with the $\gamma$-ray transparency condition. Therefore, the limits of $\delta$ and $B$ are taken as $\delta>\max[\delta_{\tau},\delta_{\rm c}]$ and $B<\min[B_{\tau}, B_{\rm c}]$.
\end{itemize}

As shown in Fig. \ref{Fig:2}, The contours of $p$ are usually thin and slanted ellipses. The slopes of the major axis of the ellipses are usually -2 to -3, indicating that $B$ and $\delta$ are dependent to each other, i.e., $B\propto\delta^{-2\sim-3}$. This relation seems to be consistent with that derived from the total luminosities of the two bumps ($L_{\rm syn}$ and $L_{\rm SSC}$; Tavecchio et al. 1998)\footnote{Note that the relation between $B$ and $\delta$ is not consistent with that derived from the two peak frequencies under the mono-frequency approximation condition, neither for Thomson regime nor for KN regime (e.g., see equations (4)(16) in Tavecchio et al. 1998).},
\begin{equation}
B\delta^{3}\geq(1+z)\left\{\frac{2L_{\rm syn}}{c^{3}t^{2}_{\rm var}L_{\rm SSC}}\right\}^{1/2}
\end{equation}
in the Thomson regime, and
\begin{equation}
B\delta^{2+\alpha_{1}}>(1+z)^{\alpha_{1}}\left[\frac{g(\alpha_1,\alpha_2)}{\nu_{\rm s}\nu_{\rm c}}\right]^{(1-\alpha_1)/2}\left\{\frac{2L_{\rm syn}}{c^{3}t^{2}_{\rm var}L_{\rm SSC}}\right\}^{1/2}\left(\frac{3mc^{2}}{4h}\right)^{1-\alpha_1}
\end{equation}
in the KN regime, where $g(\alpha_1,\alpha_2)=\exp[\frac{1}{\alpha_1-1}+\frac{1}{2(\alpha_2-\alpha_1)}]$.

Our analysis shows that the major-axis of the contours is shorter when the sampling of a SED is better, hence tighter constraints on $B$ and $\delta$ are obtained, as shown in Fig. \ref{Fig:2}(a) for 3C 66A. The best fitting parameters (or the limits) of $B$ and $\delta$ with the values of $\chi^{2}$ are listed in table 1 and the model curves derived with the parameters are also presented in Fig. \ref{Fig:1}. We find eight SEDs in our sample that are also modeled with the single-zone syn+SSC model by other authors. We compare the derived $B$ and $\delta$ with that reported in literatures for these SEDs as following:

\begin{itemize}
\item Mkn 421 in the high state

BLazejowski et al. (2005) reported a parameter set of $B=0.26$ G, $\delta=14$, and $R=7\times10^{15}$ cm (corresponding to $\Delta t=4.8$ h), but they suggested that these parameters cannot explain the TeV data well. Therefore, they used a multi-zone SSC model to fit the SED. However, by our strategy we find that the SED can be well described using the single-zone SSC model with $B=0.14$ G, $\delta=29$, $R=9.1\times10^{15}$ cm, and $\Delta t=3$ h. The most significant difference between our results with that reported by BLazejowski et al. (2005) is that the derived $\delta$ with our strategy is almost twice of their value, whereas $B$ is almost half of their value.

\item Mkn 501 in the high state

Tavecchio et al. (2001) reported $B=0.32$ G, $\delta=10$, and $R=1.9\times10^{15}$ cm (corresponding to $\Delta t=1.8$ h). We get $B=0.4$ G, $\delta=15$, $R=1.5\times10^{15}$ cm, and $\Delta t=1$ h. Our results are roughly consistent with theirs.

\item W Com in the high state

Acciari et al. (2009) reported that $B=0.24$ G, $\delta=20$, and $R=3\times10^{15}$ cm. We get $B=0.17$ G, $\delta=14$, $R=1.6\times10^{16}$ cm, and $\Delta t=12$ h. The difference between our results and theirs is the size of the radiating region. They set $R=3\times10^{15}$ cm (corresponding to $\Delta t=1.5$ h for $\delta=20$). The strong variabilities on timescales of less than one day are observed at X-rays band, so we take $\Delta t=12$ h in our calculation.

\item 1ES 1959+650

It was reported that a parameter set of $B=0.04$ G, $\delta=20$, and $R=5.8\times10^{15}$ cm (corresponding to $\Delta t=2.8$ h) can well represent the SED of the object in the high state (Krawczynski et al. 2004) and a parameter set of $B=0.25$ G, $\delta=18$, $R=7.3\times10^{15}$ cm (corresponding to $\Delta t=4$ h; Tagliaferri et al. 2008) for the low state. One can find that the reported magnetic field strength by these groups for the source in the low state is 6 times of that in the high state. We do not find such a significant difference for our results. We get $B=0.25$ G, $\delta=12$, $R=1.2\times10^{16}$ cm, and $\Delta t=10$ h for the high state and $B=1.1$ G, $\delta=11$, $R=1.1\times10^{16}$ cm, and $\Delta t=10$ h for the low state. Please note that we take $\Delta t=10$ h in our calculation, which is much larger that used by these authors.

\item PKS 2155-304 in the low state

Aharonian et al. (2009b) reported $B=0.018$ G, $\delta=32$, and $R=1.5\times10^{17}$ cm (corresponding to $\Delta t=48.4$ h), and our results are $B=0.16$ G, $\delta=50$, $R=9.7\times10^{15}$ cm, and $\Delta t=2$ h. The main difference between our results with theirs is that the adopted $\Delta t$ in our calculation is much smaller than that derived with their fitting parameters.

\item 3C 66A

Abdo et al. (2011) reported a much lower $B$ and a corresponding larger $\delta$ than the typical values of $B$ and $\delta$ for AGNs. They got $B=0.01$ G, $\delta=50$, $R=1.1\times10^{17}$ cm (corresponding to $\Delta t=29.4$ h). We get $B=0.2$ G, $\delta=24$ by taking $\Delta t=12$ h (corresponding to $R=2.2\times10^{16}$ cm in the co-moving frame). The derived $R$ from our calculation is only $1/5$ of that used by Abdo et al. (2011).

\item PKS 1424+240

Acciari et al. (2010c) reported that $B=0.18$ G, $\delta=45$, and $R=4.5\times10^{16}$ cm (corresponding to $\Delta t=13.9$ h). We get $B=0.23$ G, $\delta=33$, $R=5.7\times10^{16}$ cm, and $\Delta t=24$ h. Our results are roughly consistent with that reported by Acciari et al. (2010c).
\end{itemize}

According to the comparison discussed above and our analysis results, one can find that a given observed SED usually can be represented by the different parameter sets. They are roughly consistent in sense of a larger $\delta$ corresponding to a smaller $B$. However, the values of $B$ and $\delta$ from literatures do not exactly follow the relation of $B\propto\delta^{-2\sim-3}$ since the adopted $R$ (and the corresponding $\Delta t$) is different by different groups. The size of a radiating region is important in SED modeling. The previous authors usually took a given value of $R$ to fit the observed SEDs. In our analysis we incorporate the observed minimum variability timescale and the value of $\delta$ to constrain the size of radiation regions. Most importantly, we determine the parameter space for a SED by the minimization $\chi^2$ technique. To estimate the errors of the bolometric luminosity ($L_{\rm bol}$), jet power ($P_{\rm jet}$), and other physical parameters for our analysis below, we also derive the ranges of these parameters with the parameter sets of $B$ and $\delta$ that can represent the SED fits within 1 $\sigma$ range, and then derive the errors of these parameters and present in the following figures. Note that we do not consider the error bars in our correlation analysis since the data points usually have a large error and even only a limit is available.

\section{Physical Properties of the Radiation Region}
\subsection{Distributions of Model Parameters}
 We show the distributions of $B$, $\delta$, $\gamma_b$, and $R$ derived from the SEDs in our sample in Fig. \ref{Fig:3}, except for those that only limits are obtained from the observed SEDs. It is found that the values of $B$ are clustered at $0.1\sim 0.6$ G and the range of $\delta$ is from 5 to 50. The values of $\gamma_{\rm b}$ vary from $10^{3}$ to $10^{6}$. The radius $R$ for most sources are $(4\sim 40)\times 10^{15}$ cm.
We also show the distributions of $\nu_{\rm s}$ and $\nu_{\rm c}$ in Fig. \ref{Fig:3}(e). It is found
that $\nu_{\rm s}$ ranges from the infrared to X-ray band (from $10^{14}$ Hz to $10^{20}$ Hz). The value of
$\nu_{\rm c}$ covers a range from $10^{20}$ Hz to $10^{27}$ Hz, but most of them are narrowly clustered at
$10^{24}\sim10^{26}$ Hz.

The magnetic field of the radiation region is critical to understand radiation mechanisms of the sources. However, the origin of the magnetic field for the core regions is still uncertain. The narrow distribution of $B$ might indicate
that the magnetic field may be independent of the other jet properties. It is generally believed that the electrons are accelerated by the violent shocks via the Fermi acceleration mechanism, and then the value of $\gamma_{\rm b}$ reflects the properties of the shock to some extent. No correlation between $\gamma_{\rm b}$ and $B$ is found. Therefore, the magnetic field of the core region
may not be produced through the amplification of the interstellar medium magnetic field by the shocks. It is possible that the magnetic field comes from the vicinity around the central BH or from the accretion disk (Harris \& Krawczynski 2006). The strength of the magnetic field in the inner accretion disk region can be estimated by comparing it with that of BH binaries (e.g., Zhang et al. 2000), e.g., ${B_{\rm
AGN}}/{B_{\rm Binary}}\propto({T_{\rm AGN}}/{T_{\rm Binary}})^{2} \propto ({{10\rm \ eV}/{1\rm \ keV}})^2$,
where $T$ is the typical peak temperature of the accretion disk, and then $B_{\rm AGN}$ is $\sim10^{-4} B_{\rm
Binary}\sim 10^{4}$ G. This is much larger than what we found in the jets. It is thus plausible that the magnetic fields
in the jets are carried from the accretion flow, but diluted significantly as the jets propagate and expand
outwards. This may also explain the fact that the magnetic fields further out, e.g., in the jet knots and hot spots, are even
much weaker\footnote{Zhang et al. (2010) proposed that the magnetic field in the jet knots and hot spots may be due to amplification
of the interstellar medium magnetic field by the violent shocks.}.

\subsection {Jet Power}
Jet power ($P_{\rm jet}$) is essential to understand the physics of jet production. It can be estimated with
$P_{\rm jet}=\pi R^2 \Gamma^2 c (U^{'}_{\rm e}+U^{'}_{\rm p}+U^{'}_{B})$, where $\Gamma$ is the bulk Lorentz factor of
the radiating region and $U^{'}_{i},\ (i={\rm e,\ p,}\ B)$ are the energy densities associated with the emitting electrons
$U^{'}_{\rm e}$, the cold protons $U^{'}_{\rm p}$ and magnetic field $U^{'}_{B}$ measured in the comoving frame
(Ghisellini et al. 2010), which are given by
\begin{eqnarray}
U_{\rm e}^{'}=m_{\rm e}c^2\int N(\gamma)\gamma d\gamma,\\
U_{\rm p}^{'}=m_{\rm p}c^2\int N(\gamma)d\gamma,\\
U_B^{'}=B^2/8\pi,
\end{eqnarray}
assuming that there is one proton per radiating electron. These powers can be calculated with the SED
fitting parameters. The radiation power is estimated with the observed luminosity,
\begin{equation}
P_{\rm r}=\pi R^2\Gamma^2 cU_{\rm r}^{'}=L_{\rm obs}\frac{\Gamma^2}{4\delta^4}\approx\frac{L_{\rm obs}}{4\Gamma^2},
\end{equation}
where we take $L_{\rm obs}\sim L_{\rm bol}$ (bolometric luminosity) and $\Gamma\sim\delta$. $P_{\rm r}$ may serve as a robust
lower limit of $P_{\rm jet}$ (Celotti \& Ghisellini 2008). We calculate $L_{\rm bol}$ in the band  $10^{11}-10^{27}$ Hz based on our SED
fits, then derive $P_{\rm r}$. The calculated $L_{\rm bol}$ and $P_{\rm jet}$ are reported in Table 1 and no
statistical correlation between them is found (as shown in Fig. \ref{Fig:4}(a)). As the above statement, $P_{\rm jet}\propto \delta^{2}$, so the lower limit of $\delta$ result in the lower limit of $P_{\rm jet}$. The limits of $P_{\rm jet}$ for those sources, the SEDs of which can only provide a limit for $\delta$, are also marked in the correlation figures, but are not considered in the following distribution figures. The distributions of the powers and their ratios to the total power, i.e., $\epsilon_{\rm e}=P_{\rm e}/P_{\rm jet}$, $\epsilon_{\rm
p}=P_{\rm p}/P_{\rm jet}$, $\epsilon_B=P_B/P_{\rm jet}$, and $\epsilon_{\rm r}=P_{\rm r}/P_{\rm jet}$, are shown
in Fig. \ref{Fig:5}. One can observe that the power in the jets is carried by cold protons for most sources,
with a ratio $\epsilon_{\rm p}=P_{\rm p}/P_{\rm jet}
> 0.5$ for most sources. The portions of the power carried by the magnetic field are
comparable with $\epsilon_{\rm r}$.

It is generally believed that the jets are fed by accretion of the central massive BHs. With our results, we investigate the
relation of $P_{\rm jet}$ to the mass of the central BH with a sub-sample of 18 sources in our sample. The
masses of the BHs of these sources are collected from literature and reported in table 1. The jet power $P_{\rm
jet}$ as a function of BH mass $M_{\rm BH}$ is shown in Fig. \ref{Fig:4}(b). A tentative anti-correlation
between the two parameters is observed, i.e., $P_{\rm jet}\propto M_{\rm BH}^{-\alpha}$, $\alpha=1.22$ for high
state and $\alpha=1.21$ for low state, respectively. The correlation coefficients estimated with the Spearman
correlation analysis method are $r=-0.74$ and $p=0.04$ for high state data and $r=-0.86$ and $p=0.007$ for low state data. If the jet is
powered purely by accretion of BH and reflects the accretion rate to some extent, i.e., $P_{\rm
jet}=\eta\dot{M}c^2$, the negative correlation of jet power and BH mass would imply that the accretion rate is
correlated with $\frac{1}{M_{\rm BH}}$. That is clearly unreasonable, and thus the relationship between $P_{\rm
jet}$ and $M_{\rm BH}$ disfavors the pure accretion powered jet scenario. Therefore, we suggest that drawing the
spin energy of the central BH should play a significant role in producing these jets. On the other hand, Fig. \ref{Fig:4}(b) would imply a decrease in the BH spin with an increase in $M_{\rm BH}$. These suggest that the
growth of BH mass in our sample (all with small redshift $z<1$ ) is mostly through a series of random accretion
processes that decrease the BH spin statistically (King \& Pringle 2006; King et al. 2008, 2011). Wang et al.
(2009) and Li et al. (2010) also suggested that minor mergers are important in the BH growth at low redshift and
major mergers may dominate at high redshift, consistent with our results.

However, for either the accretion or the spin model of BH to drive the jets, both of them face the $\sigma$-problem as reported by Harris \& Krawczynski (2006), where $\sigma$ is the ratio of electromagnetic energy density to particle energy density. Both models require that the jet formation needs a Poynting flux dominated energy transport by a strongly magnetized or high-$\sigma$ plasma; however, parsec-scale observations indicate that the jets are dominated by particles and are low-$\sigma$ plasma. We also present the correlation between $P_{\rm e}$ and $P_{B}$ (excluding the data with upper or lower limits) as shown in Fig. \ref{Fig:4}(c), and find that the jets of those sources are indeed dominated by the particles too, either for the electron pair model or for the electron-proton model. Our results thus also face the $\sigma$-problem. As described in section 4.1, the magnetic field in the jets may be carried from the accretion flow (Harris \& Krawczynski 2006) and are diluted significantly as the jets propagate. It is therefore required that the majority of the Poynting flux is converted to particle energies  efficiently even beyond equipartition. However it remains unclear how and why this conversion is so efficient.

\subsection{Flux Variation}
The SEDs of ten objects in our sample are obtained in the low and high TeV states. We compare $\nu_{\rm s}$ and $\nu_{\rm c}$ between the high and low states in Fig. \ref{Fig:6}. It is clear that the SEDs in the high states shift to a higher energy band. Therefore, the flux variation is usually accompanied by a spectral shift. To investigate what may be responsible for the flux variations and the spectral shift in the low and high states, we derive the ratios of the flux density at 1 TeV $ (R_{\rm 1\ TeV})$ and the physical parameters
($R_x$) in the high state to that in the low state for the ten sources, where $x$ stands for $L_{\rm bol}$, $B$,
$\delta$, $\gamma_{\rm b}$, and $P_{\rm jet}$. Fig. \ref{Fig:7} shows $R_x$ as a function of $R_{\rm 1\ TeV}$.
Note that $\gamma_{\rm b}^2=\frac{\nu_{\rm s}(1+z)}{3.7\times10^{6}}\frac{1}{B\delta}$ and $B\propto\delta^{- 2\sim -3}$, so the lower limit of $\delta$ and the upper limit of $B$ result in a lower limit of $\gamma_b$ and the lower/upper limits of the ratios of $\gamma_b$ for high to low states. The limits of $R_x$ resulting from the limits of $\delta$ and $B$ are also marked in the following correlation figures. Here, the limits of $R_x$ are also derived by the contours of $B$ and $\delta$; taking the lower limit of $R_{\delta}$ for Mkn 421 as an example, we get $R_{\delta}=29/43$. The values of $\gamma_{\rm b}$, $L_{\rm bol}$, and $P_{\rm jet}$ of the high states are systematically higher than that of the low states. We measure the correlations between $R_{\rm 1\ TeV}$ and $R_x$ with the Spearman correlation
analysis method, and find a tentative correlation between $R_{\rm 1\ TeV}$ and $R_{\gamma_{\rm b}}$, with the correlation coefficient $r=0.66$ and the chance probability $p=0.04$. The best fit line between
$R_{\gamma_{\rm b}}$ and $R_{\rm 1\ TeV}$ is also shown in Fig. \ref{Fig:7}(e). No statistically significant
correlations between $R_{\rm 1\ TeV}$ and other parameters are found. It is well known that the electrons are accelerated by the violent
shocks via the Fermi acceleration mechanism, and then the value of $\gamma_{\rm b}$ may reflect the different microphysical properties of the
shock to some extent. As shown in Fig. \ref{Fig:3}(c), the range of $\gamma_{\rm b}$ is from $10^{3}$ to $10^{6}$ and the values of $\gamma_{\rm b}$ among sources, even among states of a given
source, are different, likely indicating that the properties of shocks significantly vary among sources (or states).
Therefore, it is possible that the flux variation is probably due to the variation of
the physical properties for the shocks, such as the obliquity of the shock and the level of magnetic field turbulence in the shock (Summerlin
\& Baring 2012). An interesting phenomenon for the flux variation of an individual source is that the increase of $\gamma_{\rm b}$ is accompanied with the decrease of $B$ as shown in Fig. \ref{Fig:7}(b). A detailed analysis on this phenomenon will be presented in our next paper.

\section{Summary}
The peak frequency and the peak luminosity of the high energy bump of SEDs for AGNs can be well determined with
simultaneous or quasi-simultaneous observations in the TeV and GeV bands. We compile the broadband SEDs of 24 BL
Lac objects that were simultaneously or quasi-simultaneously observed with {\em Fermi}/LAT and other instruments
from literature. The clean SEDs without contaminations of the accretion disks and external IC
processes of these sources are good candidates for investigating the radiation mechanisms and the physical
properties of the AGN jets. Our results are summarized as follows:
\begin{itemize}
\item  We find that the one-zone syn+SSC model can well represent the observed SEDs.
No excess due to the interaction between
the TeV photons and the extragalactic background light in GeV band over SSC component is found for the sources in our sample.

\item The distribution of $\gamma_{\rm b}$ ranges from $10^{3}$ to $10^{6}$, but the magnetic field
strength $B$ is clustered within a narrow range of 0.1-0.6 G. These results indicate that the properties of shocks significantly
vary among sources (or states), and the magnetic field may not be due to the
amplification of the interstellar magnetic field by the shocks in the jet. We propose that the magnetic field of the core region
may originate from the accretion flow.

\item The Doppler boosting factor $\delta$ of the jets ranges from 5 to 50, and the sizes of the
radiating regions are roughly $(4\sim 40)\times 10^{15}$ cm. Significant flux variations are observed
for the sources in our sample. The SEDs in the high state shift to higher frequencies. A tentative correlation between the ratio of the flux density at 1 TeV and the ratio of the $\gamma_{\rm b}$ in the low and high states is found, indicating
that the flux variation is probably due to the variation of the physical properties for the shocks, such as the obliquity of the
shock and the level of magnetic field turbulence in the shock.

\item We calculated the bolometric luminosity and the jet power for the sources in our sample.
The jet power is dominated by the kinetic energy for most sources. No correlation between $L_{\rm bol}$ and
$P_{\rm jet}$ is found. An anti-correlation between the jet power and the mass of the central BH is observed,
i.e., $P_{\rm jet}\propto M^{-1}_{\rm BH}$. This disfavors the scenario of a pure accretion-driven jet. We
suggest that extraction the rotation energy of the central BHs would be significant for these sources, and
that thus BHs with smaller masses may be spinning more rapidly in this sample.

\end{itemize}

\section{Discussion}

With a large sample of different types of blazars, Fossati et al. (1998) reported a spectral sequence of
FSRQ--LBL--HBL, i.e., along with this sequence, an increase in the peak frequencies corresponds to the
decreases of bolometric luminosity and the ratio of the luminosities for the high- and low-energy spectral components.
This spectral sequence was interpreted by Ghisellini et al. (1998) with the evolution of the external
photon fields (such as the BLRs) of the blazars, or more physically, the sequence is due to the different mass and accretion rate of the BH
(Ghisellini \& Tavecchio (2008). More recently, Chen \& Bai (2011) extended this sequence to narrow line Seyfert 1 galaxies. The significant cooling effect for the electrons by the external
field photons from the BLR of FSRQs may result in a low $\gamma_{\rm b}$ in the electron spectrum. The contribution to the luminosity
from the IC process of the external photon fields would significantly increase the total luminosities and the ratio of $L_{\rm IC}/L_{\rm syn}$. Therefore, observationally, one can expect that both
$L_{\rm bol}$ and $L_{\rm IC}/L_{\rm syn}$ are anti-correlated with $\nu_{\rm s}$, and thus
an anti-correlation between $\gamma_{\rm b}$ and $U^{'}_{\rm r}$ in the co-moving frame is also expected (Ghisellini et al. 1998). $L_{\rm bol}$ and $L_{\rm
IC}/L_{\rm syn}$ as a function of $\nu_{\rm s}$ are shown in Fig. \ref{Fig:8}. $\gamma_{\rm b}$
as a function of $U^{'}_{\rm r}$ ($U^{'}_{\rm r}=U^{'}_{\rm syn,avail}+U_B$ in this work, where $U^{'}_{\rm syn,avail}$ is the available photon energy density of synchrotron radiation for IC process) is also shown in Fig. \ref{Fig:8}. No correlation is found for pairs $L_{\rm bol}-\nu_{\rm s}$ and $\gamma_{\rm b}-U^{'}_{\rm r}$. This lack
of correlation in our sample can be explained by the fact that the EC process is negligible for the
BL Lac objects in our sample. We also show the sample of Ghisellini et al. (2010) in the $\gamma_{\rm b}$-$U^{'}_{\rm r}$ plane
(gray circles in Fig. \ref{Fig:8}(a)). Although the correlation between $\gamma_{\rm b}$ and $U^{'}_{\rm r}$ of our sample sources has large scatters, comparing our results with that of Ghisellini et al. (2010), the two results are consistent and the sources in our sample distribute
in the left top of Fig. \ref{Fig:8}(a), where the EC process is not important.

However, the ratio of $L_{\rm IC}/L_{\rm syn}$ is indeed anti-correlated with $\nu_{\rm s}$ in their
high and low states, except for PKS 2005-489 in the high state. The Spearman correlation analysis yields a correlation coefficient
$r=-0.59$ and a chance probability $p=0.07$ for the high state data, $r=-0.9$ and $p=3.4\times10^{-4}$ for low state data,
respectively. Excluding PKS 2005-489 in the high state, our best fits give $\log L_{\rm IC}/L_{\rm syn}=(3.27\pm1.)-(0.23\pm0.06)\log {\nu_{\rm s}}$ and $\log L_{\rm IC}/L_{\rm
syn}=(3.5\pm0.95)-(0.2\pm0.06)\log {\nu_{\rm s}}$ for the sources in
the low and high states, respectively. Because all the sources in our sample are BL Lac objects, the external photon fields outside their jets are much weaker than the synchrotron radiation photon field and the EC process is thus not considered in this work. Therefore, the
$L_{\rm IC}/L_{\rm syn}-\nu_{\rm s}$ anti-correlation may have a different physical origin from the blazar sequence. It is
possible that this anti-correlation is due to the KN effect. As $\nu_{\rm s}$ increases, the SEDs shift to the
higher frequency end and the KN effect should be more significant. Thus, the ratio of $L_{\rm IC}/L_{\rm syn}$
would decrease as $\nu_{\rm s}$ increases since $U_B$ is almost constant.

The TeV photons from the high redshift universe can be absorbed through interaction with the extragalactic background light (EBL), producing
electron-positron pairs. The electron-positron pairs may scatter the CMB photons to the GeV band. Therefore, the intrinsic spectrum at the TeV band should be harder than the observed one, especially for the sources at high redshift, and an excess component in the GeV band may be observed in the observed SEDs as suggested by some authors\footnote{It was also reported that the electron-positron pairs would be deflected by the intergalactic magnetic field (IGMF) from the
initial TeV photon direction and thus the secondary GeV photons produced by them would exhibit a halo around
the central bright point-source instead of a GeV excess in the SEDs (Aharonian et al. 1994; Dolag et al. 2009;
Neronov \& Semikoz 2009).}. The power of such a component depends on the EBL model\footnote{Using the multi-wavelength galaxy observation data and the luminosity functions in the near-IR band, Dominguez et al. (2011) derived the EBL
model
to redshift 4.} and the intergalactic magnetic field (IGMF). Assuming that the intrinsic spectrum of TeV
emission is the same as that of the GeV emission, one can confine the EBL model and the IGMF. Using the data of Mkn 501 during its high state, Dai et al. (2002)
calculated the spectrum under different IGMF strengths and reported that the cascade GeV emission is detectable
by {\em Fermi}/LAT if the IGMF strength is $\leq10^{-16}$ G. Comparing the predicted emission in the GeV band with the
upper limits measured by the {\em Fermi}/LAT for the blazar 1ES 0229+200, Tavecchio et al. (2010a) suggested that the
IGMF is larger than $10^{-15}$G. Our systematical analysis of the TeV BL Lac objects presented here clearly indicates that no excess emission over the SSC component is observed: this may be due to a stronger IGMF than previous predictions (Dai et al. 2002; Yang et al. 2008; Tavecchio et al. 2010a).

\section{Appendix}
\emph{Mkn 421.} It is the first TeV AGN confirmed by Whipple (Punch et al. 1992) and is a high-frequency-peaked BL Lac (HBL) object. Violent variation of the flux
at GeV-TeV regime was detected and associated with that observed at the X-ray band (BLazejowski et al.
2005). The data of high state are from BLazejowski et al.
(2005), when the source underwent an outburst in 2004 April with the peak flux $\sim135$ mCrab in the X-ray band
and $\sim3$ Crab in the gamma ray band. The variabilities on timescales of hours were observed, so we take $\Delta t=3$ h to model the SED of the high state (to see Figure 10 in BLazejowski et al.
2005). The data of the low state are taken from Paneque et al. (2009), when Mkn 421 is found in a rather low/quiet state from August 2008 to August 2009. $\Delta t$ for modeling the low state SED is also taken as 3 h. The high energy gamma-ray emission of this source has also been detected by the ARGO-YBJ detector (Aielli et al. 2010, Bartoli et al. 2011).

\emph{Mkn 501.} It was detected by Whipple during an observation of 66 h. An average flux of
$(8.1\pm1.4)\times10^{-12}$cm$^{-2}$s$^{-1}$ above 300 GeV and the variability on timescales of days were
observed (Quinn et al. 1996). From 1997 April to 1999 June, the observations with BeppoSAX showed that the peak
frequency of the synchrotron emission shifted from 100 keV back to 0.5 keV, and correspondingly the flux decreased
(Tavecchio et al. 2001). A multi-wavelength campaign in 2006 July with Suzaku and MAGIC was performed. During
this program the average VHE flux above 200 GeV is $\sim 20\%$ of Crab flux with a photon index $2.8\pm0.1$ from
80 GeV to 2 TeV, indicating that the source was in a low state (Anderhub et al. 2009a). $\Delta t$ is taken as 3 h (Figure 2 in Anderhub et al. 2009a) to fit this observed broadband SED in this paper. The data of the high state SED (with $\Delta t$=1 h) are from Tavecchio et al. (2001). The data
of {\em Fermi}/LAT observations are from Abdo et al. (2009), but they are not simultaneous with the data of low and high states.

\begin{figure*}
\includegraphics[angle=0,scale=0.30]{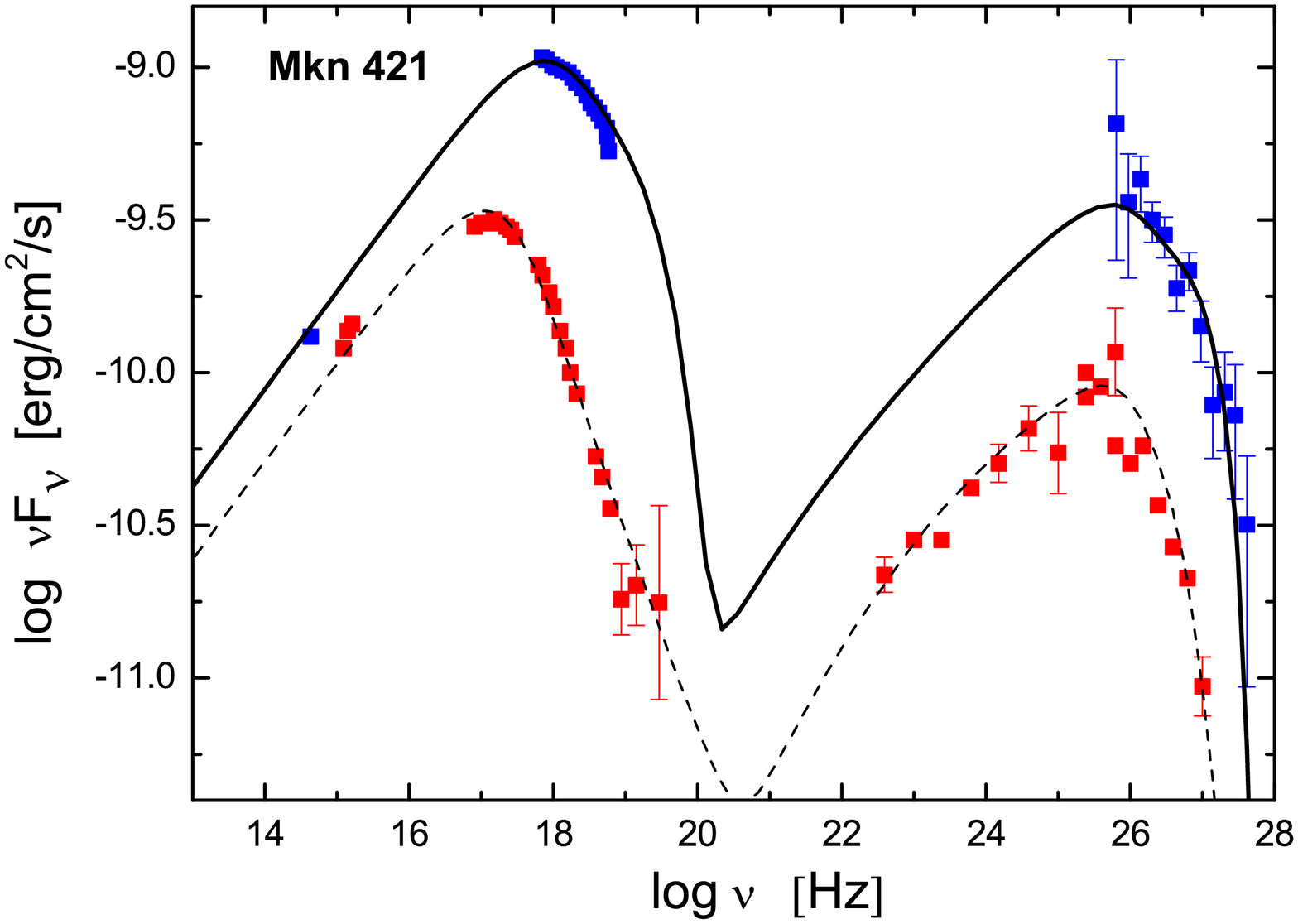}
\includegraphics[angle=0,scale=0.30]{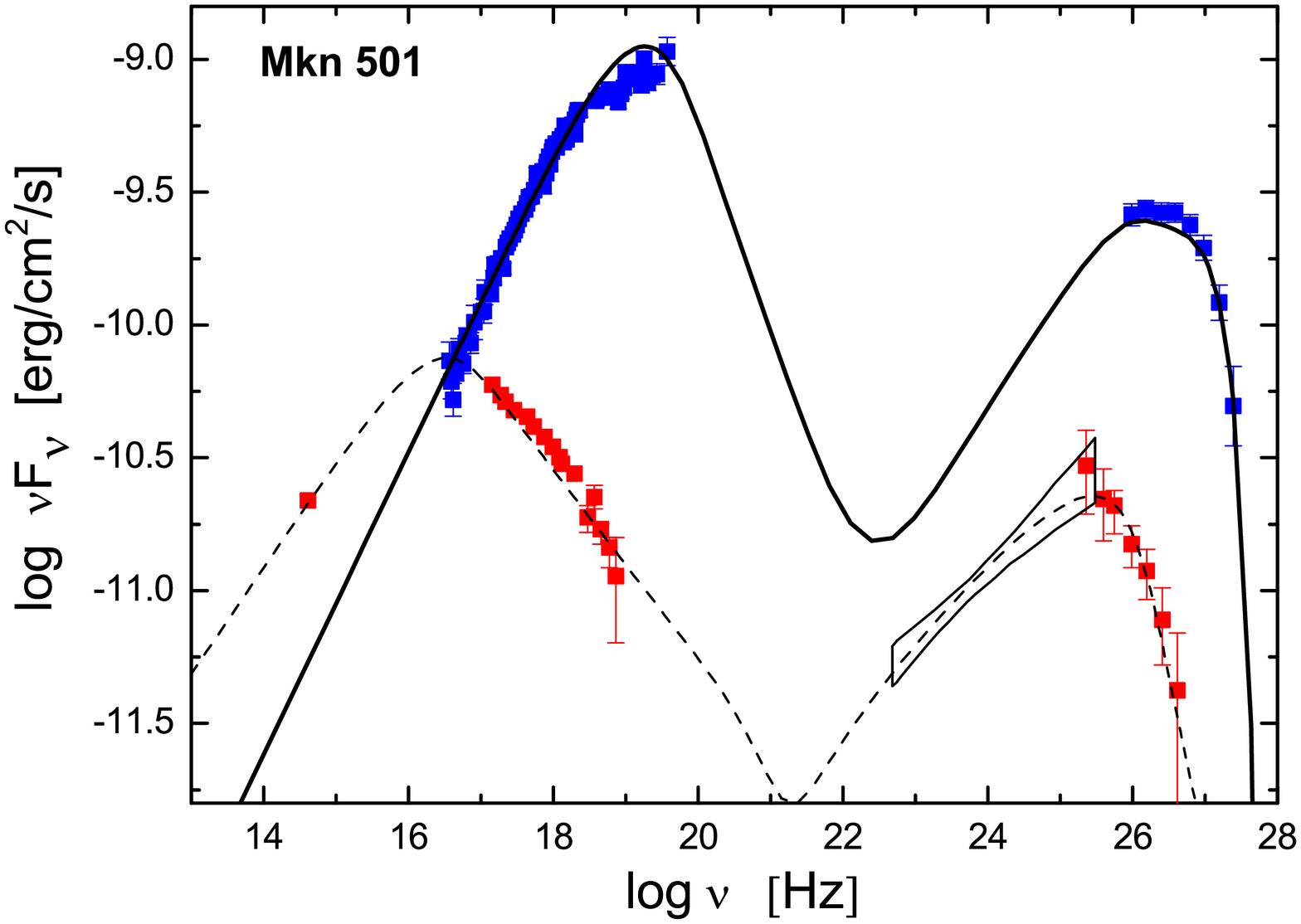}
\includegraphics[angle=0,scale=0.30]{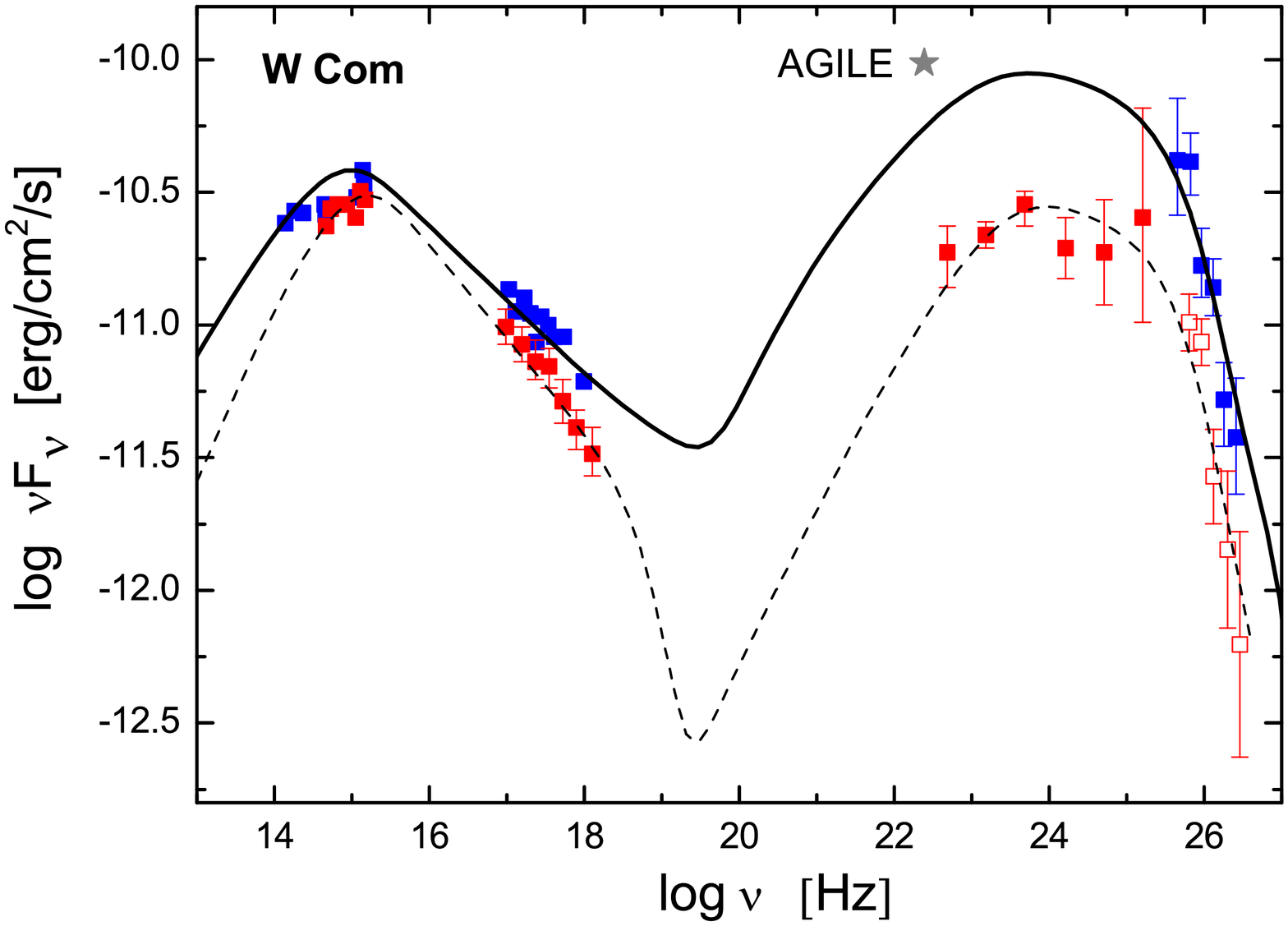}
\includegraphics[angle=0,scale=0.30]{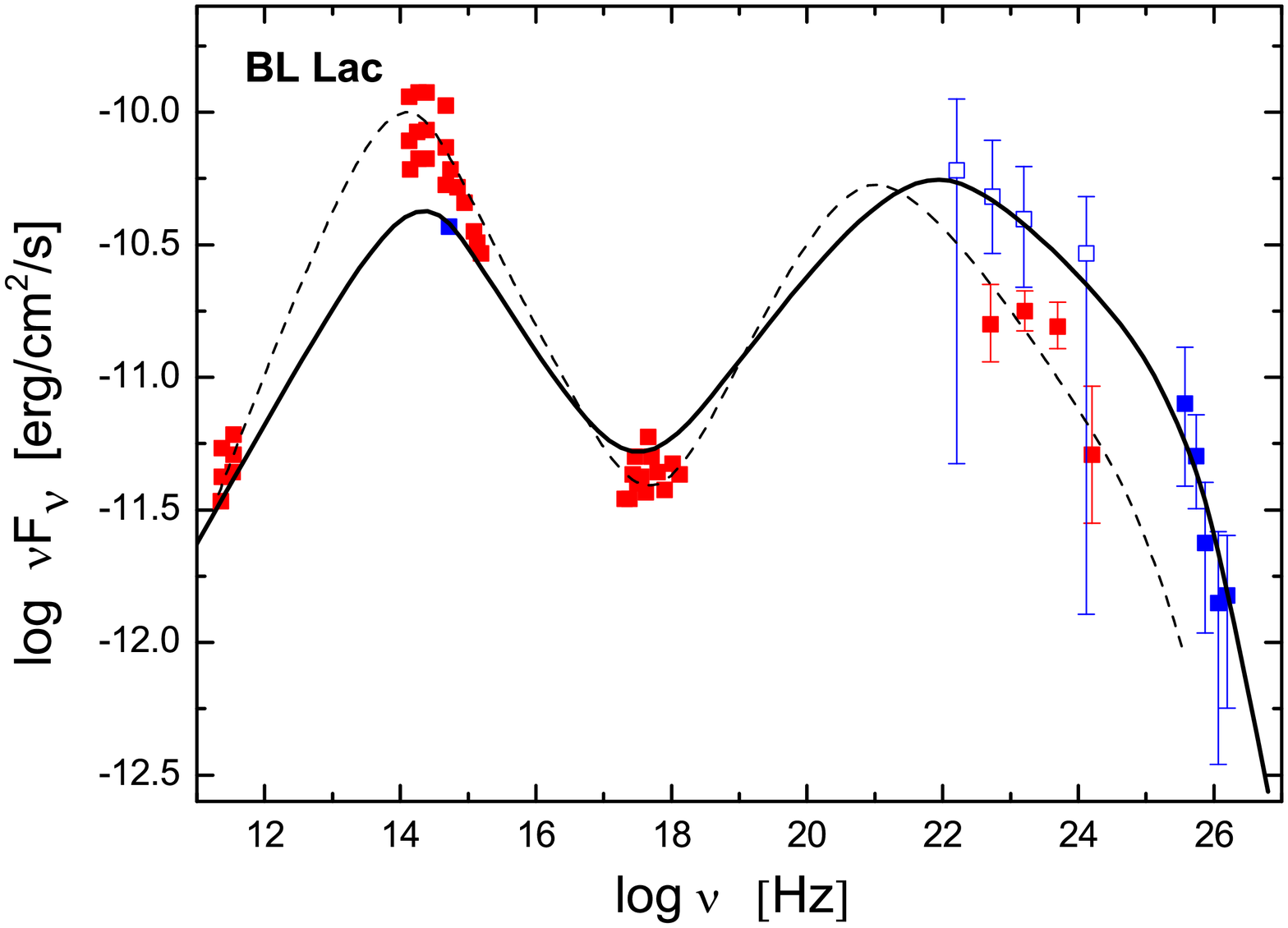}
\includegraphics[angle=0,scale=0.30]{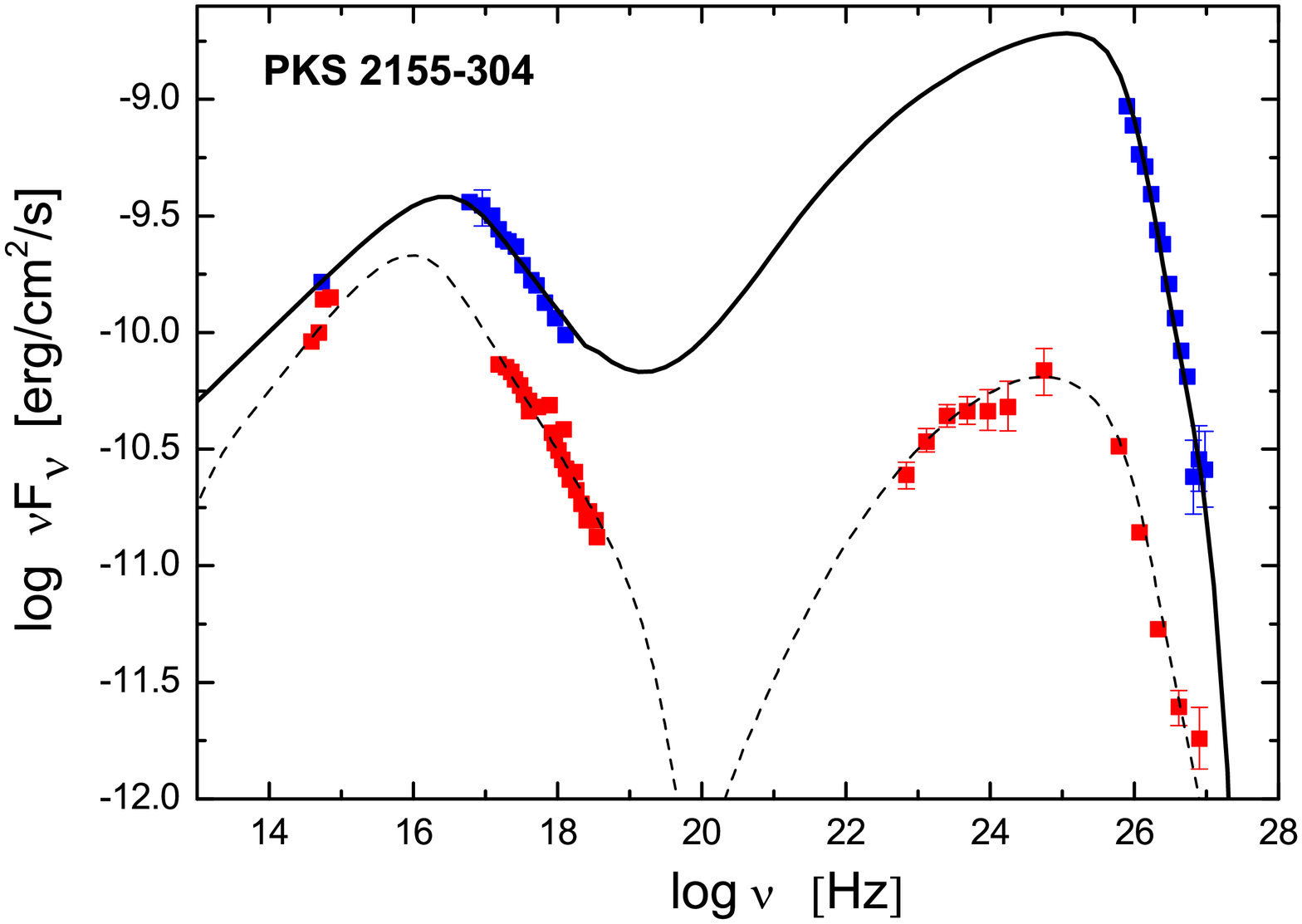}
\hfill
\includegraphics[angle=0,scale=0.30]{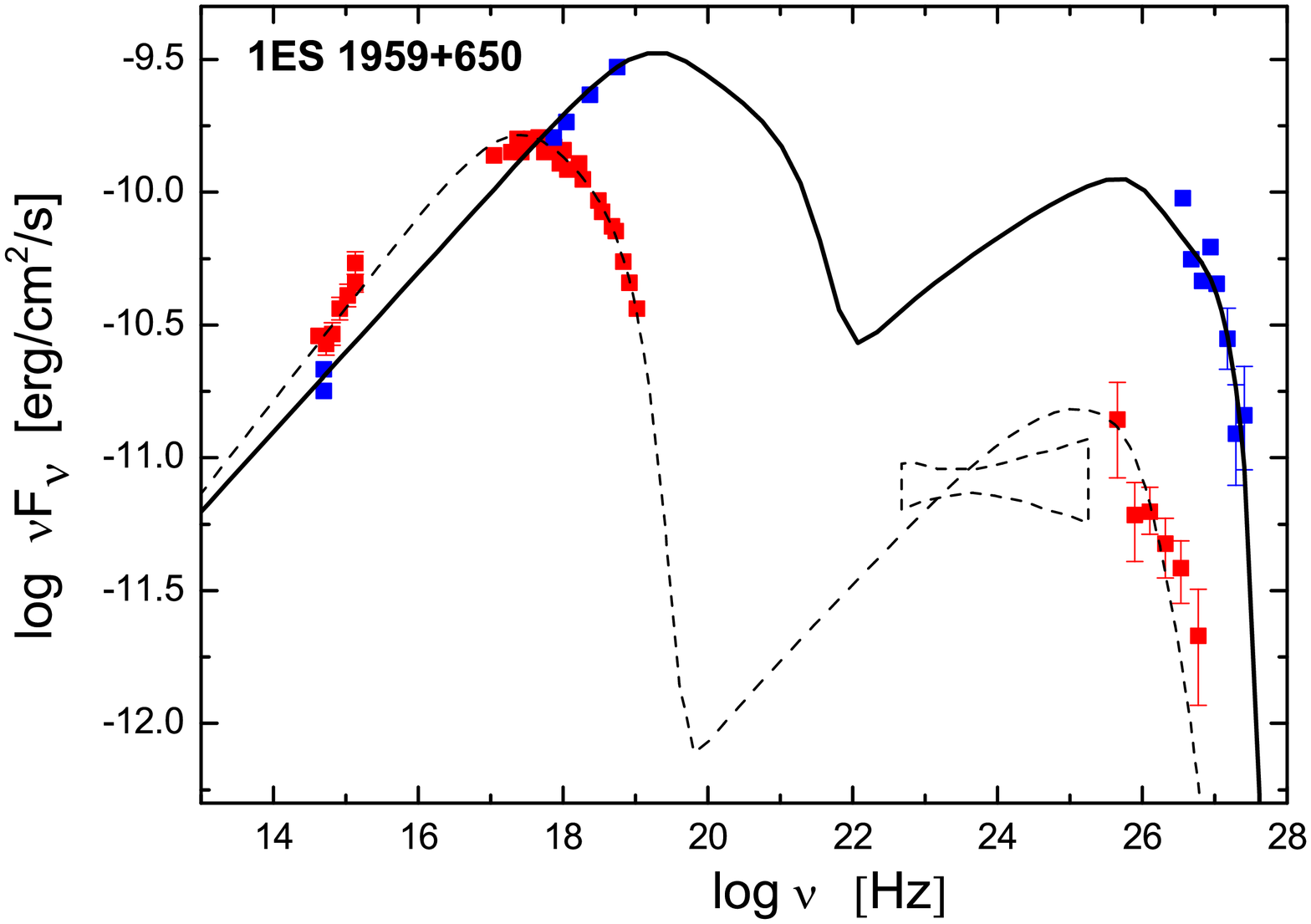}
\caption{Observed SEDs ({\em scattered data points}) with our best model fits ({\em lines}). {\em Solid squares} are for the data observed simultaneously and {\em opened squares} are for the data obtained not simultaneously. The data of high and low
states are marked with {\em blue and red} symbols, respectively. If only one SED is obtained, the data are
shown with {\em black} symbols. The {\em Fermi}/LAT
observations are presented with a {\em bow-ties} or a {\em magenta triangle} (upper-limits for six sources, namely, 1ES
1101-232, 1ES 0347-121, H 2356-309, RGB J0152+017, 1ES 0229+200, and PKS 0548-322). {\em Dashed line} of bow-tie indicates that the data are not used to calculate the values of $\chi^{2}$ with fitting results.} \label{Fig:1}
\end{figure*}

\emph{W Com.} The first TeV intermediate-frequency-peaked BL Lac (IBL) object, an evidently bright outburst in
the optical and X-ray bands was observed in 1998 (Tagliaferri et al. 2000). W Com is confirmed to be a TeV source by the observations of VERITAS during a strong TeV flare in the middle of March 2008 with an integrated photon flux above
200 GeV of $\sim9\%$ Crab (Acciari et al. 2008) and the data of the TeV observation are presented as opened squares in Fig. \ref{Fig:1}.
Both the SEDs quasi-simultaneously obtained during the TeV flare and during
the optical/X-ray outburst can be well fitted by the single-zone syn+SSC model (Zhang 2009). The broadband SED quasi-simultaneously
obtained with observations of {\em Fermi}/LAT and swift from Tavecchio et al. (2010b) is also shown in Fig. \ref{Fig:1} (red squares).
Subsequently, another outburst of very high energy gamma-ray emission was detected in 2008 June by VERITAS with
the flux of $(5.7\pm0.6)\times10^{-11}$cm$^{-2}$s$^{-1}$ (blue squares in Fig. \ref{Fig:1}, Acciari et al. 2009a), three times brighter than the observation of March
2008. The flux upper-limit derived by quasi-simultaneous observation of AGILE is also presented in Fig. \ref{Fig:1} (gray star). The strong variabilities on timescales of less than one day are observed in X-rays band (Fig. \ref{Fig:1} in Acciari et al. 2009a), so we take $\Delta t=12$ h in this paper. For the SEDs of high state, Acciari et al. (2009a) pointed out that both the SSC model and the SSC+EC model can explain the observed SED well, but the SSC+EC model makes the magnetic field strength more close to the equipartition condition.

\emph{BL Lacertae.} The first confirmed TeV low-frequency-peaked BL Lac (LBL) object, its VHE gamma-ray emission was detected
by MAGIC during 2005 August to December with an integral flux of $(0.6\pm0.2)\times10^{-11}$cm$^{-2}$s$^{-1}$,
corresponding to $3\%$ Crab flux (Albert et al. 2007a). The photon index from 150 to 900 GeV is rather steep
with $\Gamma=-3.6\pm0.5$, and the light curve shows no significant variability. The simultaneous observation for
MAGIC in optical band was performed by KVA. The broadband SED together with EGRET data (opened squares) in
1995 is shown in Fig. \ref{Fig:1} (blue squares). During an optical outburst in July 1997, BL Lacertae was detected in
X-ray band by RXTE and in gamma-ray band by EGRET, implying that the source was bright and variable at the both
bands (Madejski et al. 1999). The spectra in X-ray and gamma-ray bands are hard. Madejski et al. (1999) reported that the
X-rays are produced by SSC process while the gamma-rays are produced by Comptonization of the broad
emission line flux. They also suggested that the the emitting regions correspond to
the collision of two blobs or shells moving at slightly different velocities, therefore, the different SEDs correspond
to different collisions, which can occur at different location in the jet. If the emission regions are inside of the BLRs, they would emit most of their energy in the GeV band. The multiwavelength data during {\em Fermi}/LAT observation (red squares in Fig. \ref{Fig:1}; Tavecchio et al. 2010b) are also considered, but no simultaneous data in TeV regime are obtained. A rapid variability was detected in the soft X-ray band on timescales of 3-4 hours (Ravasio et al. 2002), therefore, we take $\Delta t=3$ h to fit the broadband SEDs of the two states.

\emph{PKS 2005-489.} A HBL object was detected by HESS in 2003 and 2004 (Aharonian et
al. 2005). A significant signal of VHE emission was detected in 2004 with the integral flux above 200 GeV of
$\sim6.9\times10^{-12}$ cm$^{-2}$ s$^{-1}$, corresponding to $2.5\%$ Crab flux. The flux level in 2003 was lower
than that measured in 2004, indicating that the activity in 2004 increased, but no significant variability on
timescale less than a year was found. The multiwavelength observation by HESS, XMM-Newton and RXTE satellites
from 2004 to 2007 was performed (HESS Collaboration 2010a), indicating that the large flux
variations at the X-ray band are coupled with weak or no variations in the VHE band. A broadband SED
with the low X-ray flux is taken as the low state and shown as red symbols in Fig. \ref{Fig:1}. A very high state
of X-ray emission for this source in 1998 was detected by BeppoSAX and RossiXTE (Tagliaferri et al. 2001). HESS
and {\em Fermi}/LAT simultaneously observed the source in 2009, when the X-ray flux was comparable to the flux level
in 1998 (Kaufmann et al. 2010). Therefore, we compile the data of the two observations together and take it as
the high state (blue symbols in Fig. \ref{Fig:1}). The variability timescale for modeling the two SEDs is taken as $\Delta t=10$ h (Tagliaferri et al. 2001).

\emph{1ES 1959+650.} A HBL object with a strong TeV outburst in May 2002 was detected by VERITAS (Holder et al.
2003), HEGRA (Horns et al. 2002) and CAT (Aharonian et al. 2003). During the outburst, the flux level reached to
3 times the Crab flux and there was evidence for strong variability. The multiwavelength campaign in radio,
optical, X-ray and TeV gamma-ray bands was performed from 2002 May 18 to August 14 (Krawczynski et al. 2004). A
time-averaged spectrum corresponding to the TeV gamma-ray high state is considered in this work (blue squares in
Fig. \ref{Fig:1}). The observations show that X-ray flux and gamma-ray flux had tentative correlation, but no
correlations of optical variability with X-ray and gamma-ray emission were found. Another multiwavelength campaign was
performed in 2006 May, when the source exhibited a high state at optical and X-ray bands and was in the lowest
level at the VHE band (Tagliaferri et al. 2008). There were variabilities in optical and X-ray bands. The spectrum butterfly of {\em Fermi}/LAT observation is also shown in Fig. \ref{Fig:1}. The fastest gamma-ray flux variability timescale of 10 hours based on the whipple data is obtained and then is taken as the value of $\Delta t$ to model the SEDs of the two states.

\emph{1ES 2344+514.} The third BL Lac object was detected at the VHE band with Whipple. The detection of VHE
emission mostly came from an apparent flare on 1995 December and the average flux above 350 GeV was
$(6.6\pm1.9)\times10^{-11}$ cm$^{-2}$ s$^{-1}$, $63\%$ of the Crab flux (Catanese et al. 1998). The observation
of MAGIC between 2005 August 3 and 2006 January 1 presented a steep spectrum with photon index $\Gamma=-2.95$
and a flux 6 times below the 1995 flare, indicating that the source was in low state (Albert et al. 2007b). No
evidence for variability was found during the MAGIC observations. The simultaneous optical observation with
MAGIC was performed by KVA. The broadband SED is shown as red solid squares in Fig. \ref{Fig:1}. The data of Swift satellite
observation (red opened squares) on 19 April 2005 (Tramacere et al. 2007) and the bow-tie of {\em Fermi}/LAT observation (Abdo et al.
2009) are also presented to constrain the radiation model. A broadband SED for a high state
quasi-simultaneously obtained by VERITAS and Swift on December 2007 is also considered in this work (blue squares), and the
data are from Acciari et al. (2011). The measured flux above 300
GeV was $(6.76\pm0.62)\times10^{-11}$ cm$^{-2}$ s$^{-1}$, corresponding to $48\%$ of Crab flux. The highest
X-ray emission ever observed was measured by Swift/XRT, and was correlated with the VHE gamma-ray emission. The TeV data of the high state in Fig. \ref{Fig:1} have been corrected for absorption by EBL. The variabilities on timescale of a few hours (Albert et al. 2007b) and on timescale of $\sim$ day (Acciari et al. 2011) are observed, so we take $\Delta t$ as 12 hours for modeling the SED of two states.

\emph{PKS 2155-304.} A HBL object, its strong VHE emission was first detected in
1997 November, at the same time the strongest X-ray emission ever observed and GeV gamma rays were also detected
by BeppoSAX and EGRET (Chadwick et al. 1999). A simultaneous observation with HESS, Chandra and the Bronberg
optical observatory was carried in the night of July 26-30 2006 when the source was in a high-activity state at gamma-ray band and the gamma ray flux reached $\sim11$ times the Crab flux (Aharonian et al. 2009a). The emission between X-ray
and VHE gamma-ray is strongly correlated. The broadband SED of this campaign is shown as blue squares in Fig. \ref{Fig:1}. The variability on timescale of 2 hours is observed at optical (and higher) frequencies (Aharonian et al. 2009a), thus we take $\Delta t=2$ h. The GeV-TeV observation with {\em Fermi} and HESS was performed between 25 August and 6 September 2008, and the
low-energy component was simultaneously covered by ATOM telescope, RXTE and Swift observations (red squares in
Fig. \ref{Fig:1}, Aharonian et al. 2009b). During that period PKS 2155-304 was in a low-activity state at X-ray and
gamma-ray bands, whereas the optical flux was much higher. The optical emission was correlated with the VHE emission,
but no correlation between X-ray and VHE was found. $\Delta t$ is also taken as 2 hours to model the low state SED.

\emph{1ES 1101-232.} A HBL object hosted by an elliptical galaxy, it was detected by HESS in March-June 2004 and
2005 with a very hard spectrum and no significant variation was found (Aharonian et al. 2007a). A
multiwavelength campaign data, including observations of VHE by HESS and X-ray by RXTE satellite in 2005 and
XMM-Newton in 2004, were obtained (red squares in Fig. \ref{Fig:1}, Costamante 2007). The Suzaku observation
simultaneously covered with the HESS measurement was carried out in 2006 May (blue squares in Fig. \ref{Fig:1}, Reimer et
al. 2008). No significant X-ray or gamma-ray variability was detected during this program, and the object was in
a quiescent state with the lowest X-ray flux ever measured. 1ES 1101-232 was not detected by {\em Fermi}/LAT
(Tavecchio et al. 2010b), and only an upper-limit was given as shown in Fig. \ref{Fig:1}. No variability timescale is found for the corresponding SEDs, so we take $\Delta t=12$ h to fit the SEDs of two states.

\emph{S5 0716+714.} It is a LBL object. The MAGIC observations were performed in November 2007 and in April 2008
and detected its TeV emission (Teshima et al. 2008; Anderhub et al. 2009b). The integral flux above 400 GeV was
$\sim7.5\times10^{-12}$ cm$^{-2}$ s$^{-1}$, corresponding to $9\%$ Crab flux. The optical emission of S5
0716+714 was simultaneously observed by KVA when the source was in a high state at optical band, and most of the
gamma-ray emission signal came from the phase of the optical high state of the object, suggesting a possible
correlation between the VHE emission and optical emission (Anderhub et al. 2009b). The data quasi-simultaneously
obtained in April 2008 by KVA, Swift and MAGIC are considered and shown as blue symbols in Fig. \ref{Fig:1}. The data of {\em Fermi}/LAT
observation with quasi-simultaneous Swift observation from Tavecchio et al. (2010b) are also considered and
shown as red symbols in this work. $\Delta t$ for the high and low states is taken as 3 hours (Foschini et
al. 2006).

\emph{3C 66A.} A IBL object, the observation of VHE was performed from September 2007 through January 2008
by VERITAS and it was confirmed to be a TeV source with an integral flux above 200 GeV $6\%$ of Crab flux
(Acciari et al. 2009b). The observed spectrum can be characterized by a soft power-law with photon index
$\Gamma=4.1$ and a variability on the timescale of days was found. The simultaneous GeV-TeV observations by
{\em Fermi}/LAT and VERITAS were performed in October 2008 (Reyes et al. 2009). The simultaneously observed broadband SED is from Abdo et al. (2011). The variability on timescales of less than a day is observed (Abdo et al. 2011) and then $\Delta t$ is taken as 12 hours.
Abdo et al. (2011) reported a SSC+EC model to describe the SED since the magnetic fields are significantly closer to equipartition
than in the pure SSC case, i.e., $L_{B}/L_{\rm e}\sim0.11$ for SSC+EC model and $L_{B}/L_{\rm e}\sim1.1\times10^{-3}$ for pure SSC model.

\emph{PG 1553+113.} Its VHE emission with a very soft spectrum (photon index of $\Gamma=4.0\pm0.6$) was detected
in 2005 by HESS, and no evidence of variability was found (Aharonian et al. 2006a). The VHE emission was
subsequently confirmed by MAGIC (Albert et al. 2007c), and the integral flux levels were consistent with that of HESS. PG 1553+113 is in the Fermi LAT bright AGN source list, but was not detected by EGRET because it was
in a low state during the observation. This source is a HBL object and a bright X-ray source with many
observations, but no strong or fast variability was detected in the X-ray band (Reimer et al. 2008). The
redshift of PG1553+113 is unknown and the VHE observation indicates that the redshift is greater than 0.25. In
this work, we take $z=0.3$. The observations by {\em Fermi}/LAT from 4 August 2008 to 22 February 2009 show that it
was a steady source with a hard spectrum in {\em Fermi}/LAT energy band (Abdo et al. 2010c). The data of broadband SED
(black squares) are from Abdo et al. (2010c), including quasi-simultaneous observations by KVA, Suzaku, MAGIC
and HESS in July 2006 and the observation of {\em Fermi}/LAT. $\Delta t$ is taken as one day.

\emph{1ES 1218+30.4.} A HBL object, it was confirmed to be a TeV source by MAGIC in 2005 January, but no variability on
timescales of days was found within the statistical errors (Albert et al. 2006b). In optical band, KVA observed
the source simultaneously with MAGIC. From 2008 December to 2009 May, VERITAS monitored the source and revealed
a prominent flare reaching $\sim20\%$ of the Crab flux (Acciari et al. 2010a). The light curve of TeV emission
for this source showed day-scale variability. Here, we take $\Delta t=12$ h. The observational flux of VERITAS is comparable to MAGIC
(Weidinger \& Spanier 2010). Swift observed this source between March and December 2005, and was
quasi-simultaneous with the observation of MAGIC (R\"{u}ger et al. 2010). In this work, the data of the broadband SED
include the observations of VERITAS, {\em Fermi}/LAT (from Abdo et al. 2009), swift and KVA.

\emph{1ES 1011+496.} A HBL object, it was observed by MAGIC from 2007 March to May after an optical outburst in March
2007, and an integral flux above 200 GeV of $1.58\pm0.32\times10^{-11}$ cm$^{-2}$ s$^{-1}$ was obtained (Albert et al. 2007d). The broadband SED is composed of the quasi-simultaneous
observations of Swift and {\em Fermi}/LAT (from Tavecchio et al. 2010b) and the observation of MAGIC in 2007 (opened squares). No corresponding variability timescale for the SED is found, so we take $\Delta t=24$ h.

\emph{PKS 1424+240.} A HBL object with an unknown redshift, its VHE emission was detected by VERITAS with a
flux normalization at 200 GeV of $\sim5.1\times10^{-11}$ TeV$^{-1}$ cm$^{-2}$ s$^{-1}$ (Acciari et al. 2010c).
The photon spectrum above 140 GeV can be described well by a power law with $\Gamma\sim3.8$. During the period
from February 2009 to June 2009, the flux of VHE emission was steady and the contemporaneous observation of
{\em Fermi}/LAT also did not detect any variability (Acciari et al. 2010c). The broadband SED is established by
simultaneous observations of VERITAS, {\em Fermi}/LAT, Swift and MDM. Considering the EBL absorption, a redshift upper
limit of 0.66 is inferred (Acciari et al. 2010c) and $z=0.5$ is taken in this work. The X-ray variability timescale of about a day (Acciari et al. 2010c) is taken as the value of $\Delta t$ to model the SED in this paper.

\emph{RGB J0710+591.} A well known HBL object, it was not detected by EGRET. It was observed at the VHE waveband by
VERITAS between December 2008 and March 2009, and was confirmed to be a TeV source (Ong et al. 2009, Acciari et al.
2010b). The observed spectrum from 0.31 to 4.6 TeV can be fit by a power law with a photon spectral index
$\sim-2.69$, and the integral flux above 300 GeV is $3.9\pm0.8\times10^{-12}$ cm$^{-2}$ s$^{-1}$, corresponding
to $3\%$ of Crab flux. The VERITAS observation was complemented by contemporaneous observations from {\em Fermi}/LAT,
Swift and Michigan-Dartmouth-MIT observatory (Acciari et al. 2010b). No corresponding variability timescale for the SED is found, so we take $\Delta t=24$ h.

\emph{1ES 0806+524.} A HBL object was detected by VERITAS in VHE gamma-ray regime between November 2006 and April
2008, and no significant variability on months timescale was found (Acciari et al. 2009c). The observed photon
spectrum from November 2007 to April 2008 can be fitted by a power law with $\Gamma\sim-3.6$ between 300 to 700
GeV. The integral flux above 300 GeV is $\sim2.2\times10^{-12}$ cm$^{-2}$ s$^{-1}$, corresponding to $1.8\%$ of
Crab flux. The data obtained quasi-simultaneously by Swift and VERITAS observations (Acciari et al. 2009c) and
the spectrum butterfly of {\em Fermi}/LAT observations (Abdo et al. 2009) are considered in this work. No corresponding variability timescale for the SED is found, so we take $\Delta t=24$ h.

\emph{Mkn 180.} A HBL object was detected with VHE gamma-ray emission by MAGIC during an optical outburst in 2006.
The integral flux above 200 GeV is $(2.3\pm0.7)\times10^{-11}$ cm$^{-2}$ s$^{-1}$ and corresponds to $11\%$ Crab
flux. The observed spectrum was rather soft with a photon index of $\Gamma\sim3.3\pm0.7$, and no variability was
found (Albert et al. 2006a). A multi-wavelength campaign in 2008 was performed (Rugamer et al. 2011). The broadband SED is compiled with the observations of swift and
Fermi (grey shaded) in 2008 and the observation of MAGIC in 2006. The bow-tie
of {\em Fermi}/LAT observation for 1-year Catalog (dashed line, Abdo et al. 2009) is also given in the SED. According to the light curve of Mkn 180 in Albert et al. (2006a), $\Delta t$ is taken as one day in this paper.

\emph{H 2356-309.} A HBL object was detected with VHE emission by HESS from June to December 2004, with a integral
flux above 200 GeV of
$4.1\pm0.5\times10^{-12}$ cm$^{-2}$ s$^{-1}$ (Aharonian et al. 2006b). A simultaneous observation with HESS in 2004
at lower energy bands was performed by ROTSE-III (optical) and RXTE (X-rays). A broadband SED obtained
simultaneously by XMM-Newton and HESS on June 2005 (HESS Collaboration et al. 2010b) is presented in this work,
in which the flux levels in X-ray and TeV bands are comparable with that measured in 2004. In the VHE band,
significant small-amplitude variations on time scales of
months and years were detected. The X-ray measurements show that it was in a low state in this band.
Unfortunately, the observations of {\em Fermi}/LAT only give an upper limit (Tavecchio et al. 2010b). $\Delta t$ is taken as 24 h in this paper.

\emph{1ES 0347-121.} A BL Lac object was detected by HESS between August and December 2006 with an integral flux
corresponding to $2\%$ of Crab flux (Aharonian et al. 2007b). The photon spectrum from 250 GeV to 3 TeV can be
described by a power law with a photon index $\Gamma\sim3.1$. During the VHE observation, no significant
variability was detected, so we take $\Delta t=12$ h. The broadband SED was compiled with the data from the
simultaneous observations of HESS, Swift and ATOM  (Aharonian et al. 2007b). The upper limit detected by {\em Fermi}/LAT is from Tavecchio et al. (2010b).

\emph{1ES 0229+200} A HBL object was observed by HESS in 2005/2006, and confirmed to be a TeV source (Aharonian et
al. 2007c). The integral flux above 580 GeV is $\sim9.4\times10^{-13}$ cm$^{-2}$ s$^{-1}$, corresponding to
$\sim1.8\%$ of Crab flux, and the observed spectrum is characterized by a hard power law with $\Gamma\sim2.5$
from 500 GeV to 15 TeV. During the observation, no significant variability on any scale was detected. Except for
the data of the HESS observation, the broadband SED considered in this work also includes the data of Swift
observation in August 2008 and an upper limit of {\em Fermi}/LAT observations (Tavecchio et al. 2010b). No corresponding variability for the SED is found, so we take $\Delta t=24$ h.

\emph{RGB J0152+017.} A HBL object was detected by HESS in late October and November 2007 (Aharonian et al. 2008).
The observed spectrum is well fit by a power law with $\Gamma\sim2.95$, and the integral flux above 300 GeV
corresponds to $\sim2\%$ of Crab flux. The data of the broadband SED also include the simultaneous observations of Swift
and RXTE (Aharonian et al. 2008), and the upper limit detected by {\em Fermi}/LAT (from Tavecchio et al. 2010b). No corresponding variability for the SED is found, so we take $\Delta t=24$ h.

\emph{PKS 0548-322.} A HBL object was observed between October 2004 and January 2008 with the HESS, and confirmed to
be a TeV source (Superina et al. 2008, Aharonian et al. 2010). The integral flux above 200 GeV is $1.3\%$ of the
Crab flux and the observed spectrum is characterized by a power-law with a photon index $\Gamma\sim2.86$.
Contemporaneous UV and X-ray observations in November 2006 were made by Swift, but it was not be detected by
{\em Fermi}/LAT (Tavecchio et al. 2010b). No significant variability was detected by HESS and Swift. In this work, the
broadband SED includes the data of the Swift and HESS observations, together with the upper limit of {\em Fermi}/LAT
observation. $\Delta t$ is taken as 24 h.

\emph{H 1426+428.} A HBL object with the strongest TeV emission was detected in 2000 and 2001 by Whipple with an
integral flux of $(2.04\pm0.35)\times10^{-11}$ cm$^{-2}$ s$^{-1}$ above 280 GeV (Horan et al. 2002). The object
was monitored by Whipple from 1995 to 1998 during a general blazar survey, but no statistical signal was
detected. No simultaneous broadband data are found and the data of broadband SED are from Wolter et
al. (2008). The spectrum butterfly of the {\em Fermi}/LAT observations (Abdo et al. 2009) is also taken into account. No corresponding variability for the SED is found, so we take $\Delta t=24$ h.

\begin{deluxetable}{llllllllllllllll}
\tabletypesize{\footnotesize} \rotate \tablecolumns{16}\tablewidth{53pc} \tablecaption{Observations and SED fit
results for the sources in our sample}\tablenum{1} \tablehead{\colhead{Source} &
\colhead{State\tablenotemark{a}} & \colhead{z\tablenotemark{b}}  & \colhead{$p_{1}\tablenotemark{\rm c}$}
& \colhead{$p_{2}$} & \colhead{$\Delta t$}  & \colhead{$\gamma_{\rm min}$} &\colhead{$\delta$} & \colhead{$B$} &  \colhead{$\chi^{2}$} & \colhead{$\delta_{\tau}$}
& \colhead{$F_{\rm 1TeV}$} & \colhead{$L_{\rm bol}$} & \colhead{$P_{\rm jet}$} & \colhead{$M_{\rm BH}\tablenotemark{\rm d}$} & \colhead{Ref}\\
\colhead{}& \colhead{} & \colhead{} & \colhead{}& \colhead{} & \colhead{(h)} & \colhead{} & \colhead{} &
\colhead{(G)} & \colhead{}& \colhead{} & \colhead{(Jy)} & \colhead{(erg/s)} &
\colhead{(erg/s)} & \colhead{$\log M_{\bigodot}$} & \colhead{}\\
\colhead{(1)}& \colhead{(2)} & \colhead{(3)} & \colhead{(4)}& \colhead{(5)} & \colhead{(6)} & \colhead{(7)} &
\colhead{(8)} &\colhead{(9)} &\colhead{ (10)} & \colhead{(11)} & \colhead{(12)}& \colhead{(13)}& \colhead{(14)}& \colhead{15}&\colhead{16}}
\startdata
Mkn 421&L&0.031&2.36&4.4&3&180&29$^{43}_{15}$&0.14$^{1.3}_{0.05}$&0.87&13&2.9E-14&6.3E45&2.4E44&8.29&1\\
--&H&&2.36&3.6&3&20&$>29$&$<0.14$&0.86&16.5&1.6E-13&2.3E46&6.2E45&\nodata&\\
Mkn 501&L&0.034&2.2&3.72&12&150&14$^{23}_{9}$&0.16$^{0.6}_{0.03}$&1.0&9&3.9E-15&1.8E45&8.4E43&9.21&2\\
--&H&&1.86&4.6&1&2&15$^{29}_{10.8}$&0.4$^{1.2}_{0.03}$&6.3&10.8&1.4E-13&2E46&8.7E44&\nodata&\\
W Com&L&0.102&1.6&3.7&12&250&15$^{23}_{14}$&0.18$^{0.3}_{0.05}$&0.69&14&1.7E-15&1E46&4.4E44&\nodata&3\\
--&H&&2&3.56&12&200&14$^{<33}_{>13.5}$&0.17$^{<0.2}_{>0.02}$&2.1&13.5&6.6E-15&2.8E46&7.3E44&\nodata&\\
BL Lacertae&L&0.069&1.64&4&3&30&19$^{26}_{18}$&0.5$^{0.8}_{0.17}$&4.8&18&9.2E-17&1E46&1.6E45&8.23&4\\
--&H&&2.1$^{(1)}$&3.8&3&30&20$^{<26}_{>14.8}$&0.2$^{<0.47}_{>0.1}$&0.15&14.8&7.5E-16&9.2E45&6.2E45&\nodata&\\
PKS 2005-489&L&0.071&1.8&4.8&10&200&$>14$&$<0.23$&5.8&14&7.3E-16&5.9E45&1.8E44&9.03&5\\
--&H&&2.5&3.2&10&150&42$^{57}_{27}$&0.09$^{0.28}_{0.03}$&2.4&14.2&2.6E-15&2.4E46&9.8E44&\nodata&\\
1ES 1959+650&L&0.048&2.3&3.4&10&2&11$^{17}_{9.2}$&1.1$^{1.82}_{0.58}$&0.84&9.2&2E-15&6.6E45&5.6E45&8.09&6\\
--&H&&2.4&3.4&10&2&12$^{<29}_{>10.7}$&0.25$^{<0.4}_{>0.02}$&0.51&10.7&5.1E-14&2.1E46&7.7E46&\nodata&\\
1ES 2344+514&L&0.044&1.78&3.84&12&2&13$^{22}_{7}$&0.12$^{0.42}_{0.05}$&1.8&7&1.7E-15&9.2E44&3.5E44&8.8&7\\
--&H&&2.24&3.4&12&2&$>10$&$<0.11$&2.8&6&1.6E-14&3.1E45&2.4E46&\nodata&\\
PKS 2155-304&L&0.116&2.2&4&2&500&50$^{68}_{32}$&0.16$^{0.6}_{0.06}$&1.3&18&8.4E-15&6.2E46&4E44&8.7$^{\rm Re}$&8\\
--&H&&2.4&3.8&2&200&26$^{<42}_{>25}$&0.4$^{0.53}_{0.14}$&0.47&25&2.9E-13&6.6E47&2.5E45&\nodata&\\
1ES 1101-232&L&0.186&1.8&3.42&12&2&12$^{17}_{11}$&1.05$^{1.8}_{0.45}$&0.93&11&3.8E-16&2.3E46&3.7E44&\nodata&9\\
--&H&&$2.06^{(2)}$&3.54&12&2&$>8.3$&$<1.1$&0.84&8.3&6.8E-16&1.6E46&2.8E45&\nodata&\\
S5 0716+714&L&0.26&$1.9^{(2)}$&4.16&3&150&$>32$&$<0.7$&3.4&13&1.3E-16&1.7E47&1.2E45&8.6$^{\rm Re}$&10\\
--&H&&$1.96^{(2)}$&3.9&3&200&$>30$&$<0.3$&1.7&30&7.6E-15&6.2E47&3.4E45&\nodata&\\
3C 66A&&0.44&2.2&4.54&12&200&24$^{28}_{20}$&0.2$^{0.31}_{0.13}$&2.&20&4.1E-15&1.2E48&5.4E45&\nodata&11\\
PG1553+113&&0.3&1.46&3.68&24&200&32$^{38}_{26}$&0.13$^{0.22}_{0.09}$&2.3&25&5.9E-15&3.1E47&7.3E44&\nodata&12\\
1ES 1218+30.4&&0.182&1.86&3.6&12&300&20$^{33}_{10.8}$&0.14$^{0.86}_{0.03}$&0.65&10.8&3.2E-15&2.3E46&1.2E44&8.58&13\\
1ES 1011+496&&0.212&2.1&4.8&24&2&13$^{25}_{11.7}$&0.7$^{1.25}_{0.09}$&5.&11.7&3.2E-15&1.1E47&1.2E46&8.32$^{\rm Wu}$&14\\
PKS 1424+240&&0.5&1.9&5.4&24&100&33$^{41}_{27}$&0.23$^{0.46}_{0.15}$&2.3&27&8.2E-16&6.6E47&1.9E45&\nodata&15\\
RGB J0710+591&&0.125&2.24&3.12&24&1000&$>17$&$<0.07$&2.4&7.7&1.8E-15&1.2E46&5.9E43&8.26&16\\
1ES 0806+524&&0.138&2.2&3.8&24&2&12$^{19}_{8}$&0.32$^{1.32}_{0.11}$&1.3&8&3E-16&6.9E45&1.7E46&8.9$^{\rm Wu}$&17\\
Mkn 180&&0.045&1.8&3.4&24&2&6$^{<12}_{>5.3}$&0.4$^{<0.7}_{>0.15}$&0.75&5.3&2.5E-16&6.7E44&2.1E44&8.21&18\\
H2356-309&&0.165&2.1&3.4&24&2&$>7.6$&$<0.5$&1.1&7.6&3.7E-16&6.4E45&3.5E45&8.6&19\\
1ES 0347-121&&0.188&2.42&3.5&12&100&$>11$&$<0.7$&1.5&11&1.5E-15&2.1E46&2.8E44&8.65&20\\
1ES 0229+200&&0.14&2.08&3.16&24&2&$>8.4$&$<0.48$&10.8&8.4&2.1E-15&2.4E46&2.2E45&9.24&21\\
RGB J0152+017&&0.08&2.1$^{(2)}$&3.4&24&2&5$^{<18}_{>5}$&0.28$^{<0.28}_{>0.01}$&3.3&5&8.1E-16&1.2E45&2.8E45&\nodata&22\\
H1426+428&&0.129&2.4$^{(2,3)}$&3.2&24&200&8.5$^{<15.5}_{>8.4}$&0.1$^{<0.14}_{>0.018}$&1.1&8.4&1.5E-14&2E46&7.5E44&9.13&23\\
PKS 0548-322&&0.069&2.1&3.8&24&2&6$^{<20}_{>5.6}$&0.6$^{<0.6}_{>0.02}$&8.5&5.6&5.2E-16&1.6E45&1.1E45&8.15&24\\

\enddata
\tablenotetext{a}{The state of each source in TeV band. ``H" indicating ``high state" and ``L" indicating ``low
state".}
\tablenotetext{b}{z: redshift.}
\tablenotetext{\rm c}{$p_1$ for six SEDs is cited from the references. Corresponding to: (1) Albert et al. (2007a); (2) Tavecchio et al. (2010b); (3) Wolter et al. (2008).}
\tablenotetext{\rm d}{BH mass for 18 sources, among which fourteen are from Woo \&
Urry (2002), two are from Wu et al. (2002), and two are from other references. The superscript ``Re" denotes the
reference given in column (16) and the superscript ``Wu" denote the reference Wu et al. (2002).}

\tablecomments{Columns: (4) (5) The energy indices of electrons below and above the break; (6) The minimum variability timescale; (7) The minimum Lorentz factor of electrons; (8) The beaming factors $\delta$; (9) The inferred magnetic field strength; (10) $\chi^{2}$ of SED fitting results; (11) The lower limit of $\delta$ derived by the $\gamma$-ray transparency condition; (12) Flux density at 1 TeV;
(13) The bolometric luminosity of each SED; (14) The total jet power; (15) The mass of the black hole in the center of each AGN;
(16) The references.}

\tablerefs{ (1) BLazejowski et al. 2005; Paneque et al. 2009; (2) Tavecchio et al. 2001; Anderhub et al. 2009a; (3) Tagliaferri et al. 2000; Acciari et al. 2008; Acciari et al. 2009a; Tavecchio et al. 2010b; (4) Albert et
al. 2007a; Ravasio et al. 2002; Tavecchio et al. 2010b; (5) HESS Collaboration 2010a; Tagliaferri et al. 2001;
Kaufmann et al. 2010; (6) Krawczynski et al. 2004; Tagliaferri et al. 2008; (7) Albert et al. 2007b; Tramacere et al. 2007; (8) Aharonian et
al. 2009a; Aharonian et al. 2009b; Rieger \& Volpe 2010; (9) Costamante 2007; Reimer et al. 2008; (10) Anderhub et al. 2009b; Vittorini et al. 2009; (11) Abdo et al. 2011; (12) Abdo et al. 2010c; (13) R\"{u}ger et al. 2010; Abdo et al. 2009; (14) Tavecchio et al. 2010b; (15) Acciari et al. 2010c; (16) Acciari et al. 2010b; (17) Acciari et al. 2009c; Abdo et al. 2009; Tavecchio et al. 2010b; (18) Rugamer et al. 2011; Abdo et al. 2009; (19) HESS Collaboration et al. 2010b; (20) Aharonian et al. 2007b; Abdo et al. 2010b; (21) Tavecchio et al. 2010b; (21) Aharonian et al. 2008; (23) Wolter et al. 2008 ; Abdo et al. 2009; (24) Aharonian et al. 2010.}
\end{deluxetable}

\clearpage

\begin{figure*}
\includegraphics[angle=0,scale=0.30]{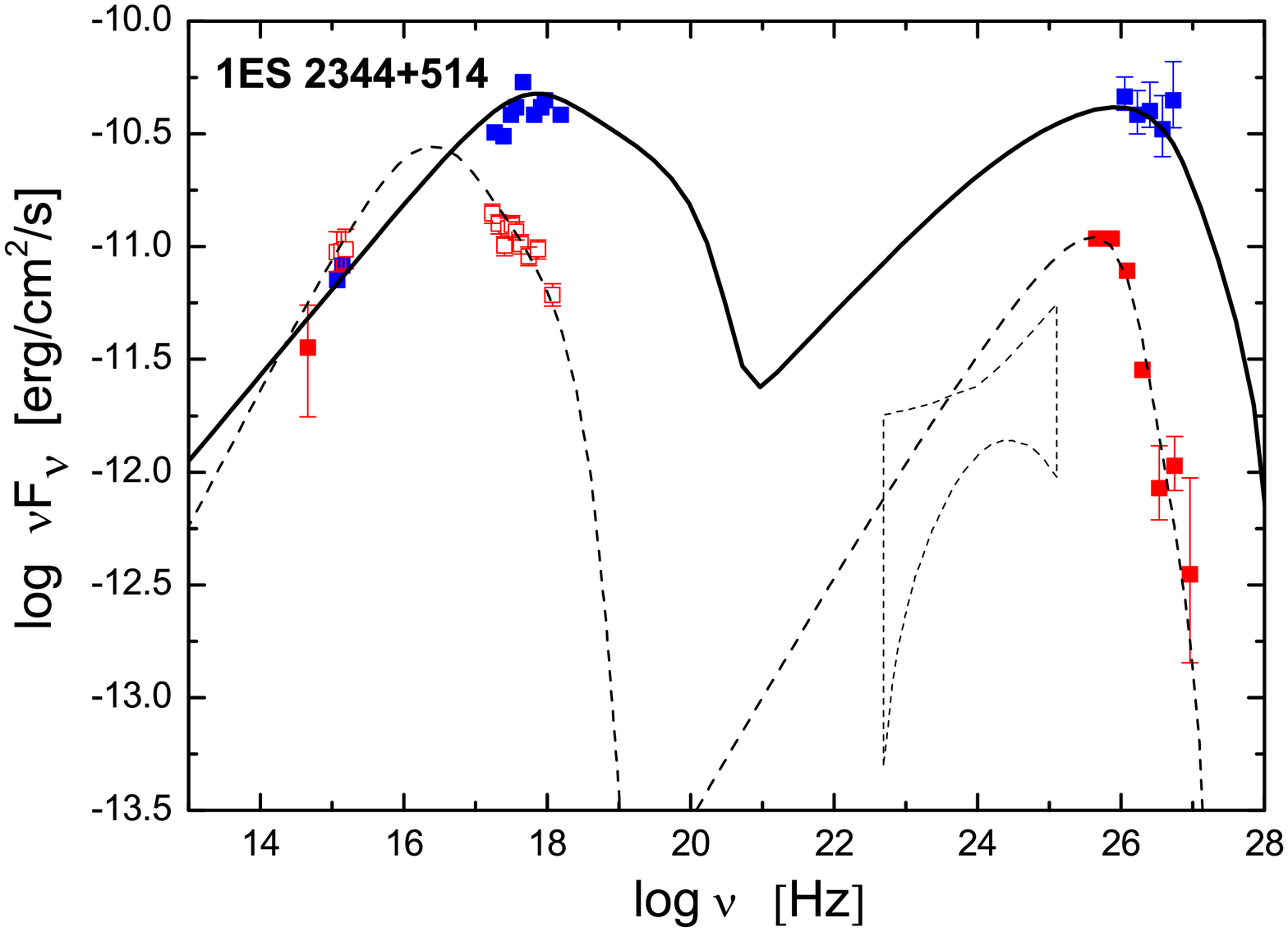}
\includegraphics[angle=0,scale=0.30]{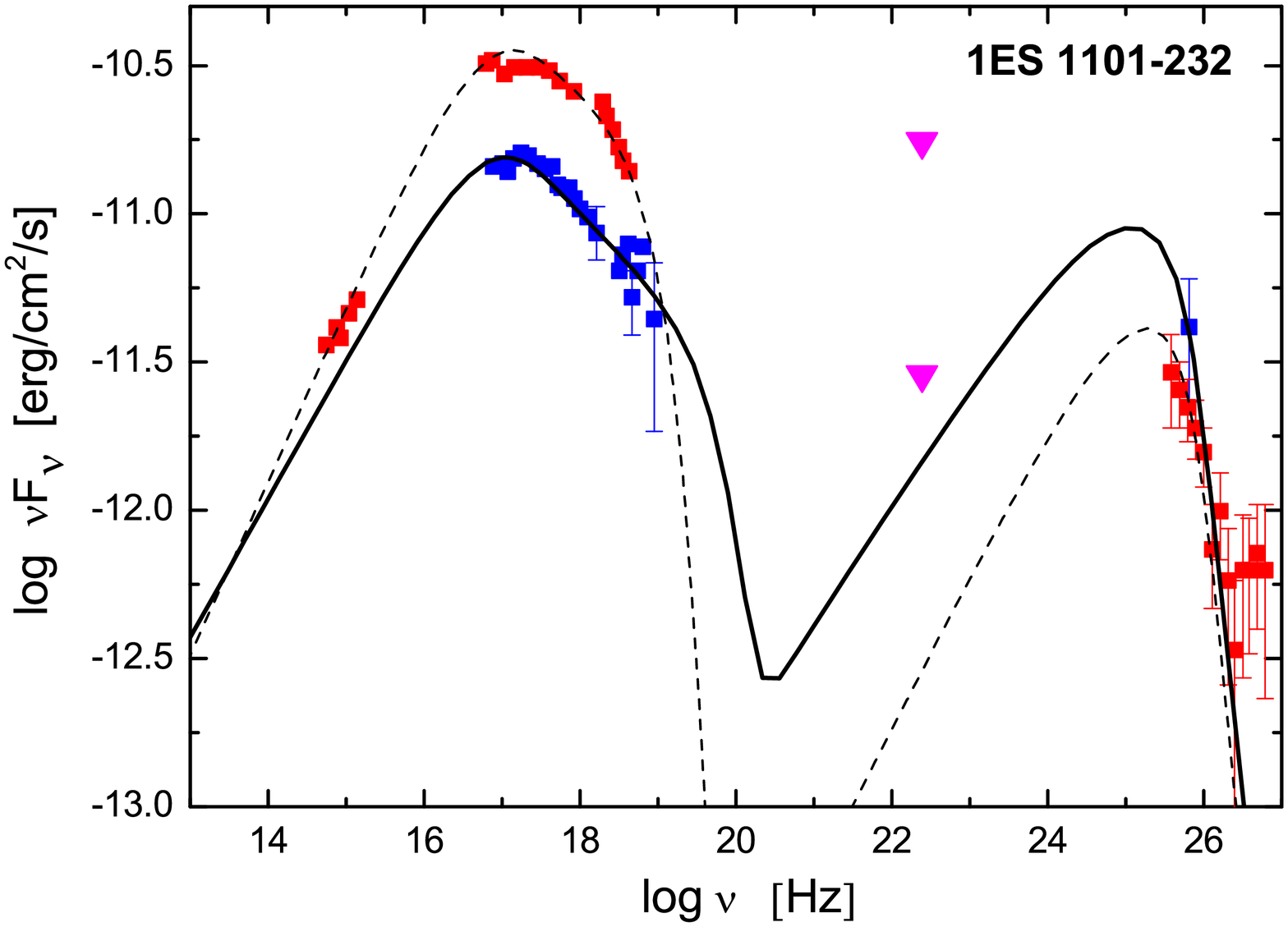}
\includegraphics[angle=0,scale=0.30]{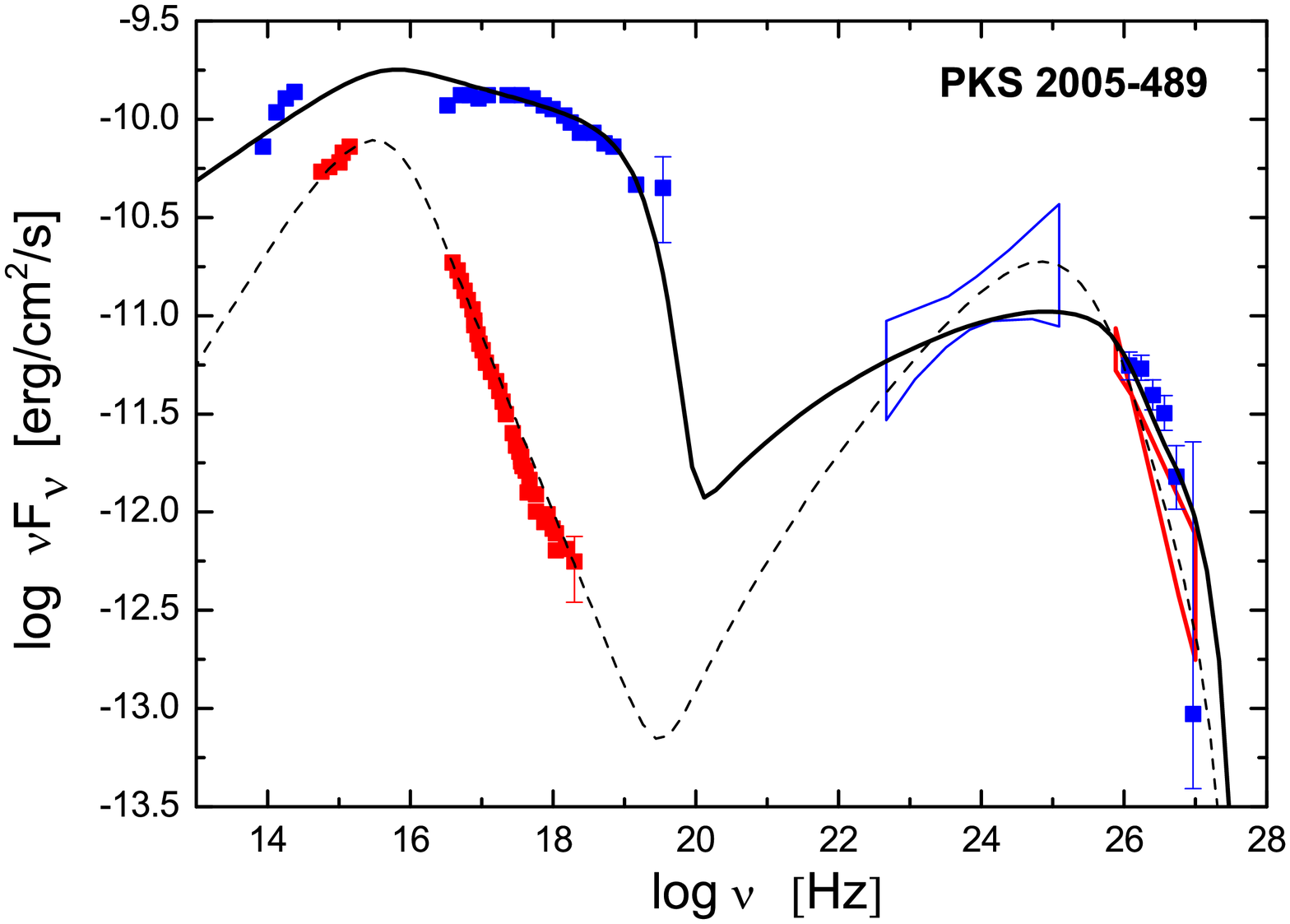}
\includegraphics[angle=0,scale=0.30]{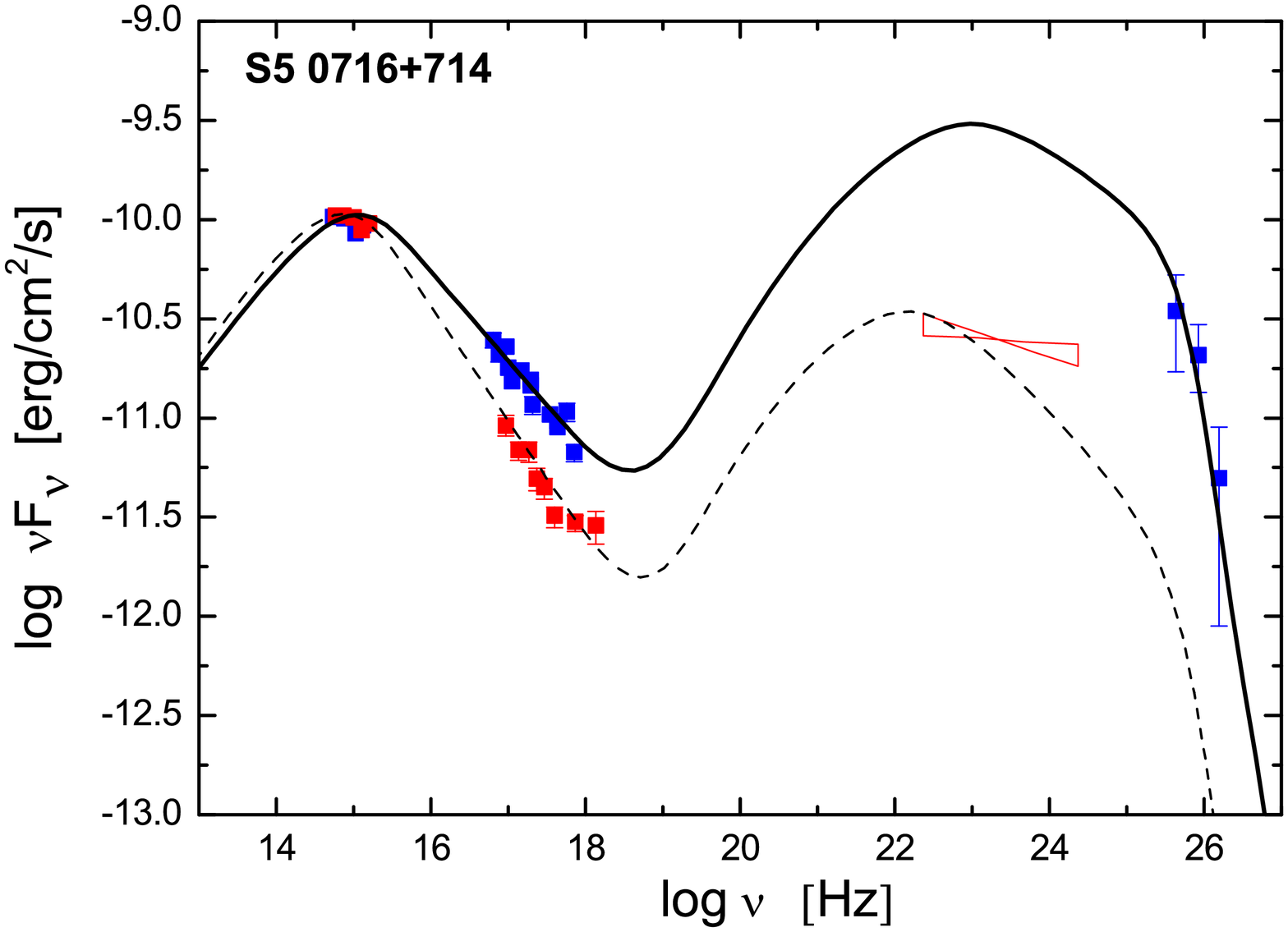}
\includegraphics[angle=0,scale=0.30]{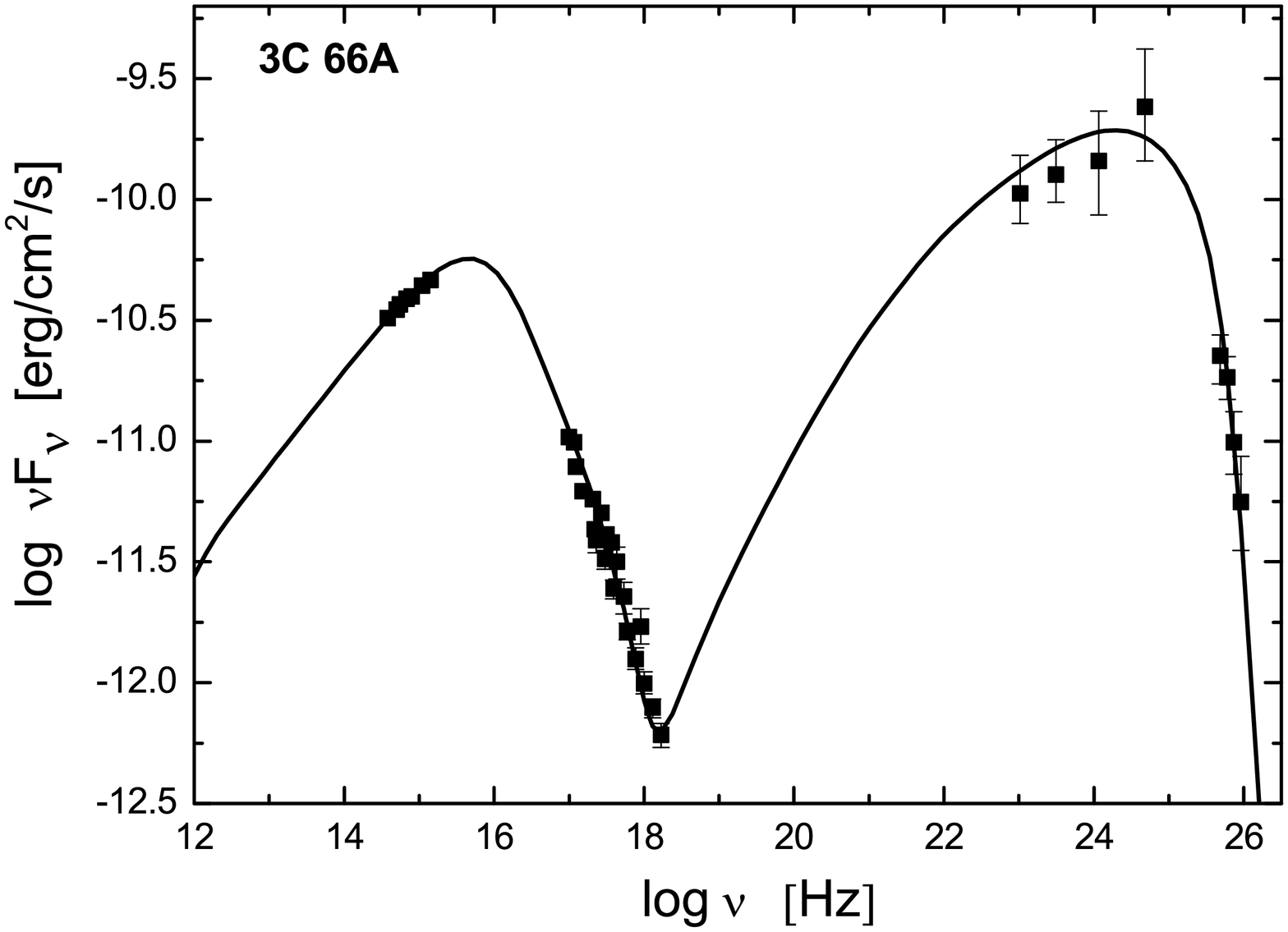}
\includegraphics[angle=0,scale=0.30]{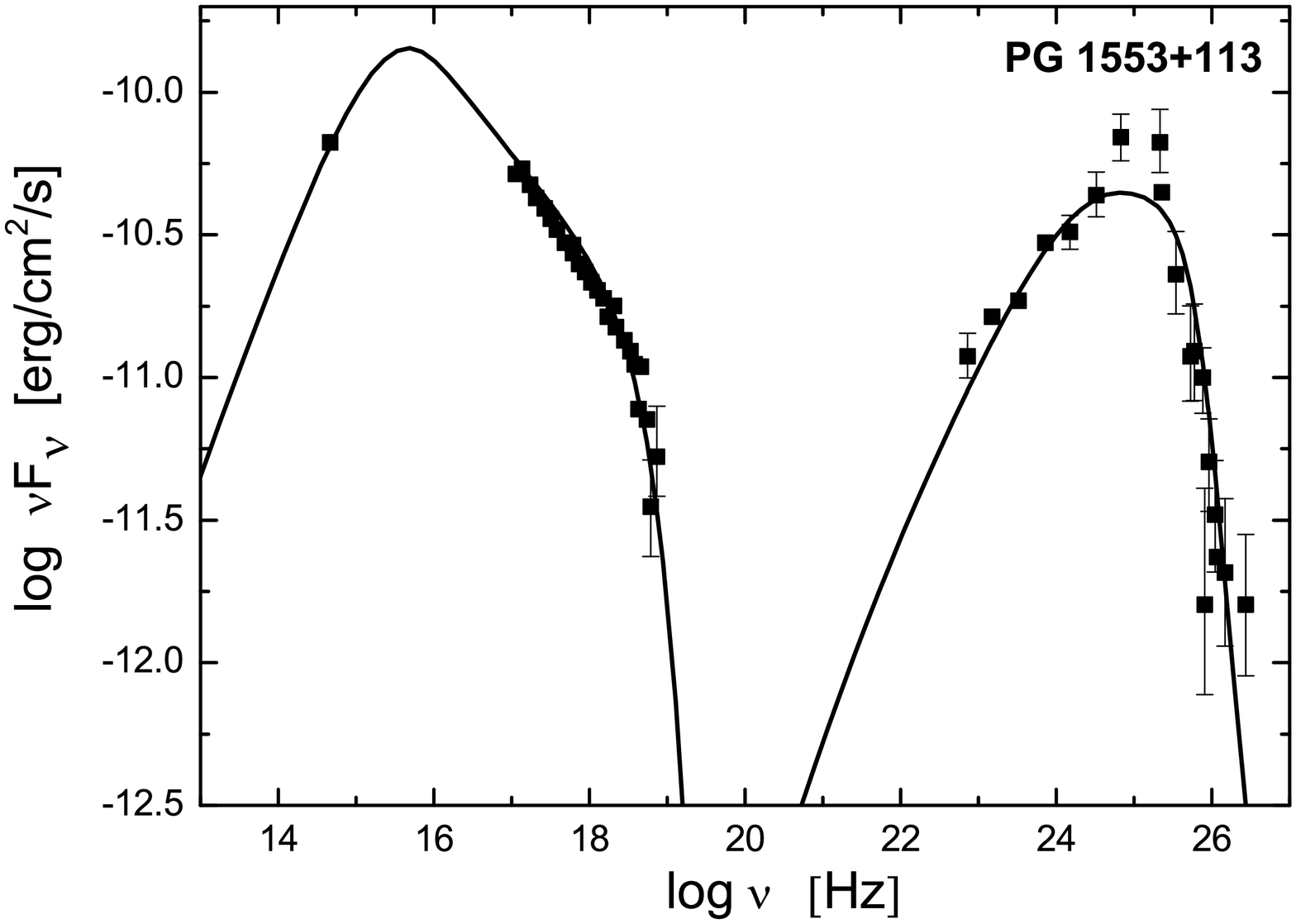}
\includegraphics[angle=0,scale=0.30]{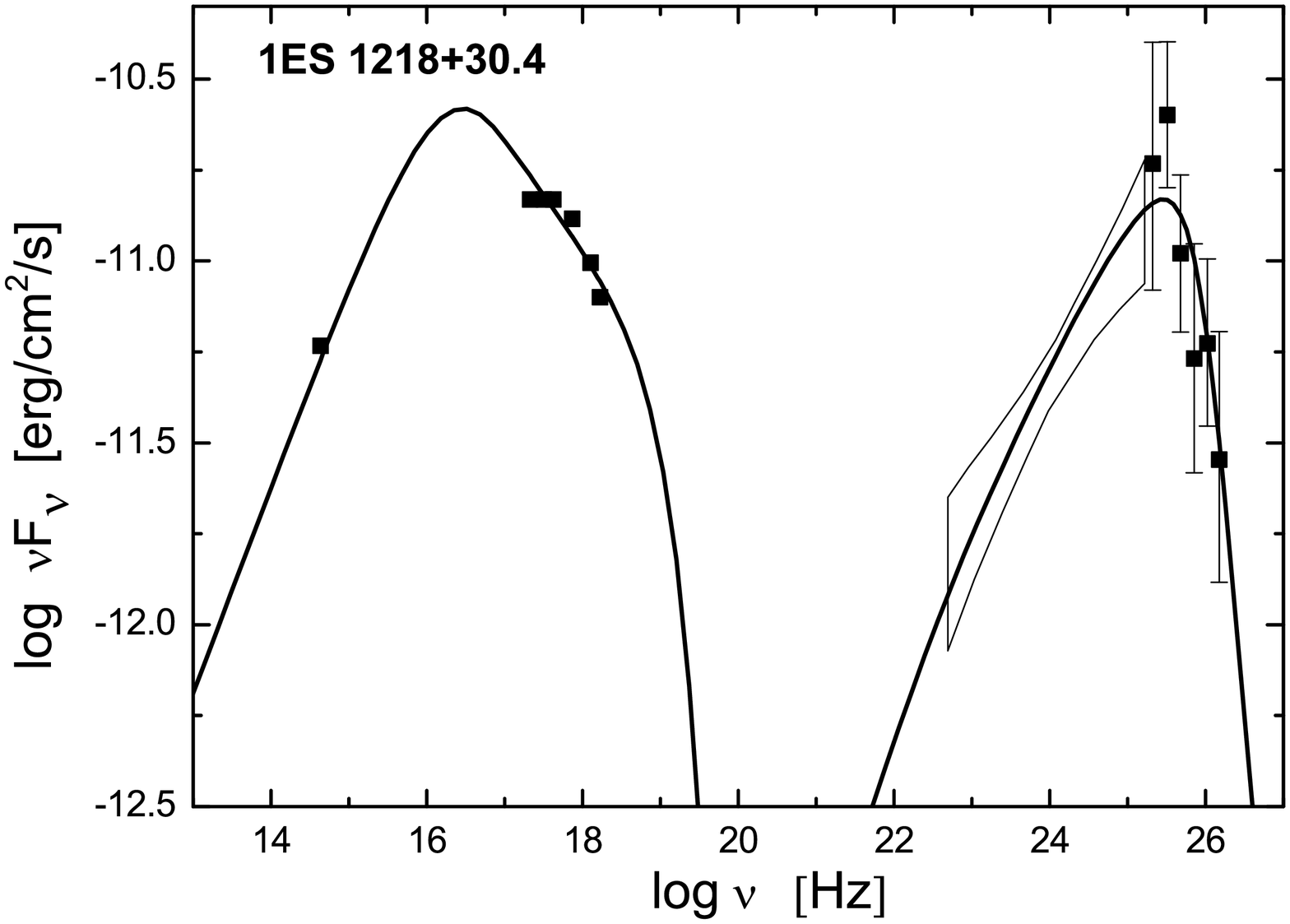}
\hfill
\includegraphics[angle=0,scale=0.30]{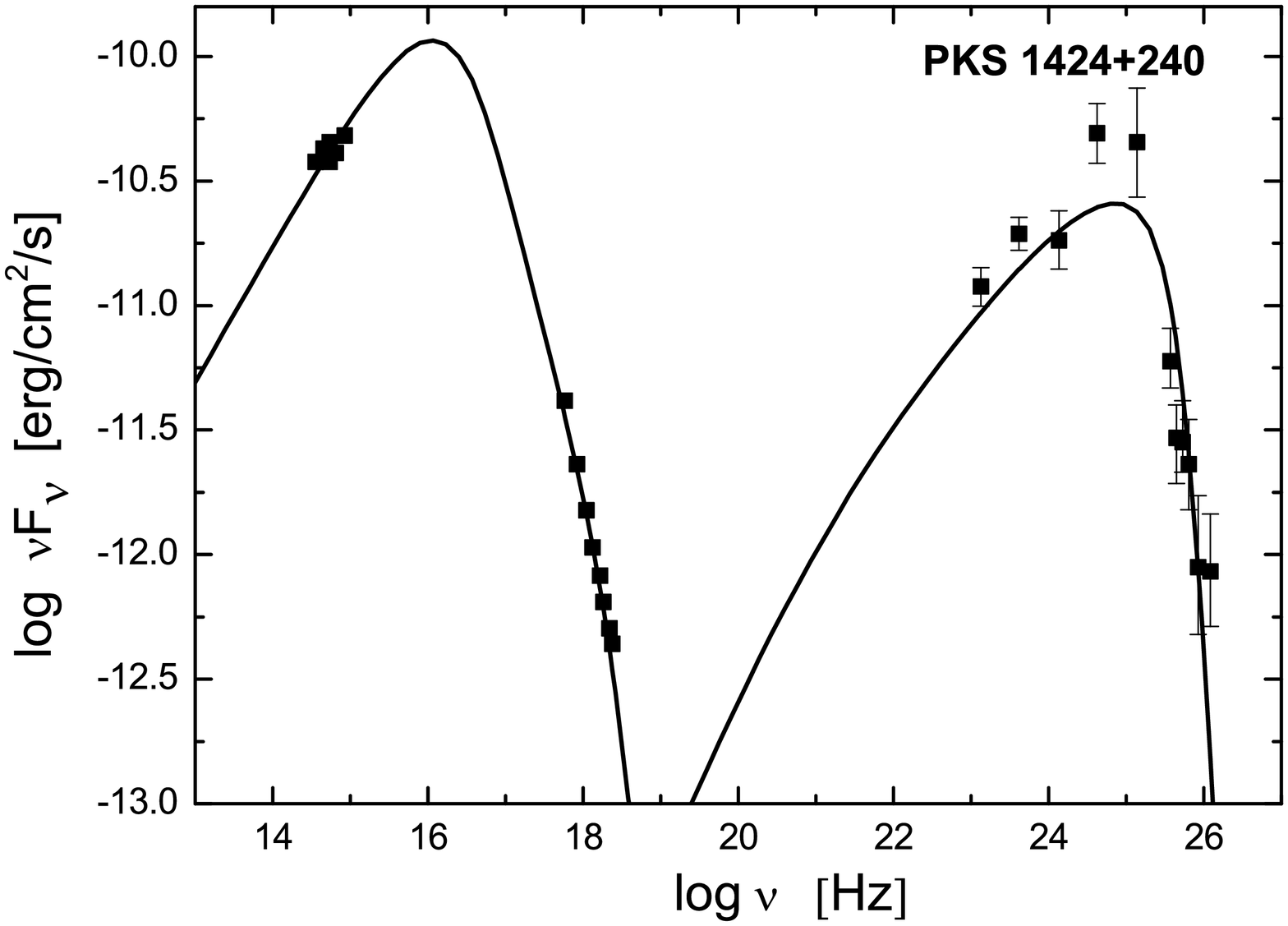}

\center{Fig. 1---  continued}
\end{figure*}

\begin{figure*}
\includegraphics[angle=0,scale=0.30]{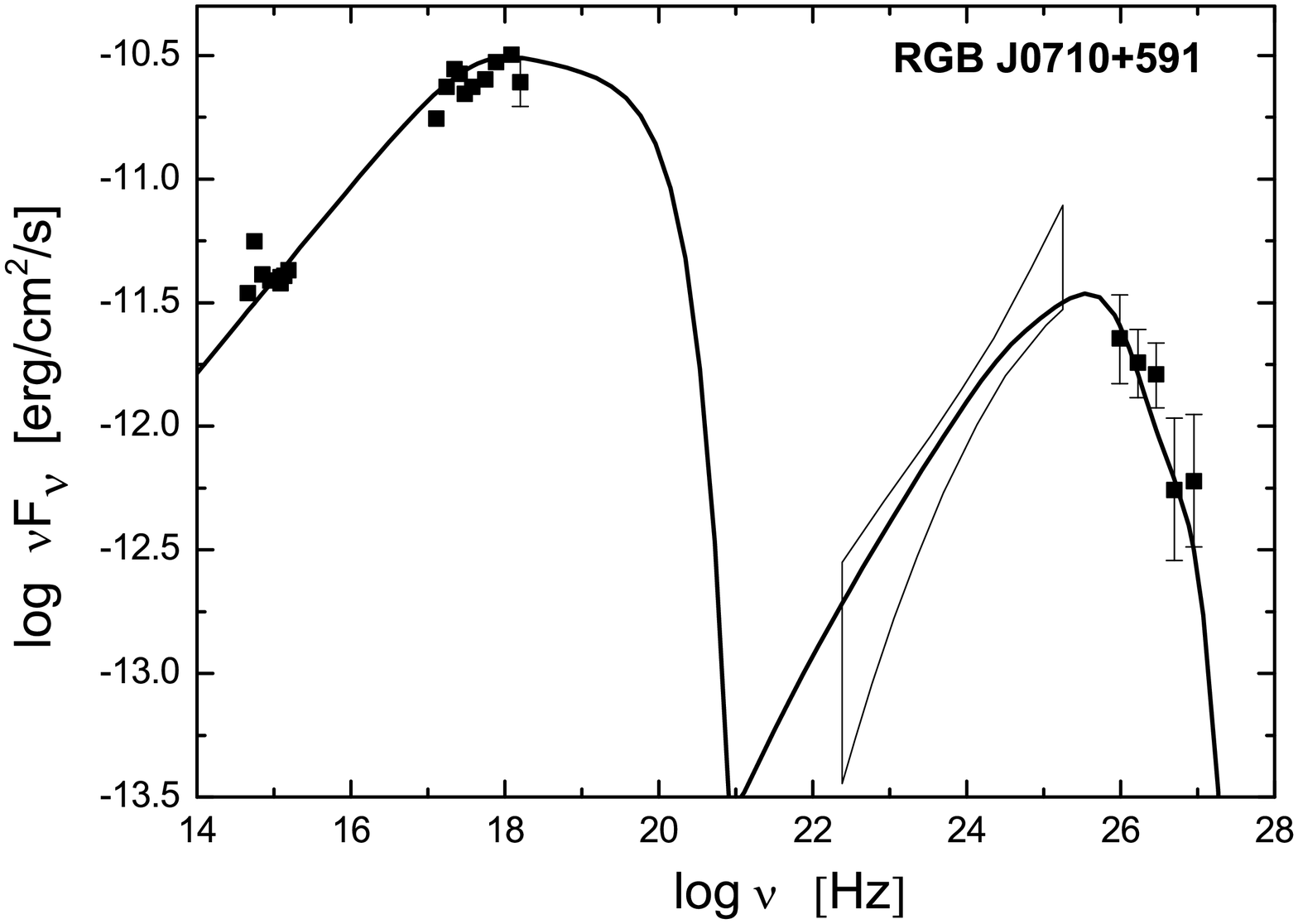}
\includegraphics[angle=0,scale=0.30]{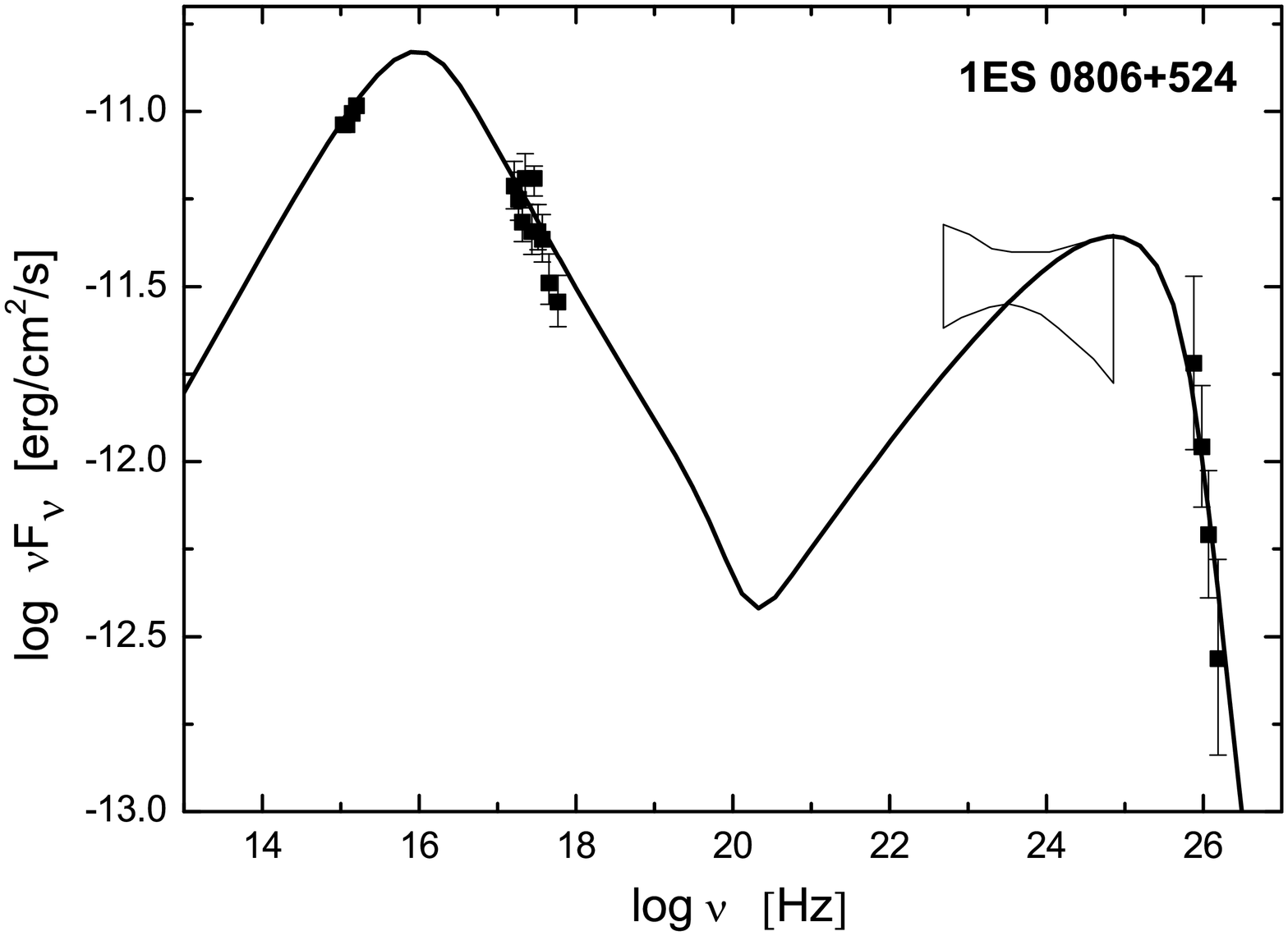}
\includegraphics[angle=0,scale=0.30]{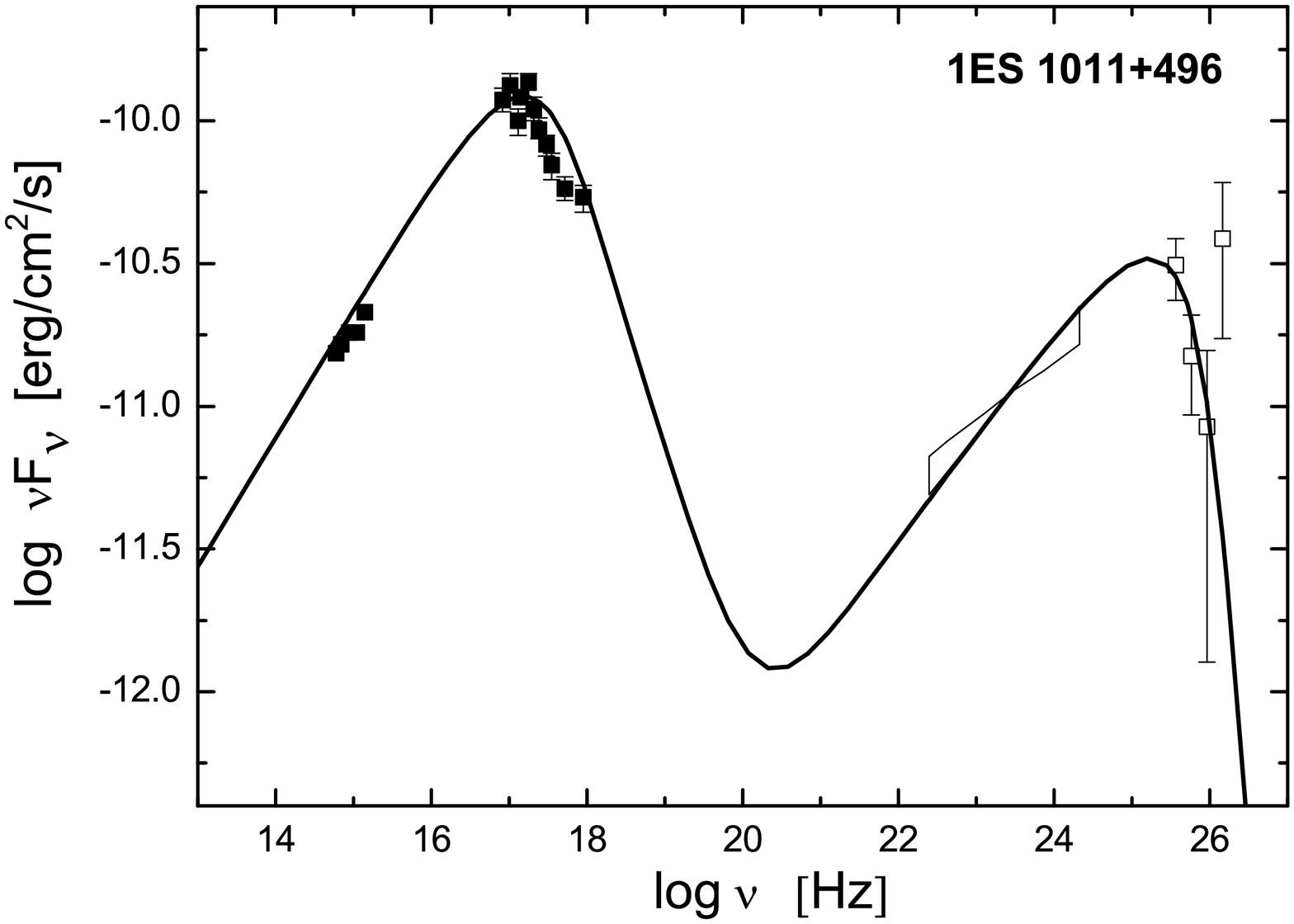}
\includegraphics[angle=0,scale=0.30]{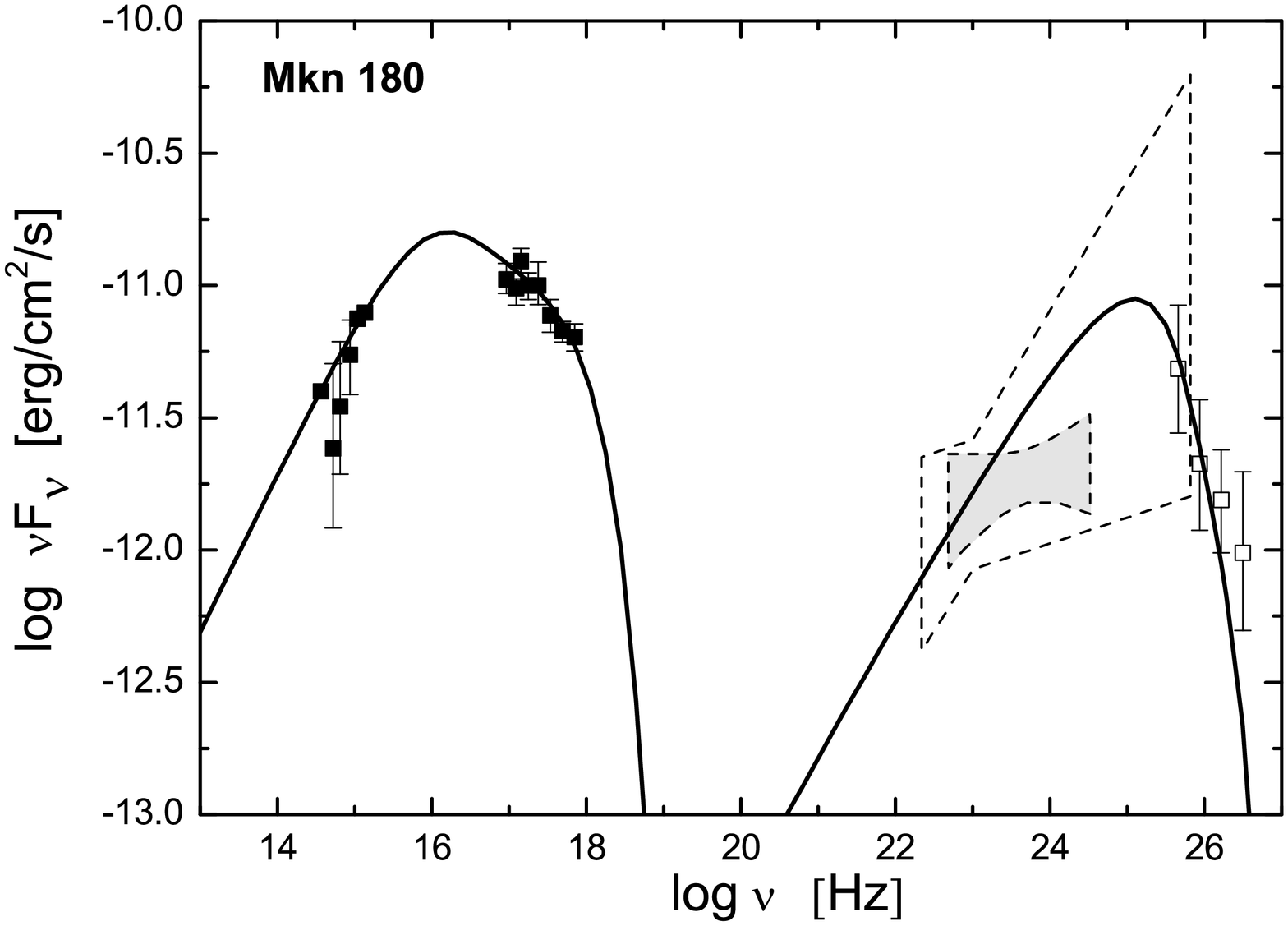}
\includegraphics[angle=0,scale=0.30]{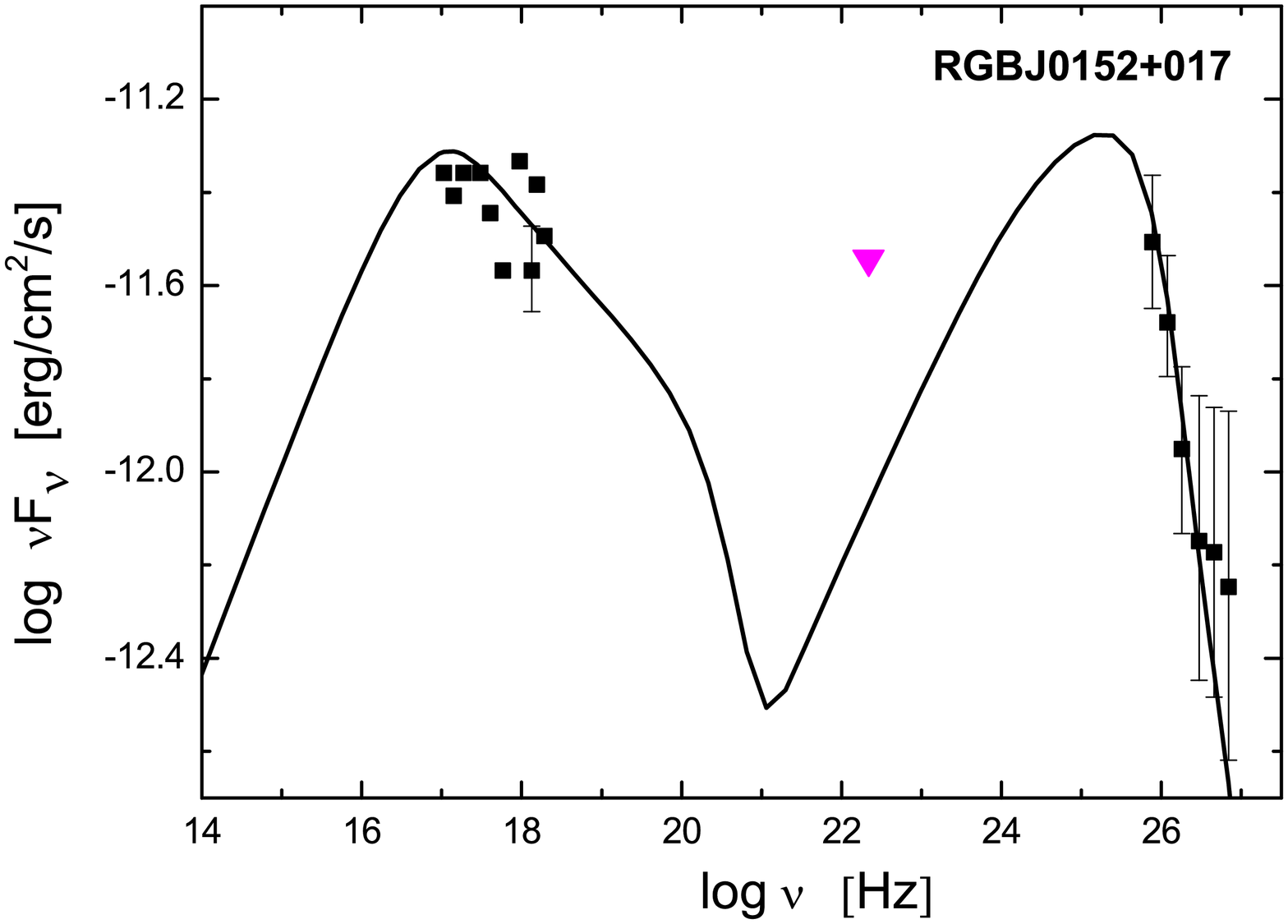}
\includegraphics[angle=0,scale=0.30]{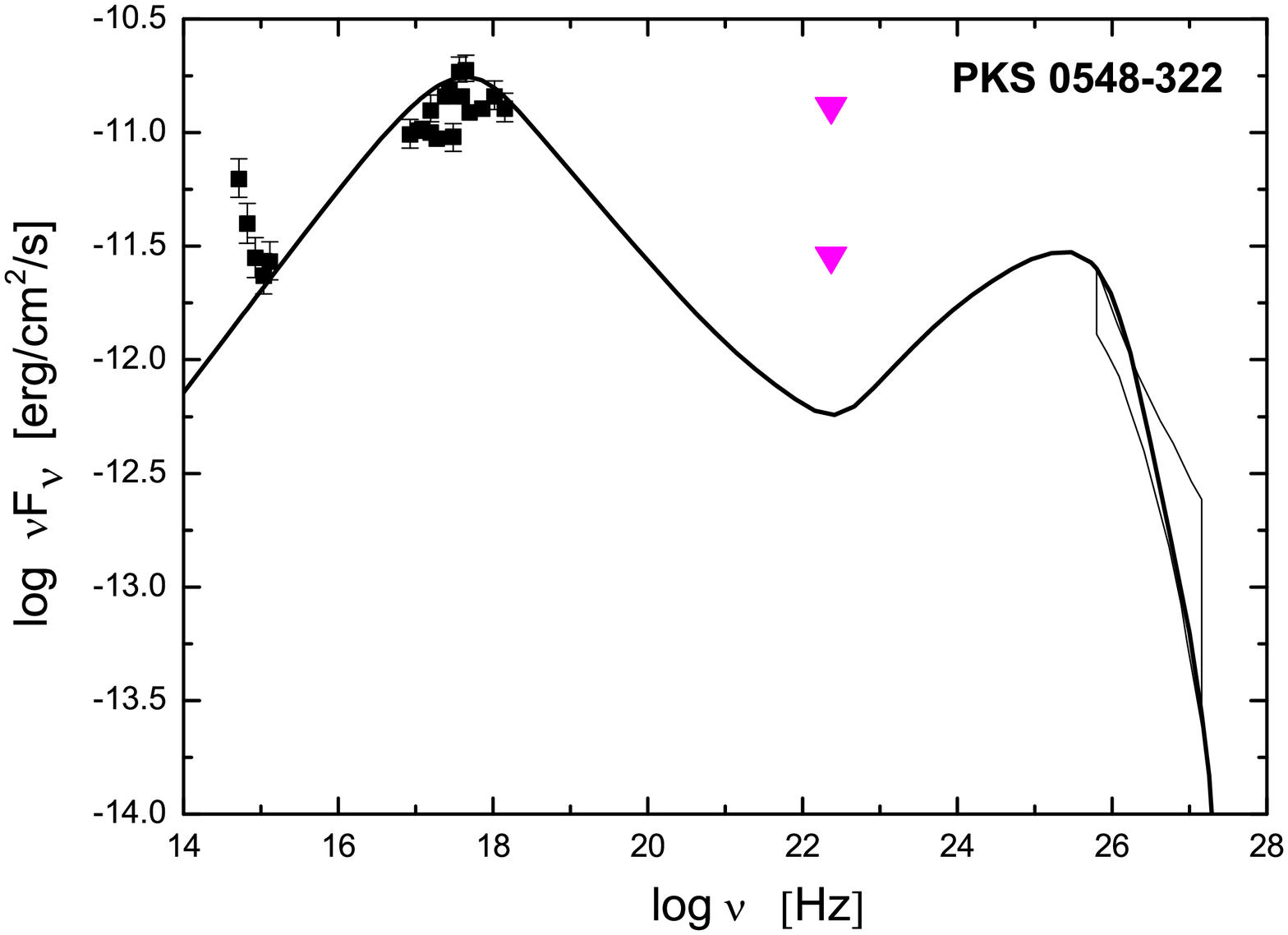}
\includegraphics[angle=0,scale=0.30]{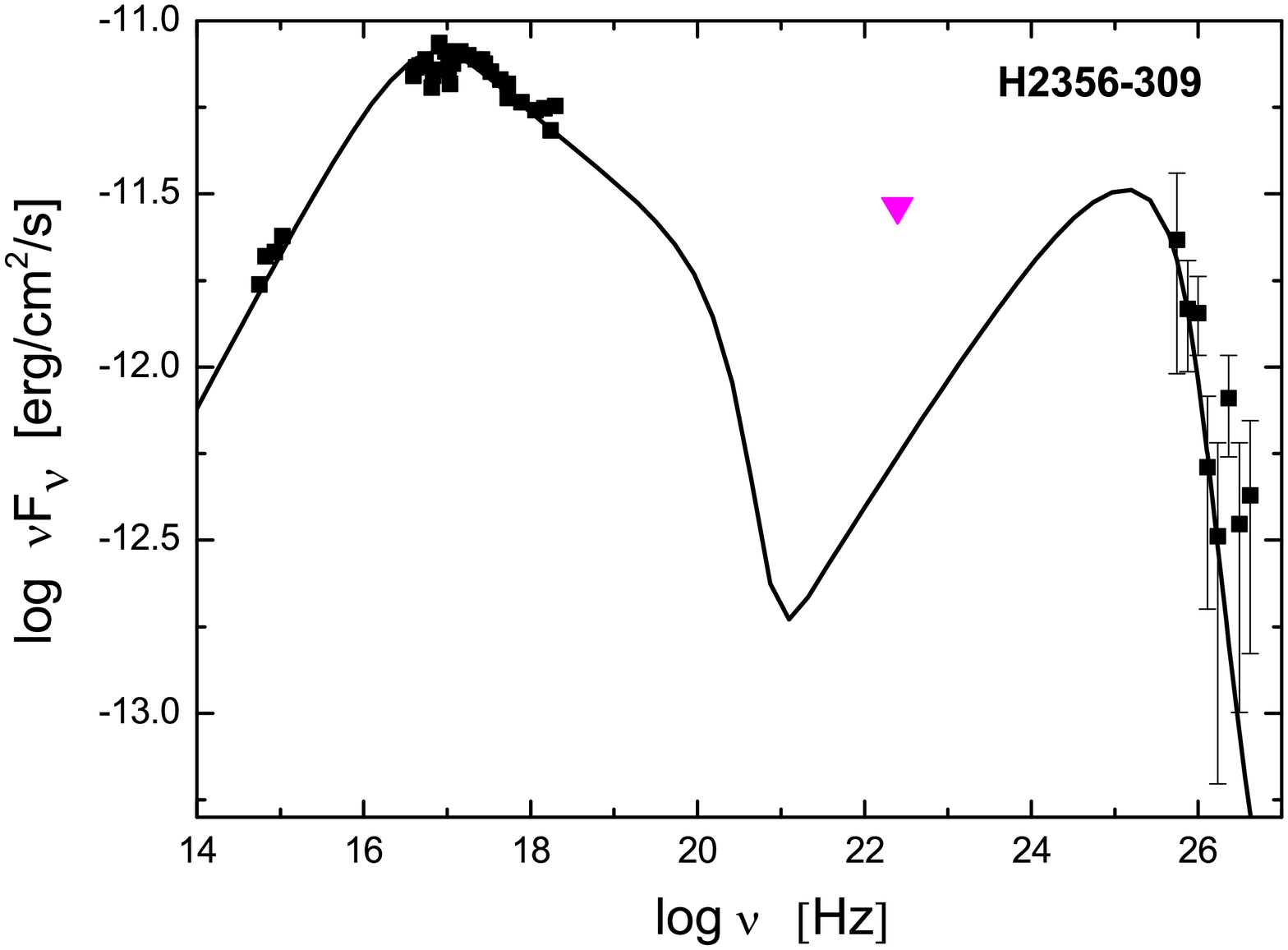}
\hfill
\includegraphics[angle=0,scale=0.30]{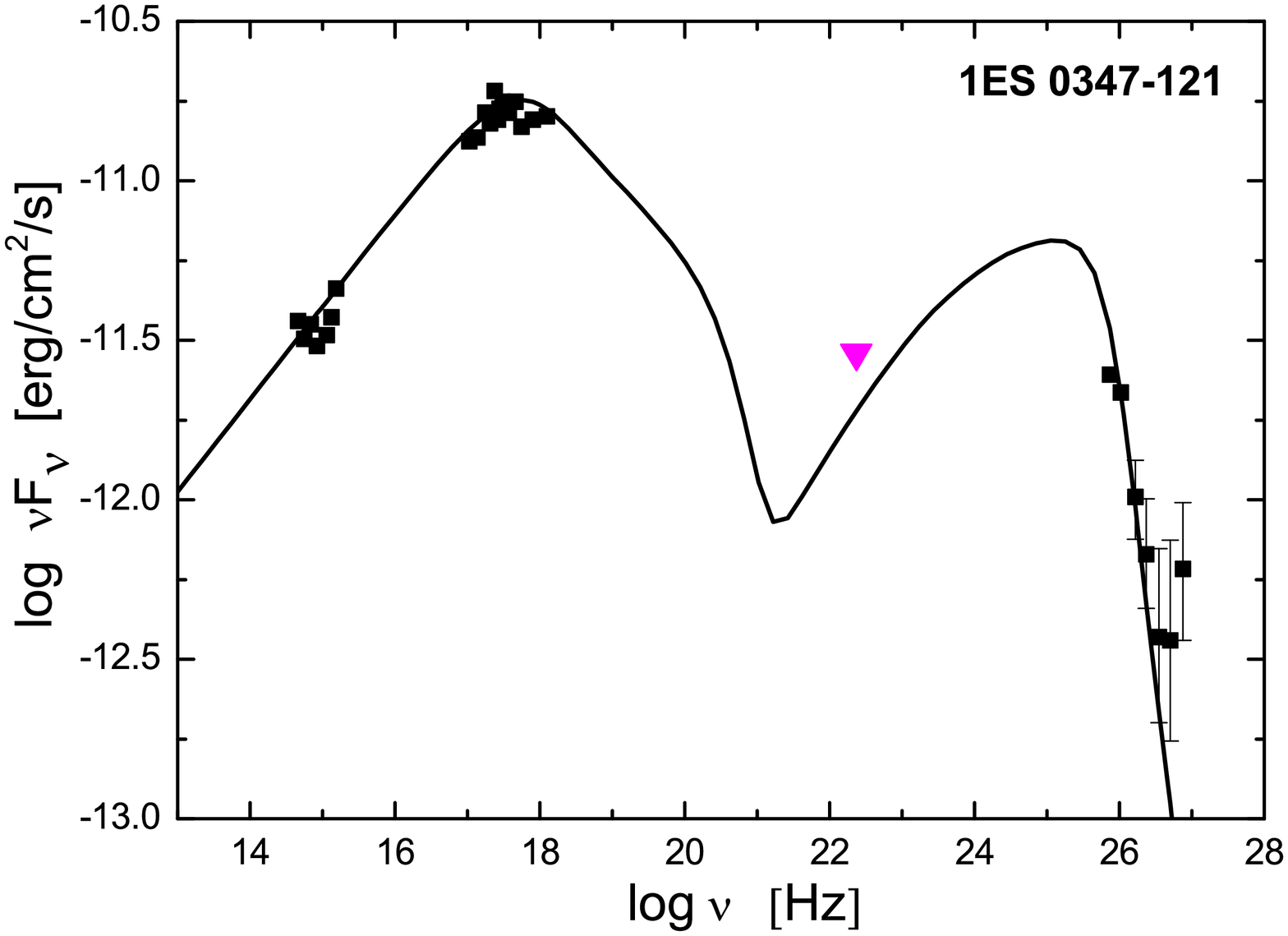}
\center{Fig. 1---  continued}
\end{figure*}

\begin{figure*}
\includegraphics[angle=0,scale=0.30]{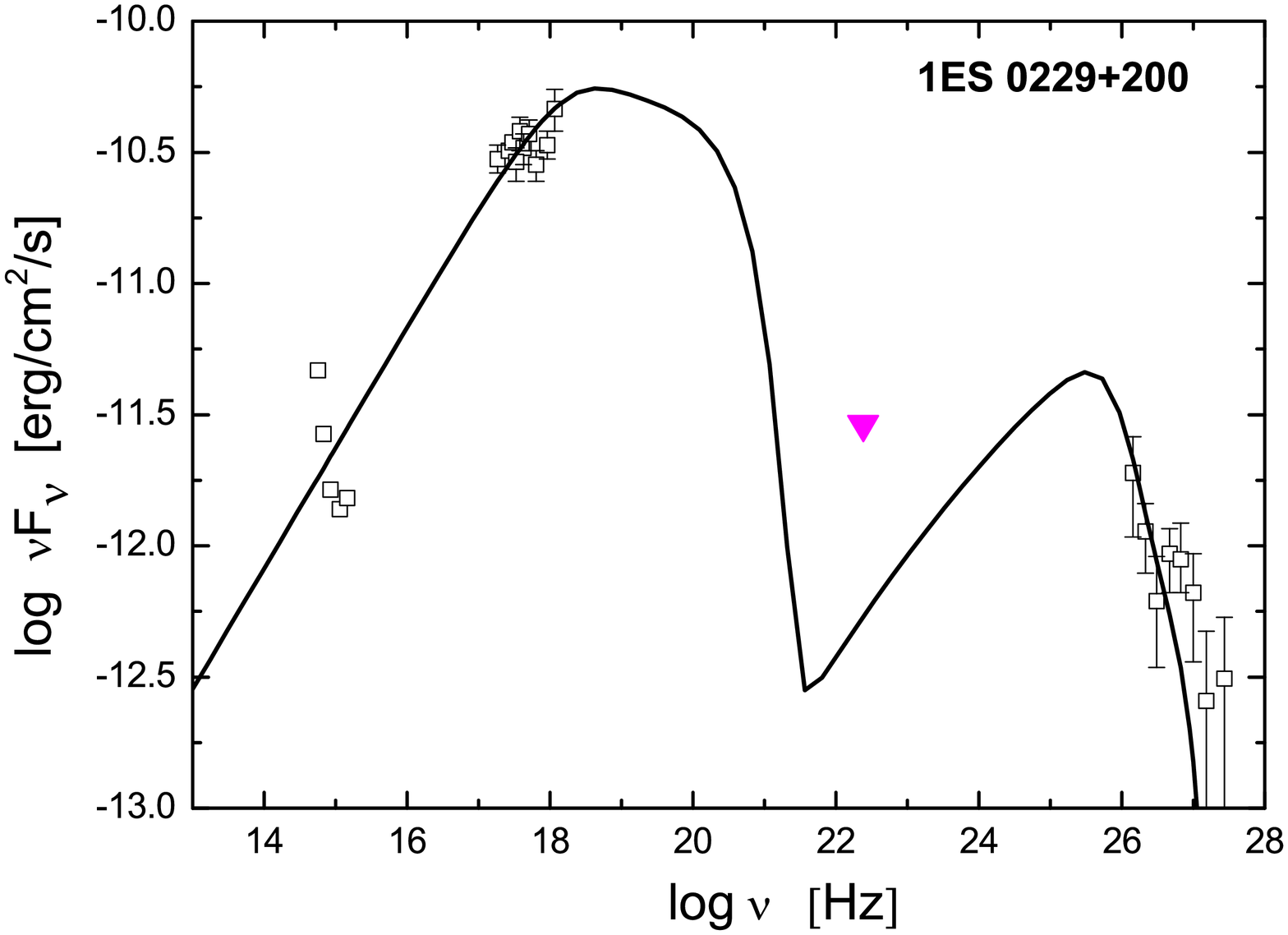}
\hfill
\includegraphics[angle=0,scale=0.30]{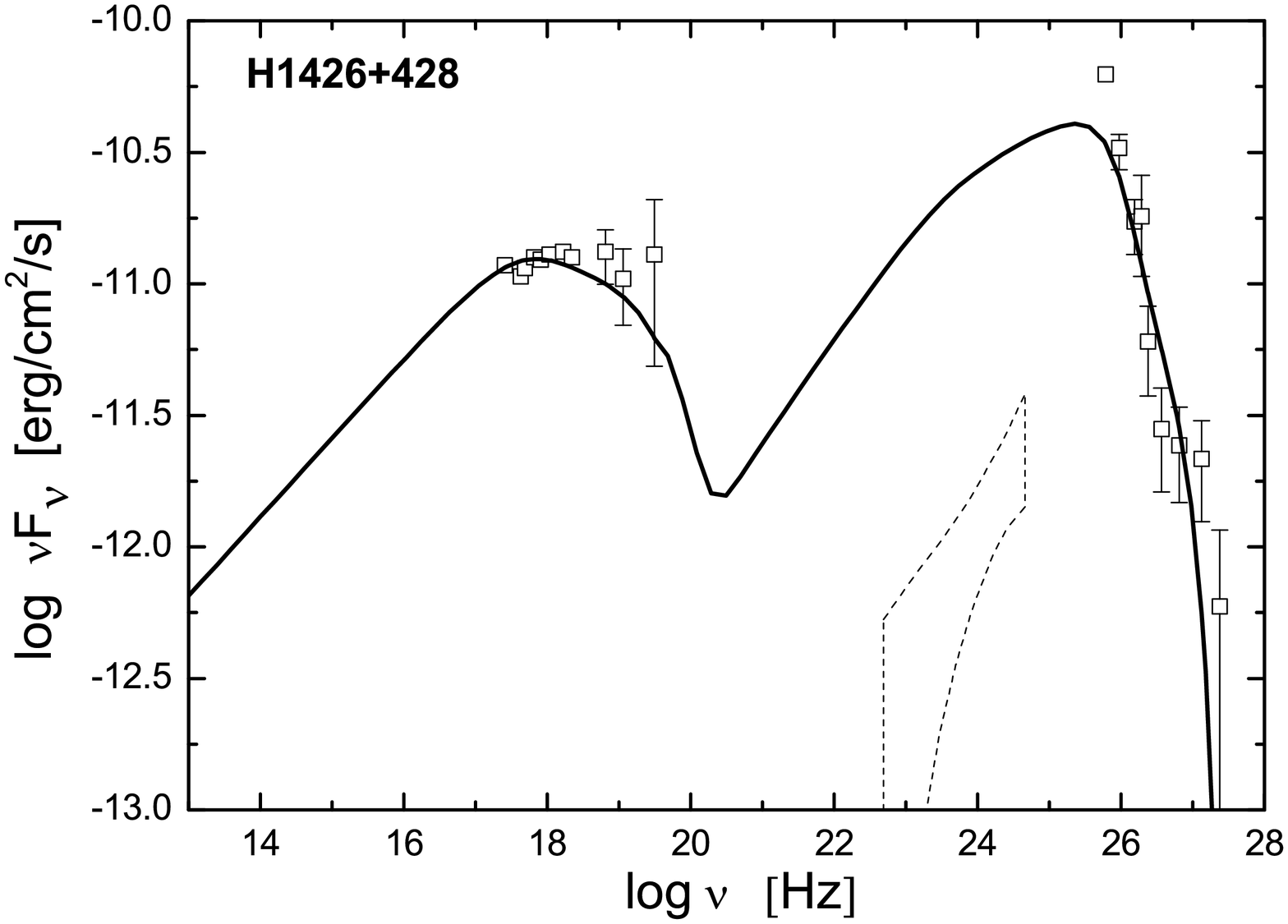}
\center{Fig. 1---  continued}
\end{figure*}

\begin{figure*}
\includegraphics[angle=0,scale=0.26]{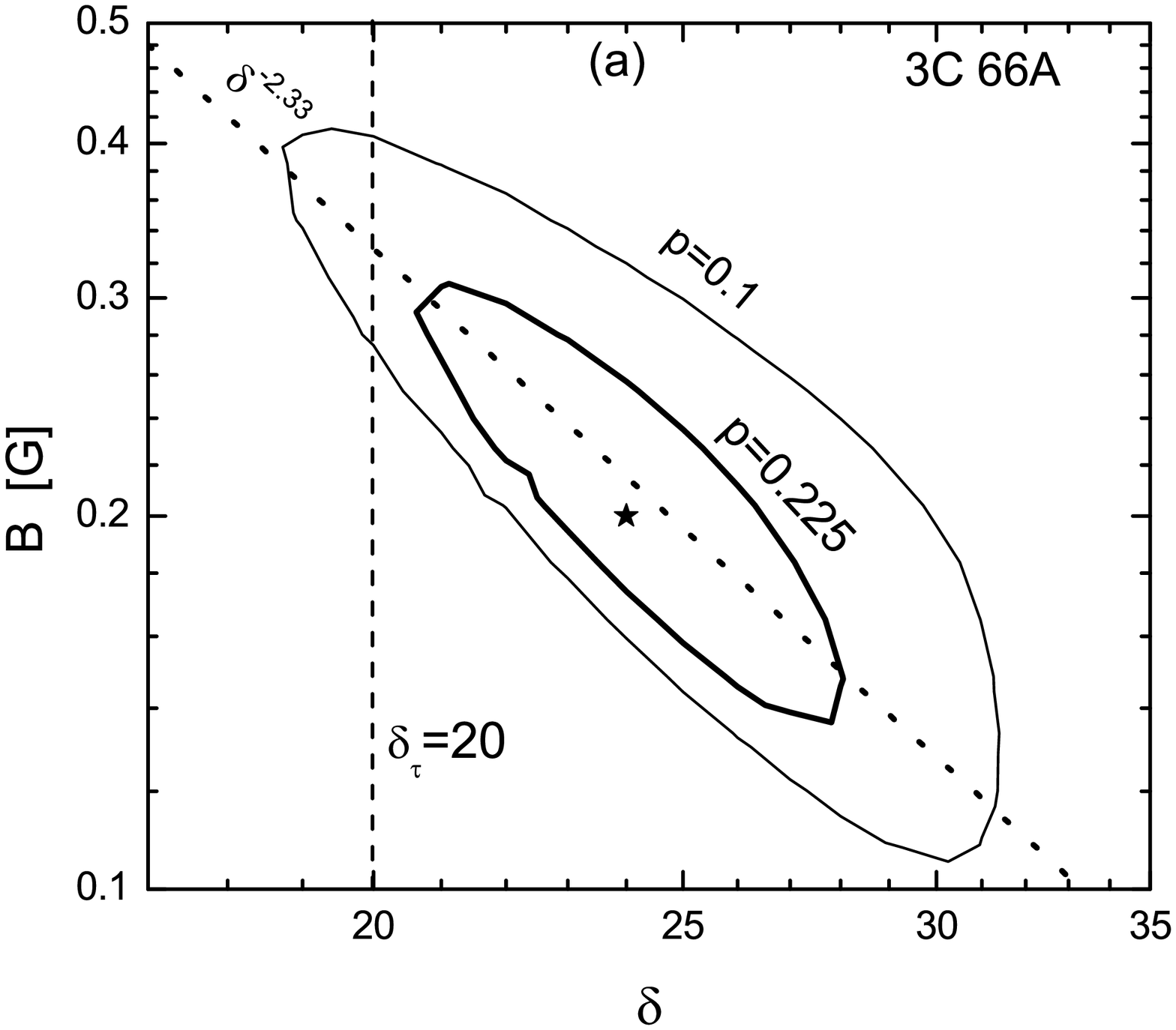}
\includegraphics[angle=0,scale=0.26]{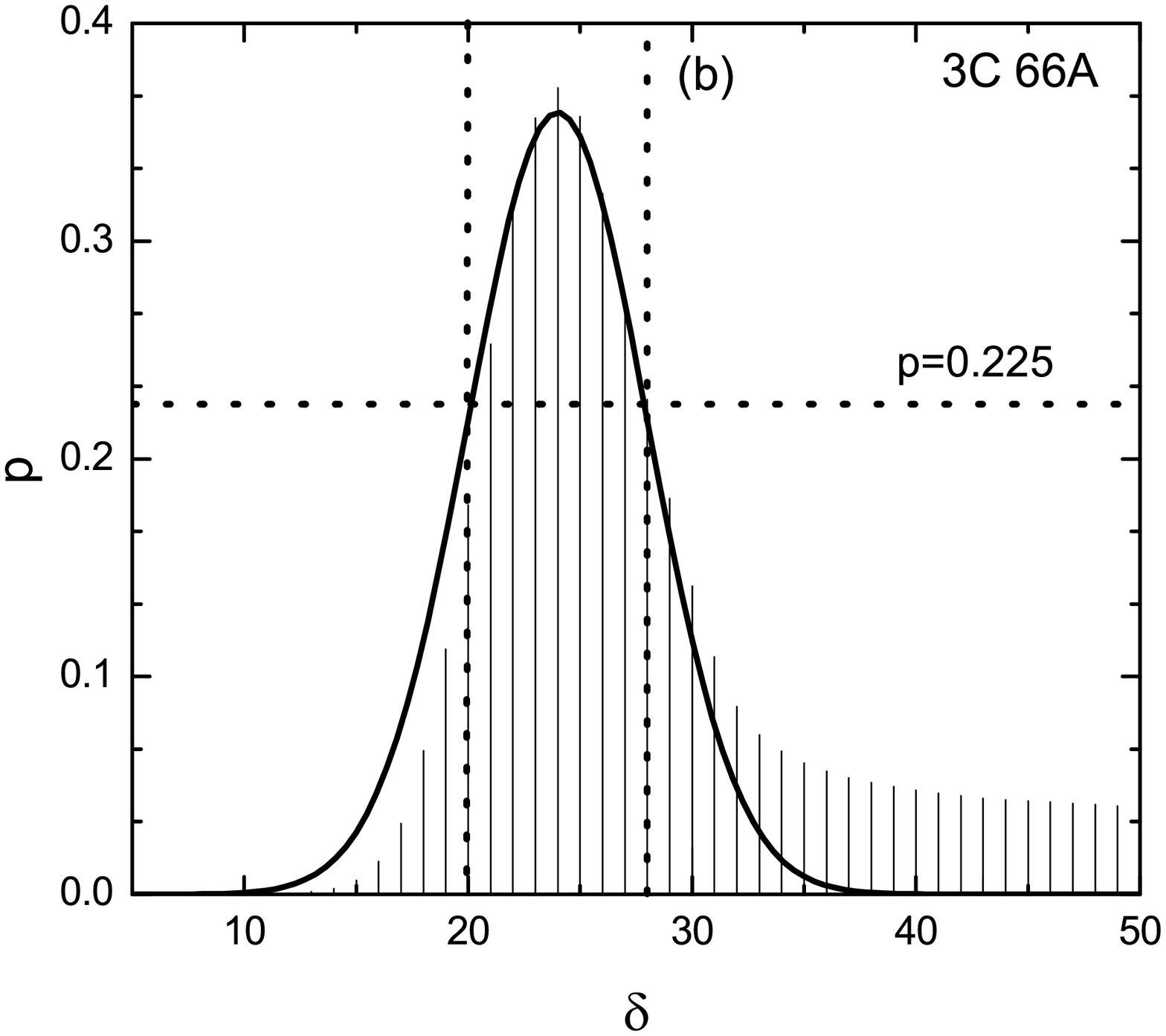}
\includegraphics[angle=0,scale=0.26]{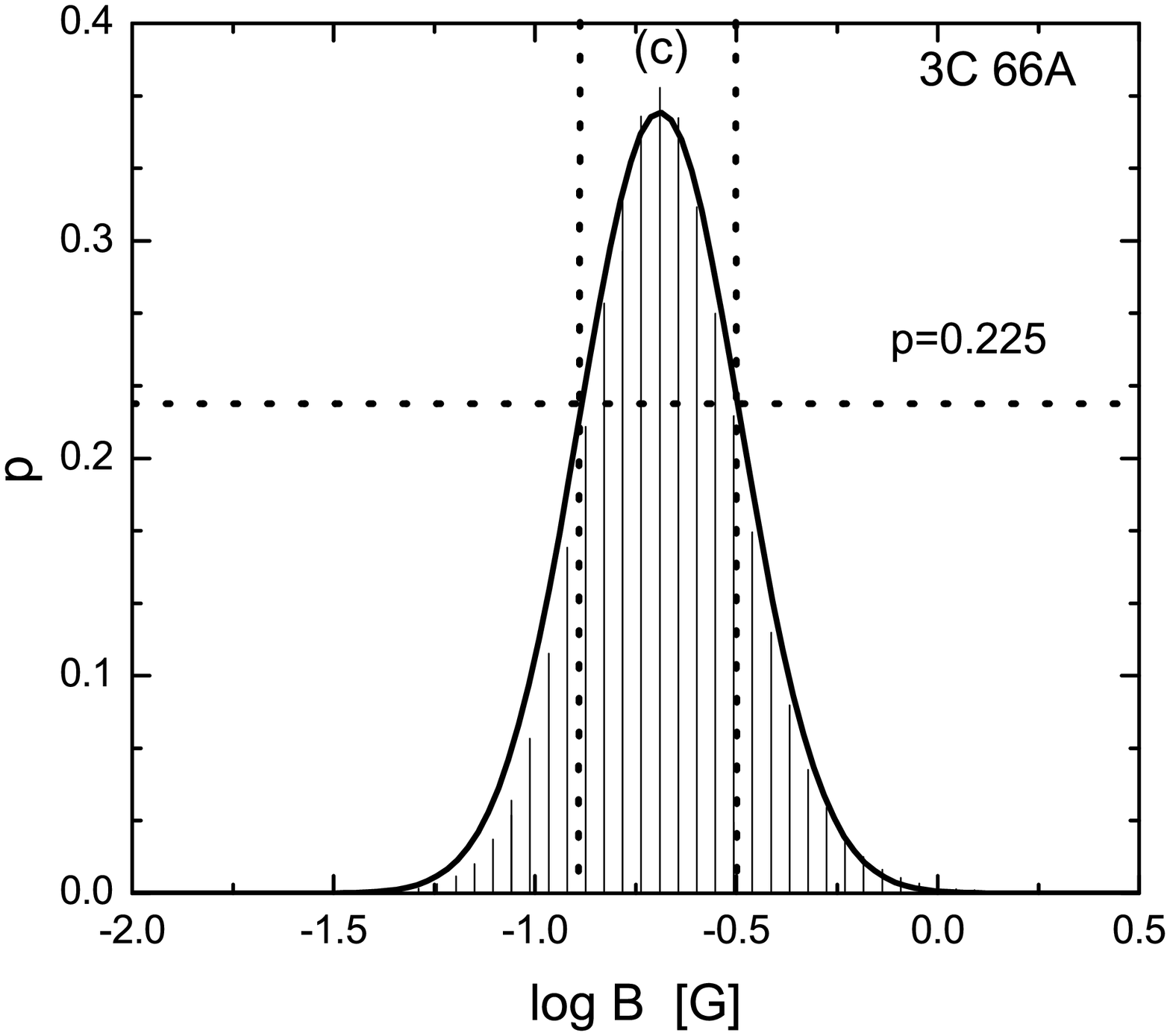}
\includegraphics[angle=0,scale=0.26]{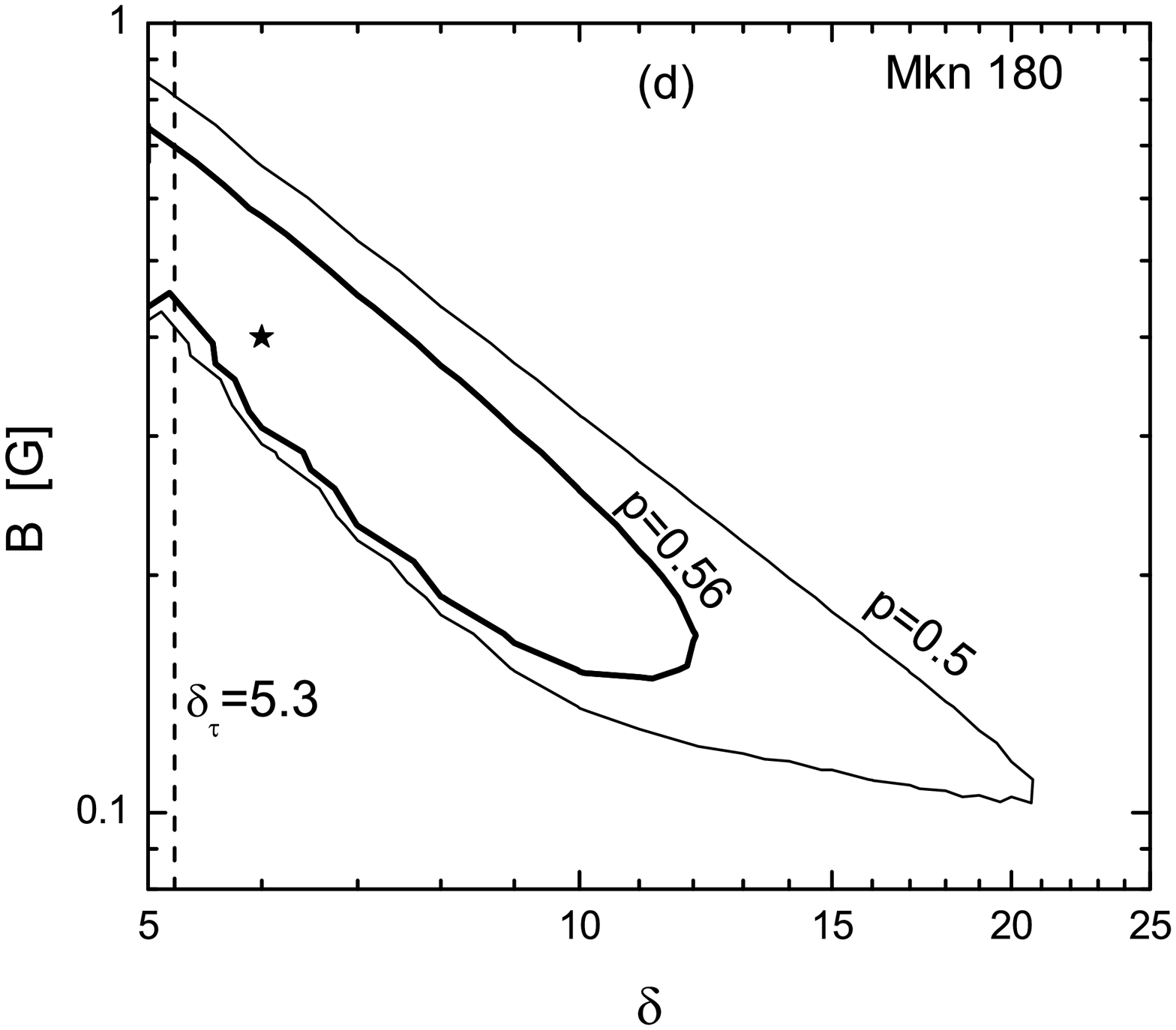}
\includegraphics[angle=0,scale=0.26]{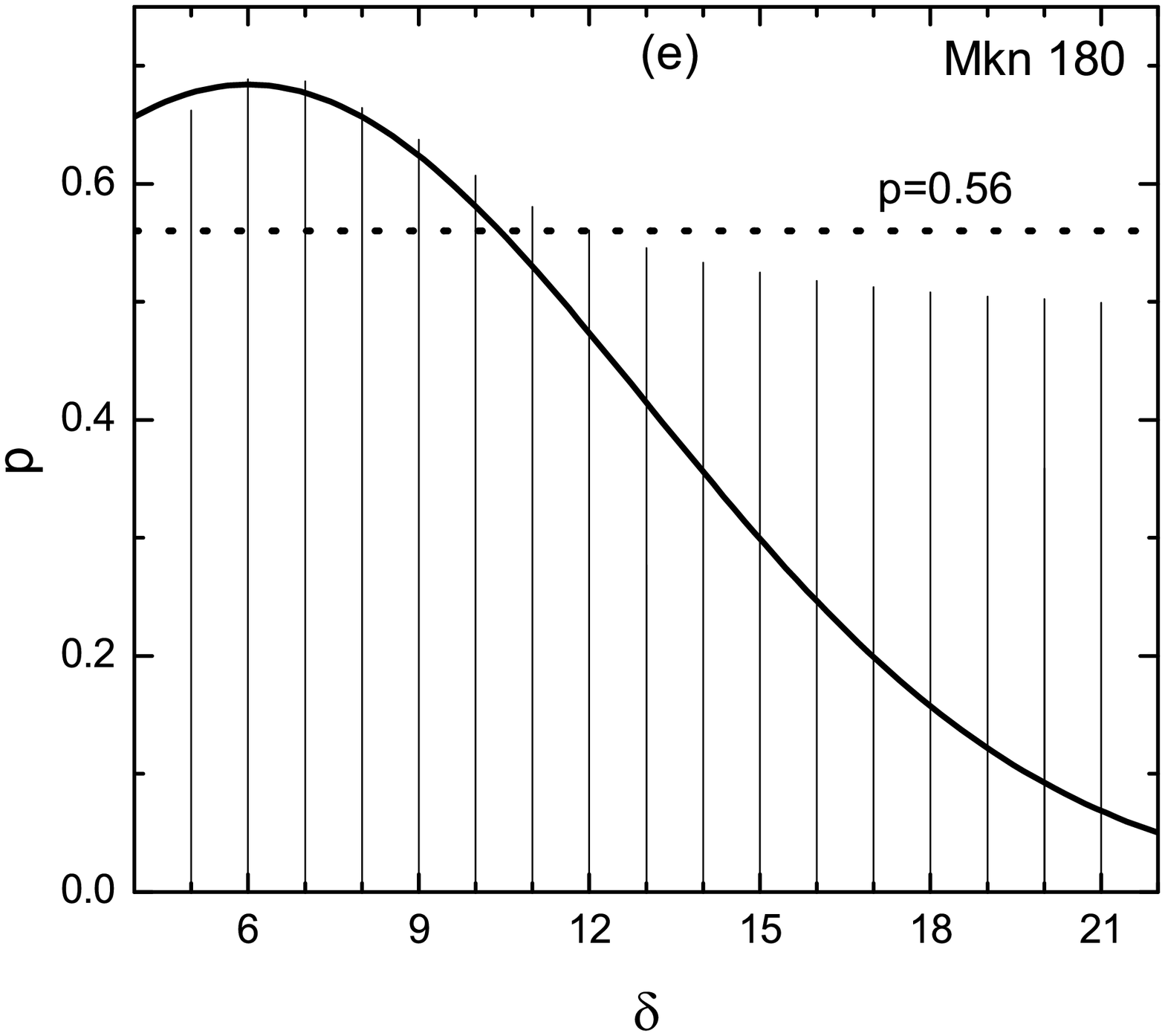}
\includegraphics[angle=0,scale=0.26]{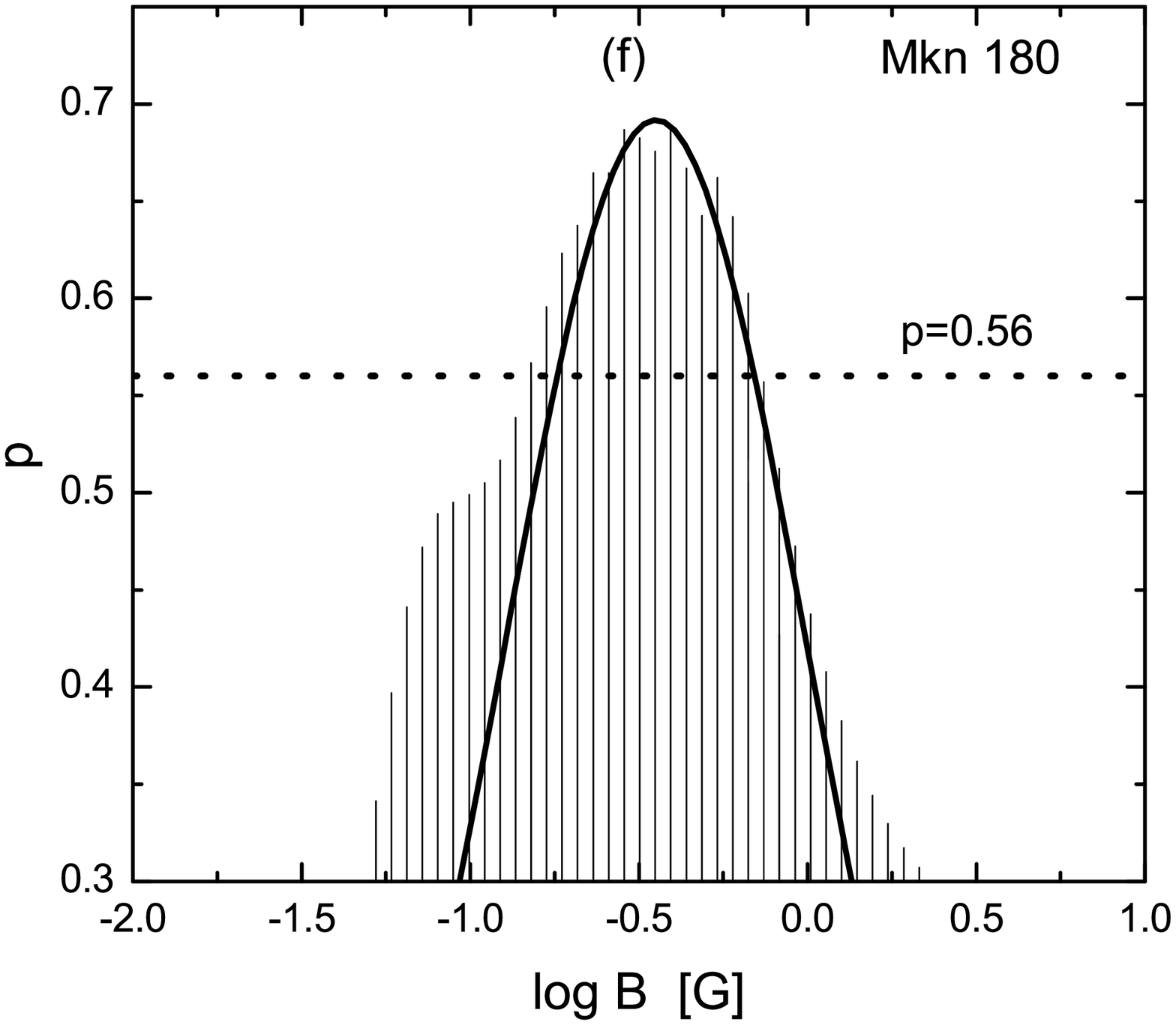}
\includegraphics[angle=0,scale=0.26]{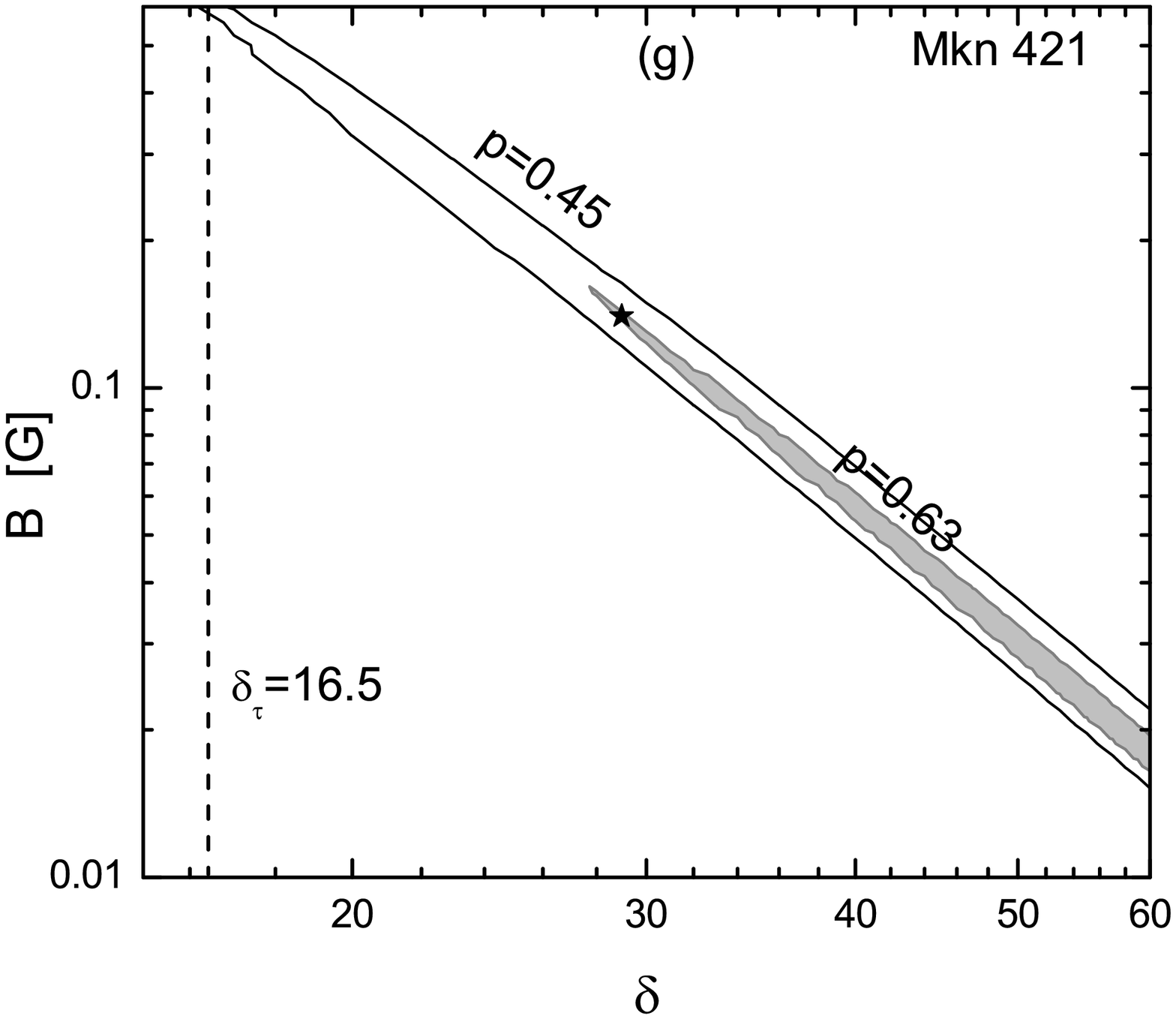}
\hfill
\includegraphics[angle=0,scale=0.26]{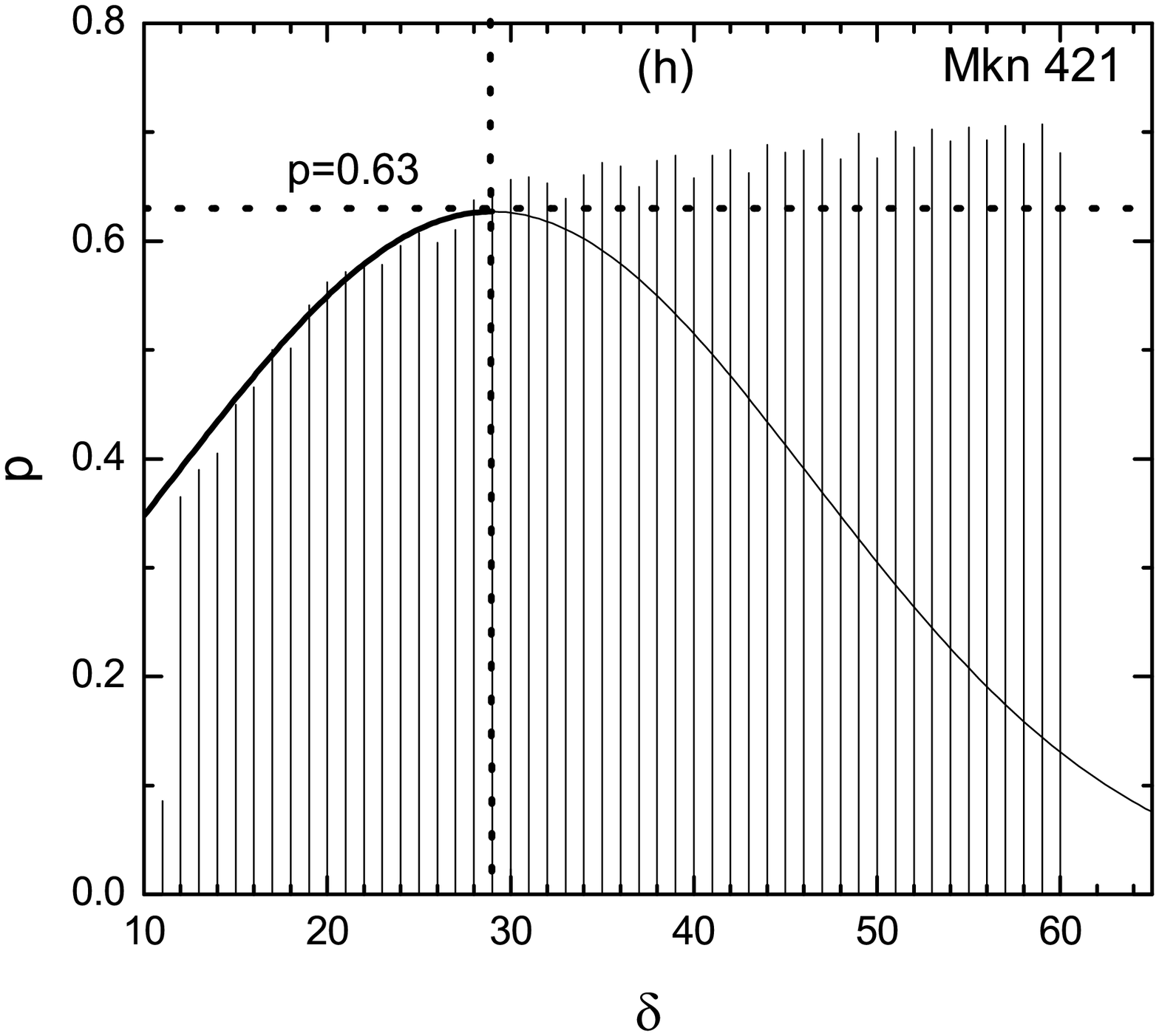}
\hfill
\includegraphics[angle=0,scale=0.26]{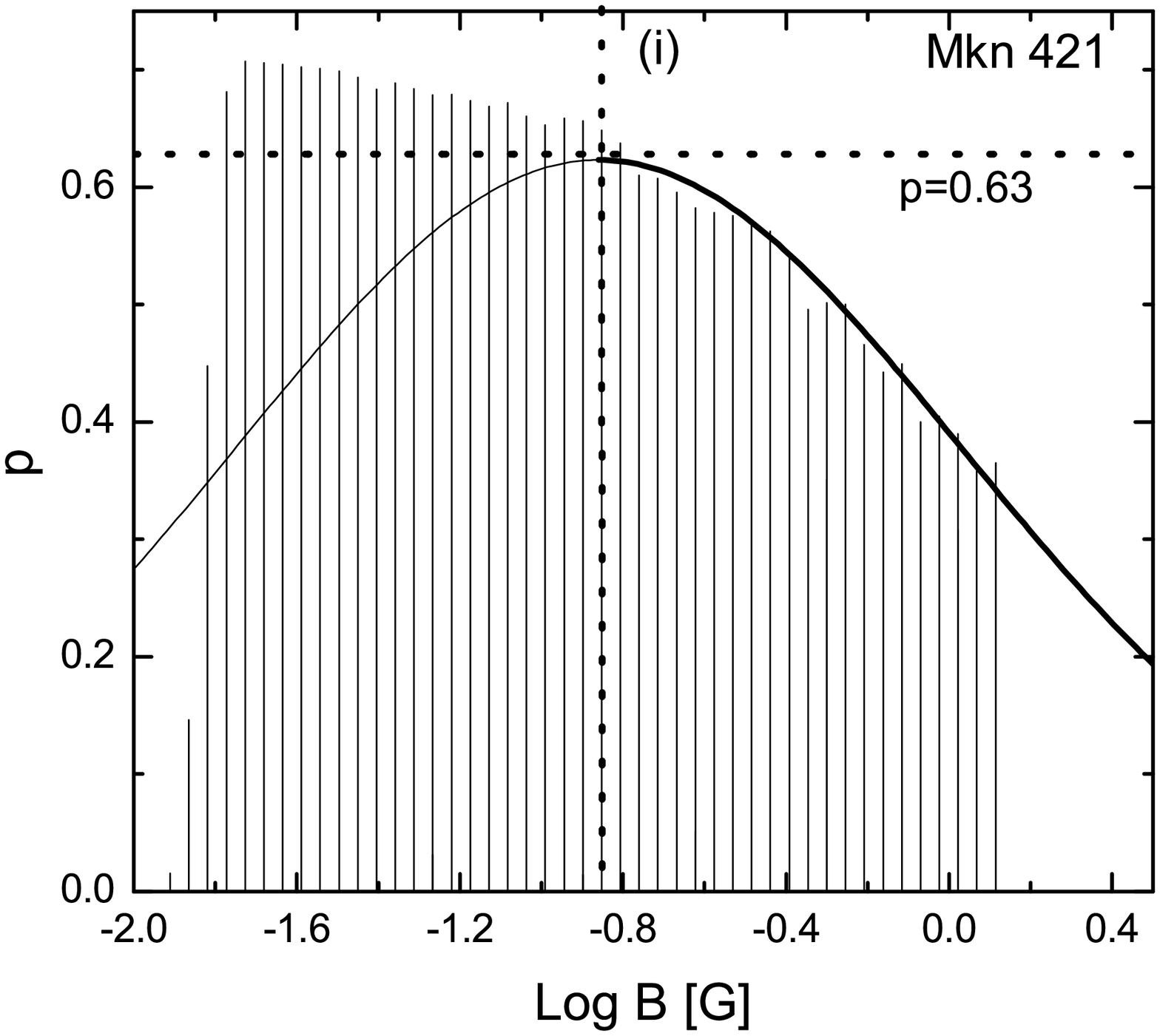}
\caption{Contours of the probability $p$ in the $B-\delta$ plane ({\em panels a, d, g}) along with the one-dimensional distributions of $p$ for $\delta$ and $B$ ({\em panels b, c, e, f, h, i}). Gaussian or one-sided Gaussian function fits to the one-dimensional distributions are shown with thick solid lines ({\em panels b, c, e, f, h, i}). The values of $p$ that correspond to the 1 $\sigma$ region of the Gaussian fits ({\em panels b, c, e, and f}) or the peak of the one-sided Gaussian function ({\em panels h and i}) are shown with dotted lines. The contours of $1\sigma$ are marked with the thick lines in {\em panels a and d}, and the filled contour in {\em Panel g} is for $p$ that corresponds to the peak value of the one-sided Gaussian function. The stars stand for the best fitting parameter set of $\delta$ and $B$. The vertical dashed lines in {\em panels a, d, g} mark the lower limit of $\delta$ derived from the $\gamma$-ray transparency condition. The dotted line in {\em Panel a} is the best fitting line for the 1 $\sigma$ region and the slope is $\sim -2.33$.}\label{Fig:2}
\end{figure*}

\begin{figure*}
\includegraphics[angle=0,scale=0.35]{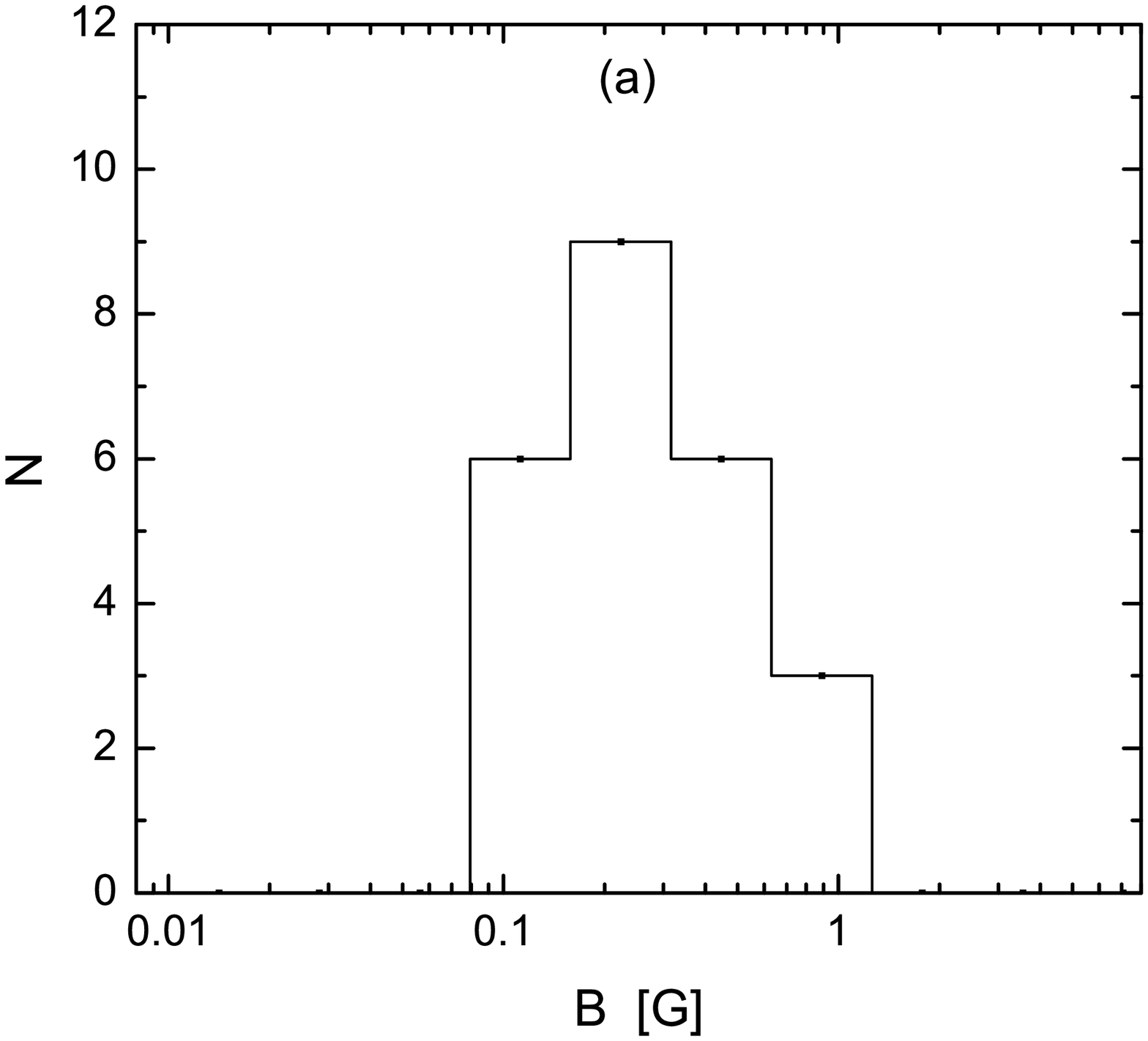}
\includegraphics[angle=0,scale=0.35]{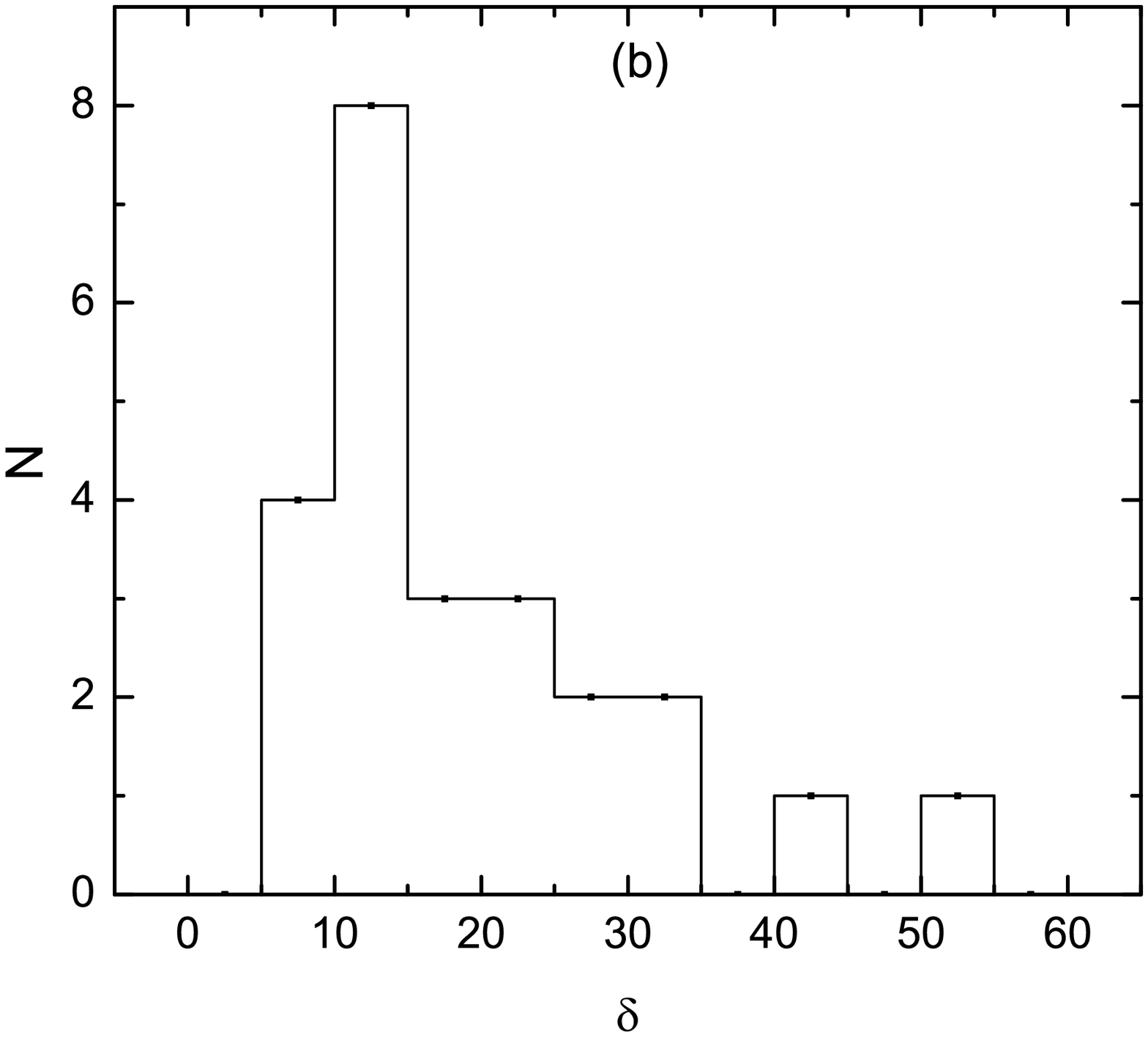}
\includegraphics[angle=0,scale=0.26]{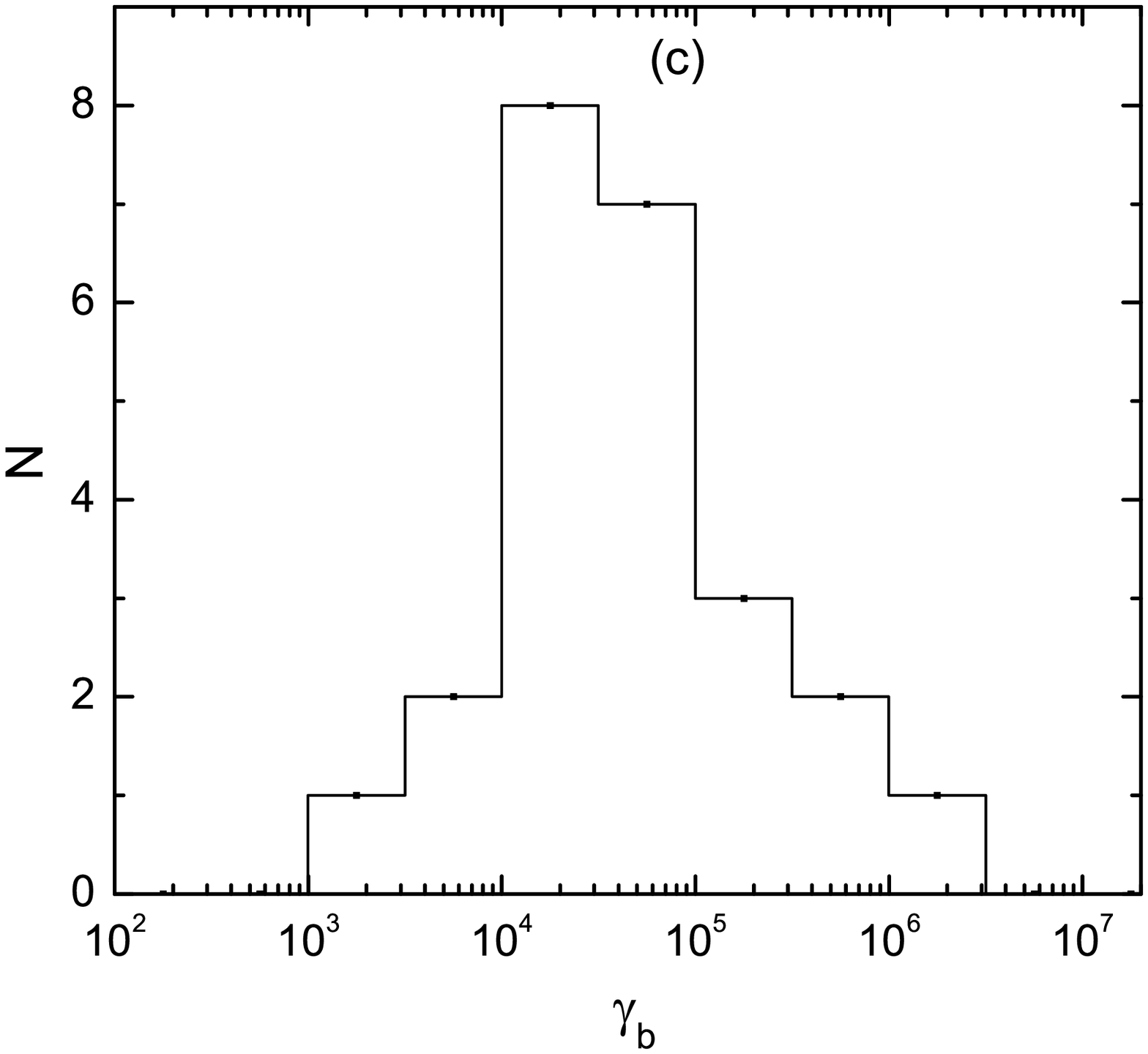}
\includegraphics[angle=0,scale=0.26]{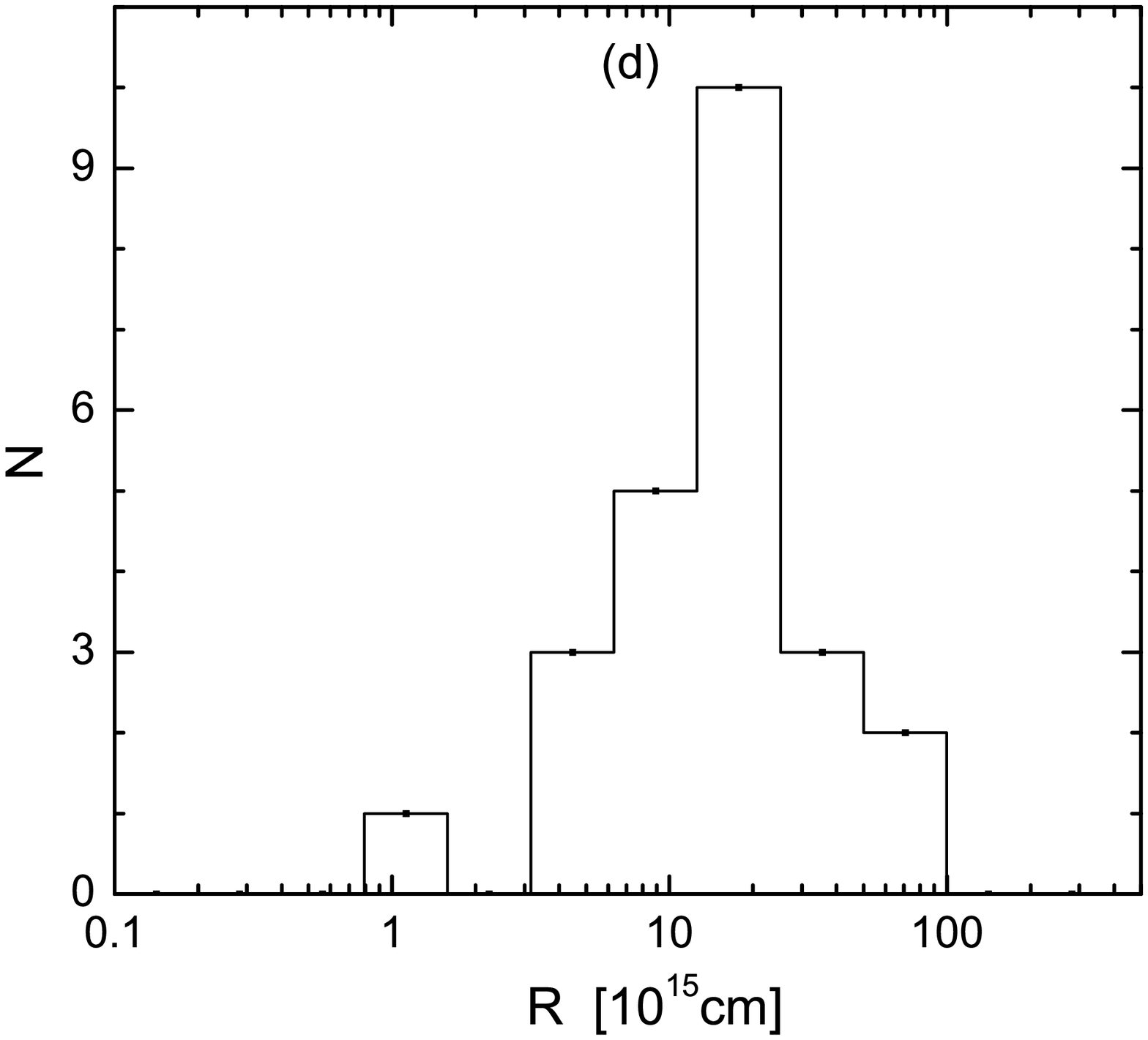}
\includegraphics[angle=0,scale=0.26]{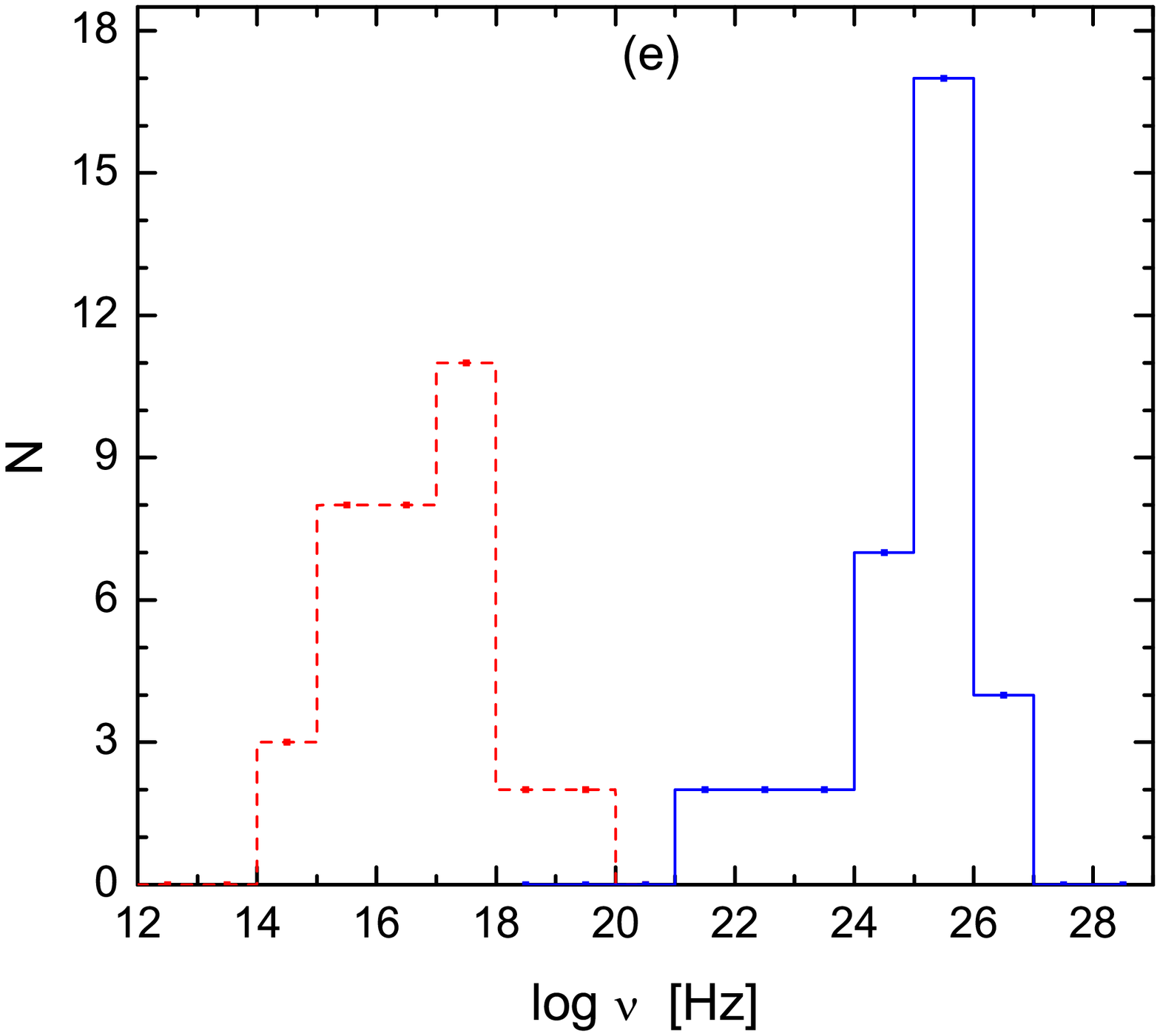}
\caption{Distributions of the magnetic field $B$ ({\em Panel a}), the beaming factor $\delta$ ({\em Panel b}), the break Lorentz factor $\gamma_{\rm b}$ of the electron distribution ({\em Panel c}), and the size of the radiating
region ({\em Panel d}) for the 24 SEDs. {\em Panel e}---Distributions of the synchrotron radiation peak frequency $\nu_{\rm s}$ ({\em red dashed line}) and the inverse Compton scattering peak frequency $\nu_{\rm c}$ ({\em blue solid} line) for the 34 SEDs derived from our theoretical model fits. }\label{Fig:3}
\end{figure*}

\begin{figure*}
\includegraphics[angle=0,scale=0.26]{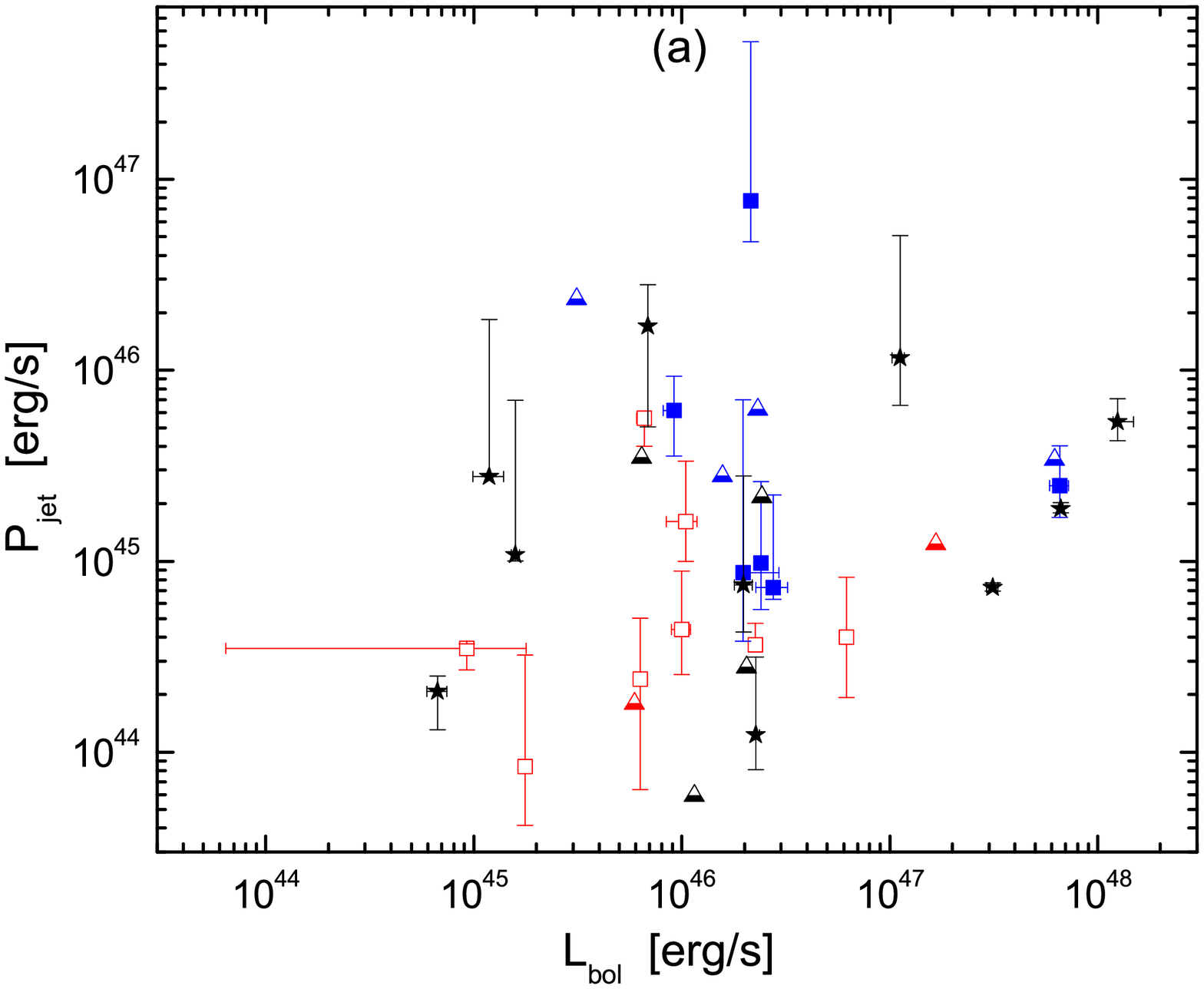}
\includegraphics[angle=0,scale=0.26]{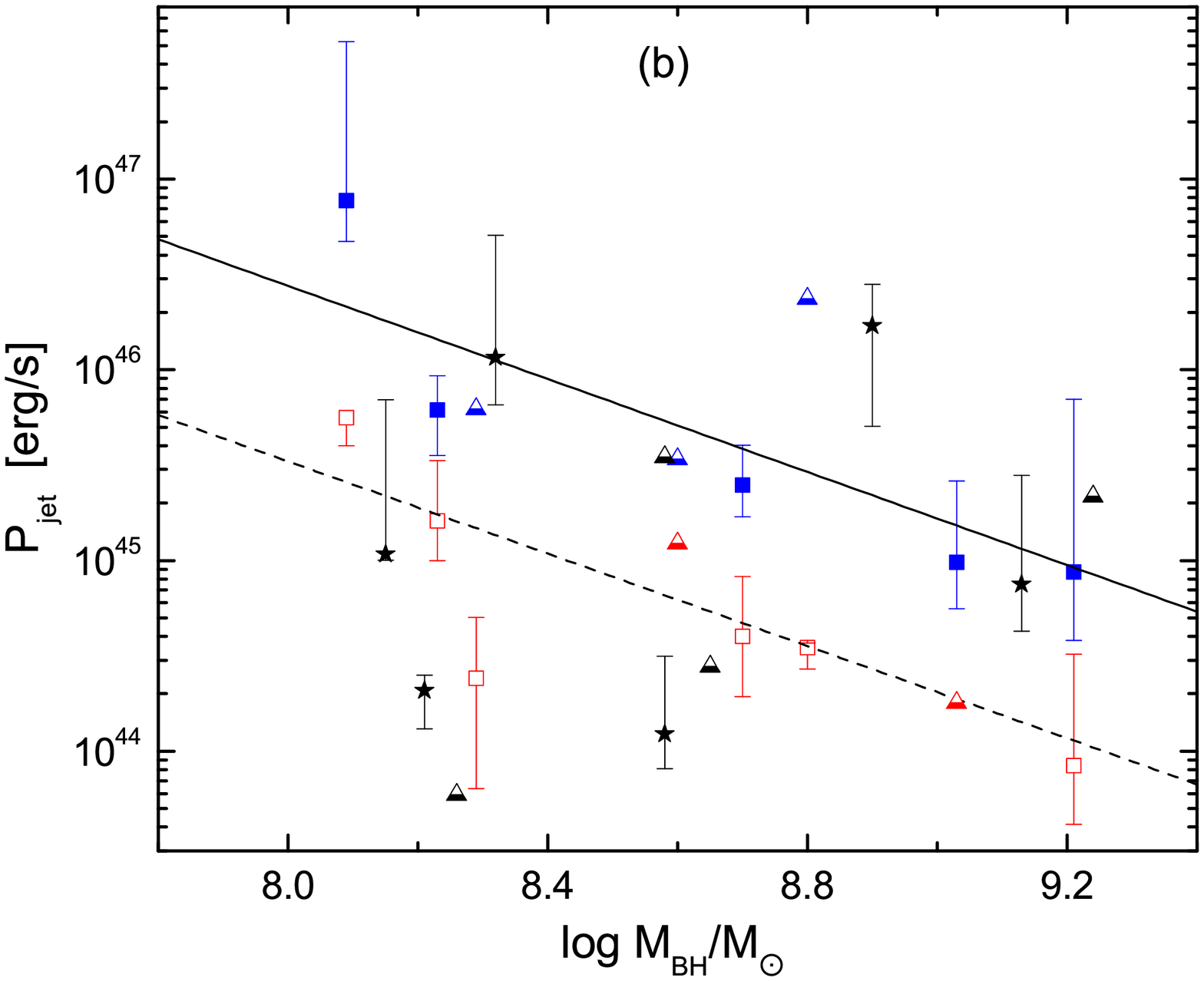}
\includegraphics[angle=0,scale=0.26]{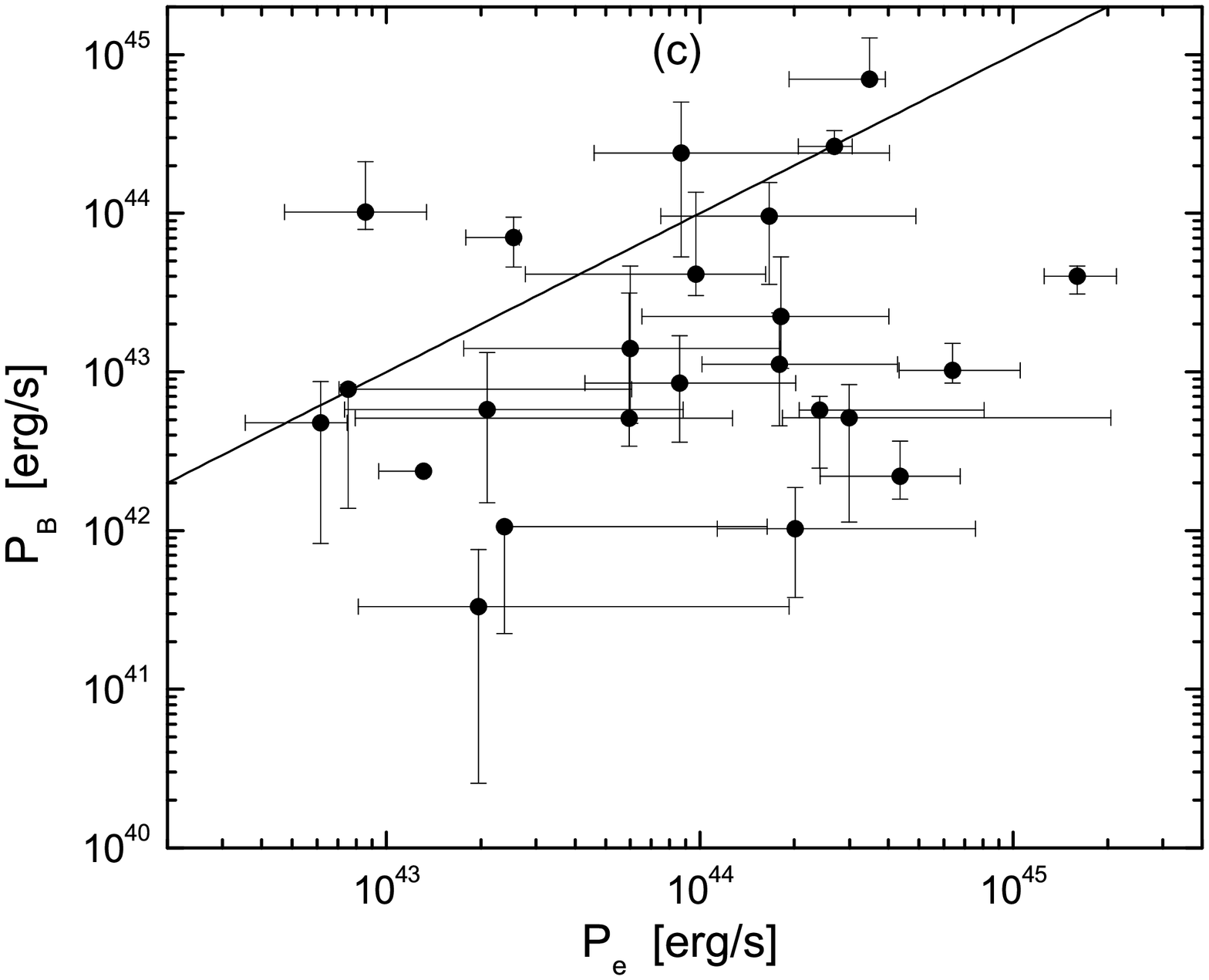}
\caption{{\em Panel a, b}---Jet power as a function of the bolometric luminosity and the BH mass. {\em Blue and red} squares are for the sources in the high and low states, respectively, and {\em black} stars are for the sources with only one SED available. The up-triangle indicates that it is a lower limit of $P_{\rm jet}$ and no error of $L_{\rm bol}$ is available.
The lines are the best fit lines (including the data of limits), which are $\log P_{\rm jet}=(55.2\pm3.1)-(1.21\pm0.35)\log M_{\rm BH}$
for the low state data ({\em dashed line}) and $\log P_{\rm jet}=(56.2\pm4.1)-(1.22\pm0.47)\log M_{\rm BH}$ for
the high state data ({\em solid line}). {\em Panel c}---The powers associated with relativistic electrons ($P_{\rm e}$) as a function of the powers of Poynting flux ($P_B$) in the jets. The line is the equality line.}\label{Fig:4}
\end{figure*}

\begin{figure*}
\includegraphics[width=2.5in,height=7.0in]{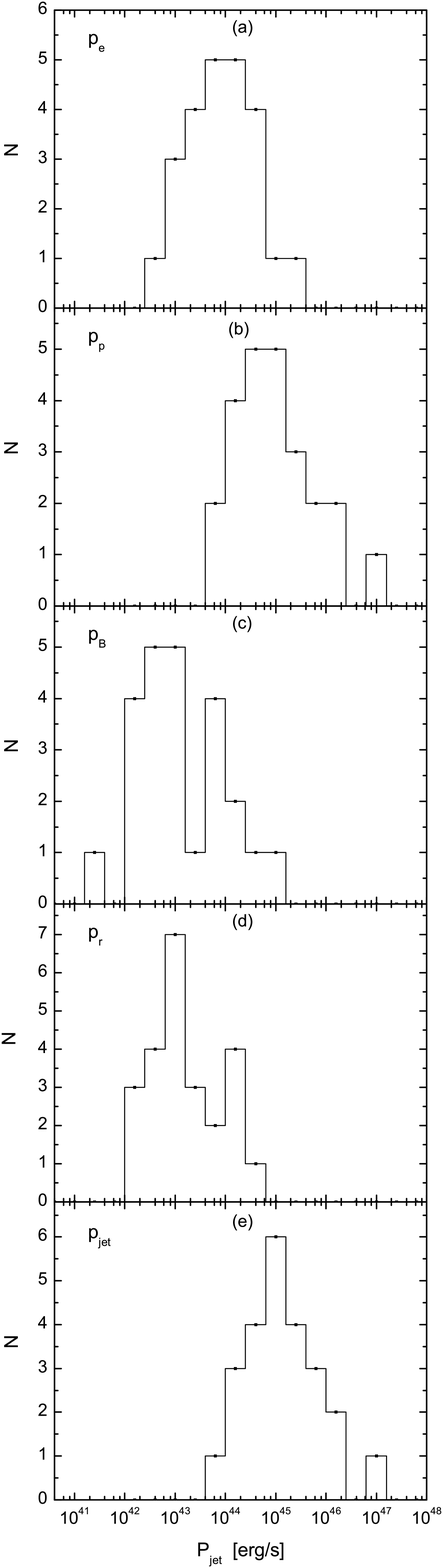}
\includegraphics[width=2.5in,height=7.0in]{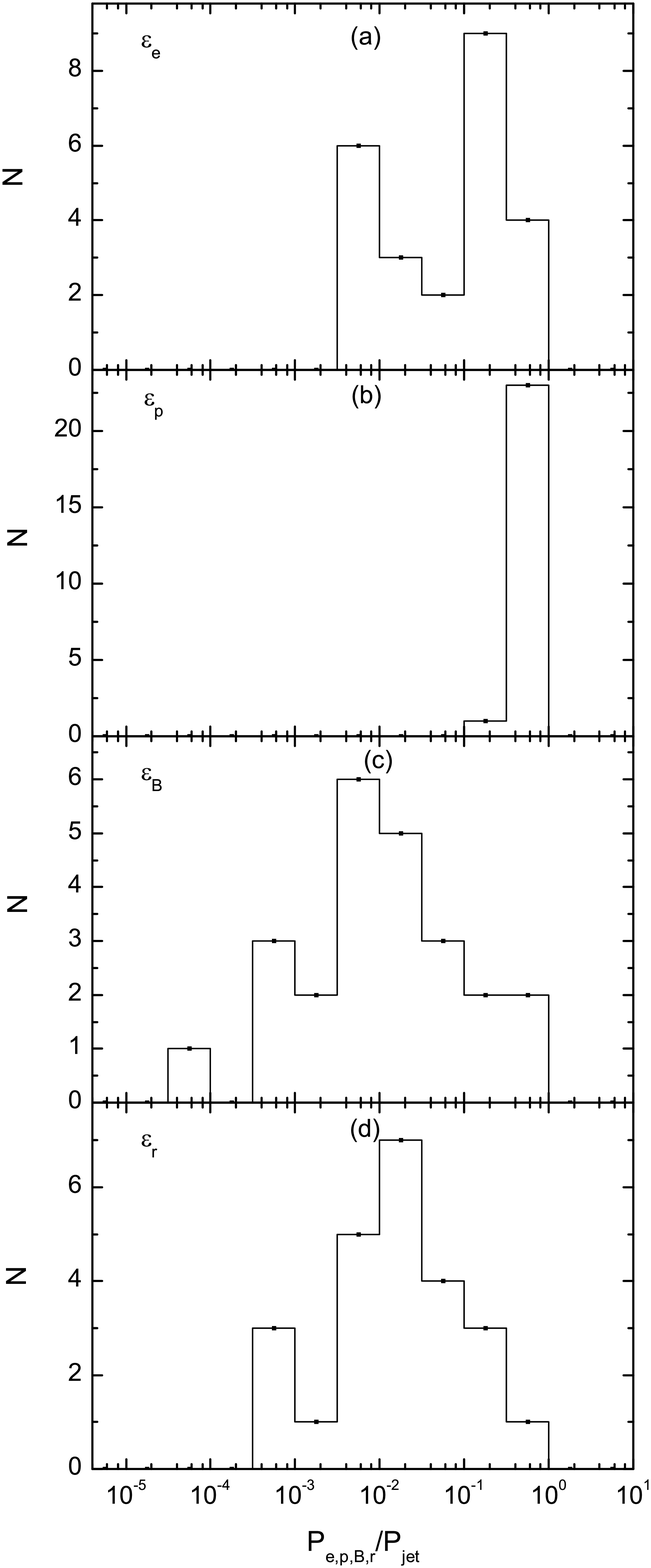}
\caption{Distributions of the powers associated with relativistic electrons $P_{\rm e}$, cold protons $P_{\rm p}$,
Poynting flux $P_B$, radiation component $P_{\rm r}$, and the total power $P_{\rm jet}$ of the jets ({\em left panel}),
together with the distributions of the ratios of these powers to the total jet power ({\em right panels}) for the 24 SEDs.}\label{Fig:5}
\end{figure*}

\begin{figure*}
\includegraphics[width=3.in,height=2.5in]{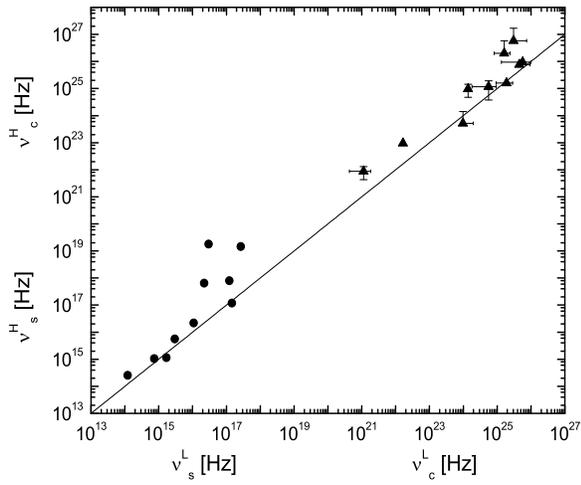}
\caption{Comparison of the peak frequencies $\nu_{\rm s}$ (and $\nu_{\rm c}$) between the high
and low states. {\em Circles} are for $\nu_{\rm s}$ and {\em triangles} are for $\nu_{\rm c}$. Since $\nu_{\rm s}$ is a input parameter, no error of $\nu_{\rm s}$ is available. The {\em solid} line is the equality
line.}\label{Fig:6}
\end{figure*}

\begin{figure*}
\includegraphics[width=2.5 in,height=7.3 in]{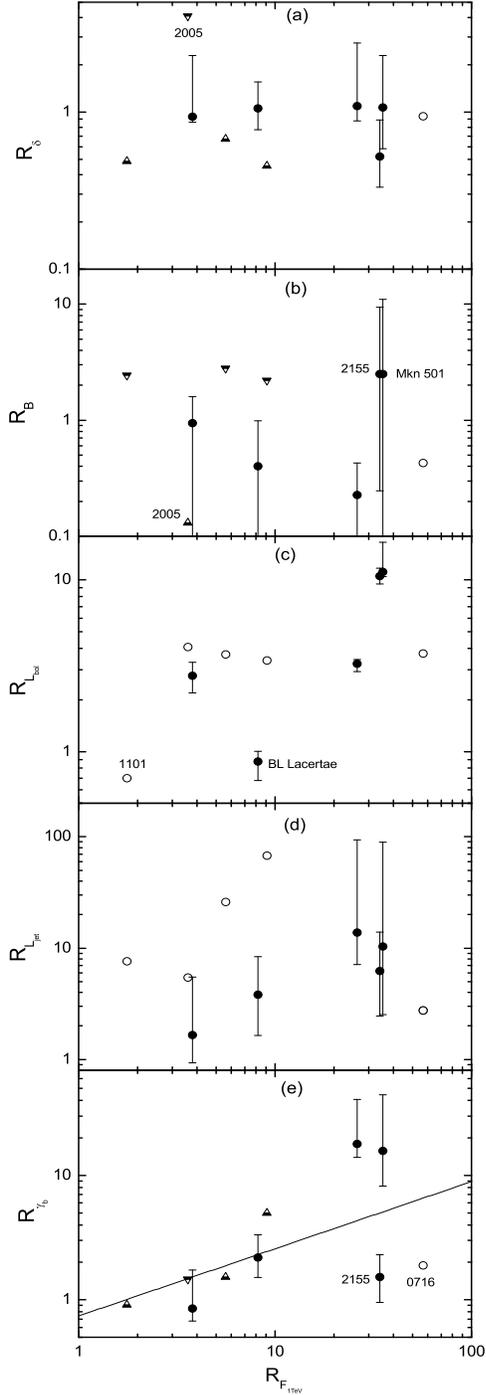}
\caption{Ratios of $R_{\rm x}=x^{\rm H}/x^{\rm L}$ as a function of the ratio $R_{\rm 1\ TeV}=F^{\rm H}_{\rm 1\rm TeV}/F^{\rm L}_{\rm 1 \rm TeV}$, where x=[$\delta$, $B$, $L_{\rm bol}$, $P_{\rm jet}$, $\gamma_{\rm b}$], $F$ is the flux density at 1
TeV, and the letters ``H" and ``L" mark the high and low states, respectively. The up-triangles and down-triangles indicate that they are the lower limits and the upper limits, and the opened circles indicate that no errors are available. The best fit line (including the data of limits) for $R_{\gamma_{\rm b}}-R_{\rm 1\ TeV}$ is $\log R_{\gamma_{\rm b}}=(-0.13\pm0.3)+(0.54\pm0.27)\log R_{\rm 1\
TeV}$.}\label{Fig:7}
\end{figure*}

\begin{figure*}
\includegraphics[width=3.in,height=2.5in]{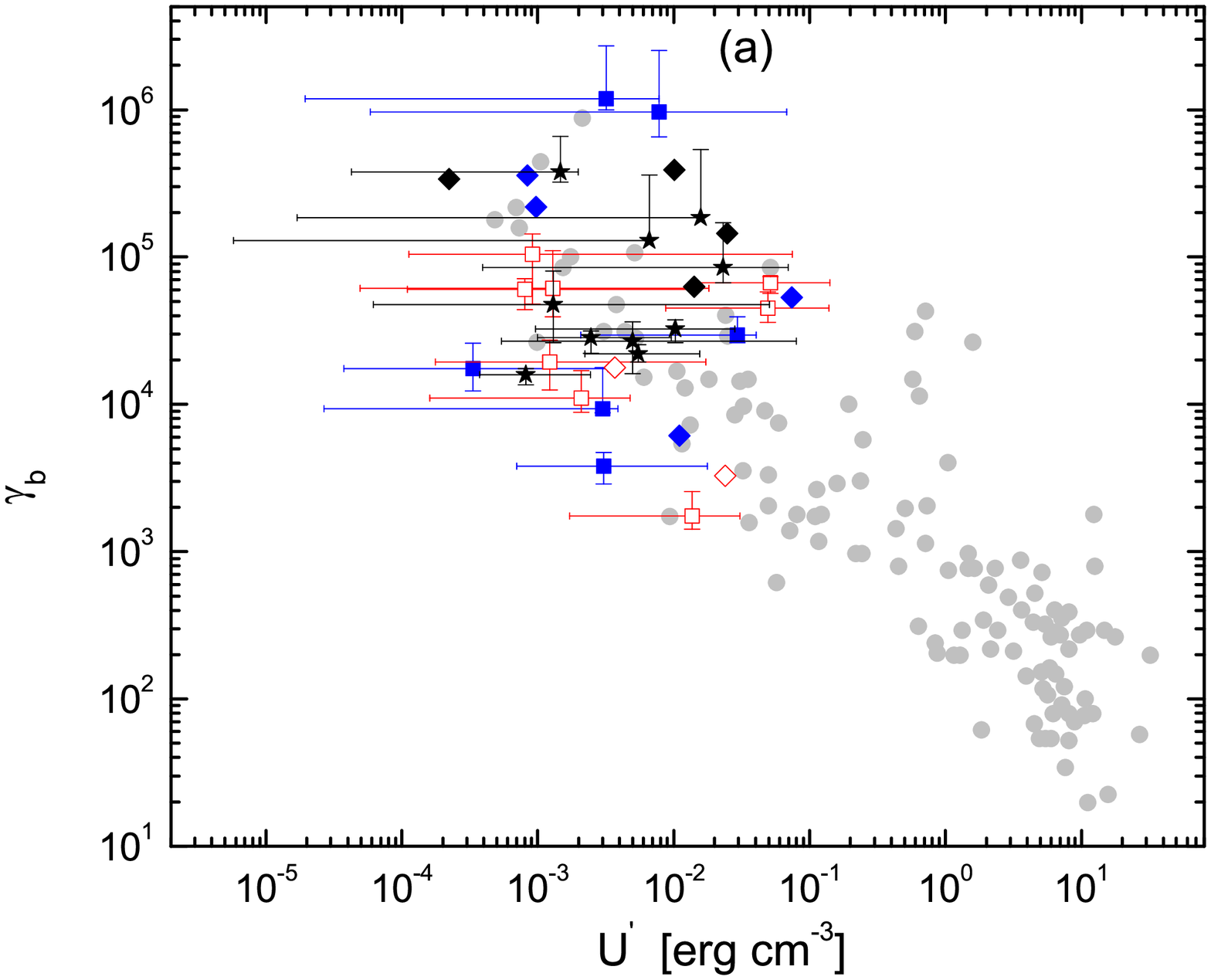}
\includegraphics[width=3.in,height=4.in]{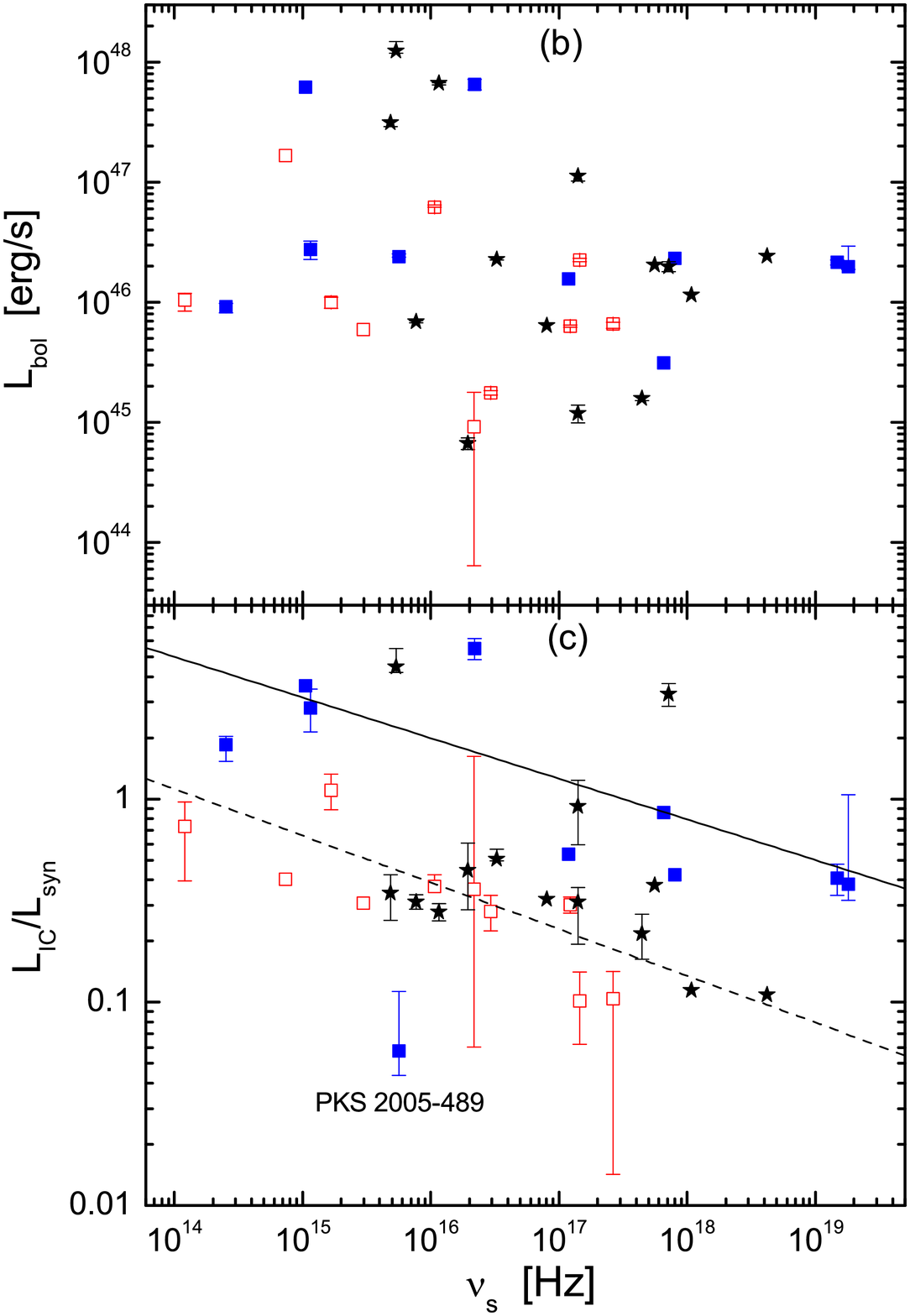}
\caption{{\em Panel a}---$\gamma_{\rm b}$ as a function of energy density $U^{'}$. {\em Panel b} and
{\em Panel c}---Bolometric luminosity ($L_{\rm bol}$) and the ratio of the two bump luminosities $L_{\rm
IC}/L_{\rm syn}$ as a function of synchrotron radiation peak frequency $\nu_{\rm s}$. The {\em gray circles} in {\em Panel a} are
the sample data from Ghisellini et al. (2010), other symbols with other color are the same as in Fig. \ref{Fig:4}, and the diamonds indicate that those data points should move toward the left top of Figure. The best fit lines are $\log L_{\rm IC}/L_{\rm syn}=(3.27\pm1.0)-(0.23\pm0.06)\log {\nu_{\rm
s}}$ for the low state data ({\em dashed line}) and $\log L_{\rm IC}/L_{\rm
syn}=(3.5\pm0.95)-(0.2\pm0.06)\log {\nu_{\rm s}}$ for the high state data (excluding the source PKS 2005-489 in the high state; {\em solid line}).
}\label{Fig:8}
\end{figure*}

\acknowledgments
We thank the anonymous referee for his/her valuable suggestions. This work is supported by the National Natural Science Foundation of China (Grants 11078008, 11025313, 10873002, 10821061, 11133002, 10725313, 10973034, 11133006, 11103060), the National Basic Research Program (973 Programme) of China (Grant 2009CB824800), China Postdoctoral Science Foundation, and Guangxi Science Foundation (2011GXNSFB018063,
2010GXNSFC013011). EWL also acknowledges the special funding from Guangxi Science Foundation with Contract No. 2011-135 and Guangxi SHI-BAI-QIAN project (Grant 2007201).

\label{lastpage}

\end{document}